\documentclass[titlepage,apalike,11pt]{article}

\usepackage[dvipdfmx]{graphicx}
\usepackage[dvipdfmx]{color}

\usepackage[myheadings]{fullpage}
\usepackage{pmetrika}
\usepackage{pmbib}
\usepackage{submit}

\usepackage{bm}
\usepackage{amsmath, amssymb}
\usepackage{bbm}
\usepackage{mathtools}
\usepackage{booktabs}
\usepackage{algorithm}
\usepackage{algorithmic}
\usepackage{subcaption}
\usepackage{here}
\usepackage{rotating}
\usepackage{adjustbox}
\usepackage{empheq}
\usepackage{multirow}
\usepackage{tabularx}
\usepackage{arydshln}

\def\eqref#1{(\ref{#1})}

\begin{document}

\begin{titlepage}


\title{Dynamical Non-compensatory Multidimensional \\
IRT Model Using Variational Approximation}

\author{Hiroshi Tamano}
\affil{The Graduate University for Advanced Studies, SOKENDAI}

\author{Daichi Mochihashi}
\affil{The Institute of Statistical Mathematics}



\linespacing{1}

\end{titlepage}

\setcounter{page}{2}
\vspace*{2\baselineskip}

\RepeatTitle{
Dynamical Non-compensatory Multidimensional IRT Model \\
Using Variational Approximation
}\vskip3pt

\linespacing{2.0} 

\abstracthead
\begin{abstract}

Multidimensional item response theory (MIRT) is a statistical test theory
that precisely estimates multiple latent skills of learners from the responses 
in a test. Both compensatory and non-compensatory models have been proposed
for MIRT: the former assumes that each skill can complement other skills,
whereas the latter assumes they cannot.
This non-compensatory assumption is convincing in many tests that measure
multiple skills; therefore, applying non-compensatory models to such data
is crucial for achieving unbiased and accurate estimation.
In contrast to tests, latent skills will change over time in daily learning.
To monitor the growth of skills, dynamical extensions of MIRT models
have been investigated. However, most of them assumed compensatory
models, and a model that can reproduce continuous latent states of skills
under the non-compensatory assumption has not been proposed thus far.
To enable accurate skill tracing under the non-compensatory assumption,
we propose a dynamical extension of non-compensatory MIRT models by
combining a linear dynamical system and a non-compensatory model.
This results in a complicated posterior of skills, which we approximate
with a Gaussian distribution by minimizing the Kullback–Leibler
divergence between the approximated posterior and the true posterior.
The learning algorithm for the model parameters is derived through
Monte Carlo Expectation Maximization. Simulation studies verify
that the proposed method is able to reproduce latent skills accurately,
whereas the dynamical compensatory model suffers from significant
underestimation errors.
Furthermore, experiments on an actual dataset demonstrate that our
dynamical non-compensatory model can infer practical skill tracing and
clarify differences in skill tracing between non-compensatory and
compensatory models.

\end{abstract}\vspace{\fill}\pagebreak

\section{Introduction}

\begin{figure}
 \centering
 \begin{tabular}{cc}
  \begin{minipage}{0.4\linewidth}
  \includegraphics[width=\linewidth]{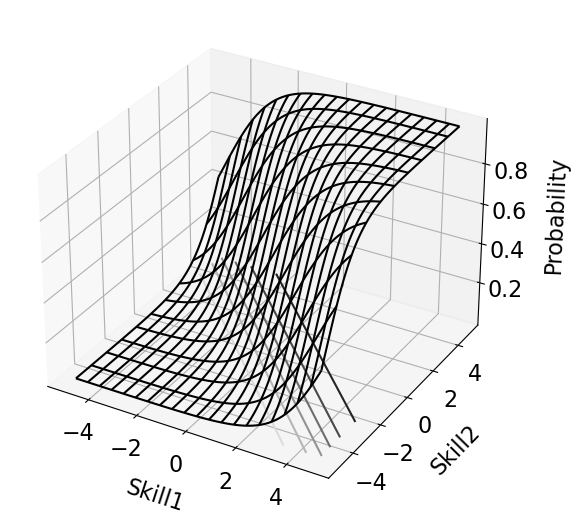}
  \end{minipage}
  &
  \begin{minipage}{0.4\linewidth}
  \includegraphics[width=\linewidth]{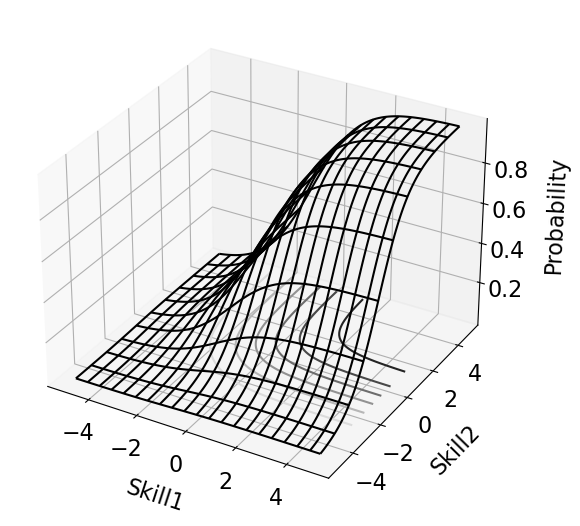}
  \end{minipage}\\
  (a) & (b) \\
  \end{tabular}
\caption{Comparison of item response surfaces between a compensatory
model and a non-compensatory model. (a) represents the compensatory
model and (b) represents the non-compensatory model. The probability
is high when either skill is high in (a), whereas it is high
when both skills are high in (b).
}
\label{pic:cm_ncm_diff}
\end{figure}

Multidimensional item response theory
(MIRT; Sympson 1978; Reckase 1985; Ackerman 1996; Reckase 2009)
is a multidimensional extension of item response theory
(IRT; Rasch 1960; Lord 1980; Embretson and Reise 2013), a statistical test
theory that accurately estimates latent skills of learners from binary responses
avoiding the biases of test designers. MIRT models
can deal with a question associated with multiple skills and are divided
into compensatory models (Bogan and Yen 1983; Reckase 1985; Reckase 2009) and
non-compensatory models
(Sympson 1978; Whitely 1980; Embretson 1984; Embretson and Yang 2013),
according to the relationship between the skills.
Compensatory models assume that each skill can complement other skills, and
learners can thereby answer questions correctly when the sum of the skills
exceeds a threshold.
In contrast, non-compensatory models assume that each skill
cannot complement other skills; thus, each skill must independently exceed
a threshold for learners to be able to answer questions correctly.
Figures \ref{pic:cm_ncm_diff} (a) and (b) show the
item response surfaces for the compensatory and non-compensatory models, respectively.
In this example, the probability of correctly answering a question is high
for compensatory models when either skill level is high.
In contrast, the probability is high for non-compensatory models
only when both skill levels are high.

The assumption of non-compensatory MIRT models is convincing in many tests
associated with multiple skills. For example, understanding or solving an equation
such as $1/5\, x + 3/10 = 2x$ requires both skills of fraction and equation;
learners who only have either one cannot solve this equation.
As can be seen from this example, tagging a problem with multiple skills under
``and'' conditions corresponds to the non-compensatory assumption,
meaning that all the attached skills must be mastered to solve the problem.
(On the other hand, ``or'' conditions correspond to the compensatory assumption.)
Open datasets (Feng et al. 2009; Stamper et al. 2010)
provide sequential logs of item responses answered by learners wherein
problems are tagged with multiple skills. In these datasets,
many problems are tagged under ``and'' conditions.
Because tagging skills correctly is a difficult and time-consuming task,
there are studies to estimate the assignment of skills from data
(Oka and Okada 2021), and one can utilize cloud sourcing as well
(e.g., Amazon Mechanical Turk).

Although this non-compensatory assumption fits many situations,
it has not been frequently
employed thus far. A key reason for this is that achieving accurate estimations
of item response parameters is more difficult in such models.
Bolt (2013) employed Markov chain Monte Carlo (MCMC) to estimate
a model proposed by Whitely (1980) and found that non-compensatory models
require many samples and the estimation is less precise than that of
compensatory models even under such conditions.
Wang and Nydick (2015) compared MCMC and Metropolis– Hastings Robbins–Monro
(MH-RM) and revealed that MCMC is better able to estimate model parameters
across a variety of conditions than MH-RM.
More importantly, compensatory and
non-compensatory models were reported to yield quite similar results
regarding skill estimation (Spray et al. 1990; DeMars 2016).
DeMars (2016) applied both models to the
synthetic item responses generated via non-compensatory models and found that
the difference in the estimation errors was quite small.
Furthermore, the difference vanished when the correlation between the skills
increased.

However, a recent study revealed that compensatory models significantly
underestimate skills (Buchholz and Hartig 2018), highlighting the
importance of using a correct model when the non-compensatory assumption
holds on data. They generated item response data via a non-compensatory
model, assuming a uniform distribution of skills. Subsequently,
they fitted a compensatory model to the data and investigated the
difference between the estimated and actual skills. Their finding was
that the compensatory model largely underestimated high skills for
a specific subgroup of examinees whose actual level for one skill is
high and that for the other is low. Such a subgroup of examinees can
be significantly disadvantaged when a compensatory model is
incorrectly assumed.

In contrast to tests, latent skills change over time in daily learning;
thus, dynamical models have been studied to trace the growth of skills.
They are known as {\it knowledge tracing}
(Corbett and Anderson 1994; Piech et al. 2015), which assumes that
the skill changes even after a question has been answered.
Therefore, the responses collected in learning drills,
where feedback is available after answering each question, are examples
of appropriate target data.
Knowledge tracing models tend to place greater importance on
accurately predicting next responses than estimating the true latent skills.
These prediction models are used in adaptive learning systems to recommend
questions that learners should try next.
Owning to the recent successes of deep learning, many deep neural network
models have been proposed and sophisticated the network structure to fit the
complicated human skill-growth
(Piech et al. 2015; Zhang et al. 2017; Pu et al. 2020).
Since the interpretability of those neural models is low, recent studies
(Yeung 2019; Su et al. 2021) integrated IRT/MIRT and deep neural models
to achieve high accuracy and interpretability.

However, most of the dynamical models employ compensatory models;
a model that can reproduce continuous latent states of skills under
the non-compensatory assumption has not been proposed thus far.
Applying the dynamical extensions of compensatory MIRT models to
situations wherein the non-compensatory assumption holds is expected
to result in significant underestimation of skills, according to
the results in the non-dynamical situation (Buchholz and Hartig 2018).
Non-compensatory models are also employed in
{\it cognitive diagnostic models}
(Leighton and Gierl 2007; Templin and Henson 2010),
which are the statistical models to diagnose skills in an
educational test. Their dynamical extensions have been developed
(Li et al. 2016; Wang et al. 2018; Zhan et al. 2019); however, their
latent skills are binary, i.e., mastery or non-mastery.
Chen et al. (2018) employed a non-compensatory model in a deep
neural network; however, they did not evaluate the accuracy of
latent skills because their main concern was in the accuracy of
predicting next responses.
The dynamical model that can reproduce the latent skills
under the non-compensatory assumption is crucial for accurately 
tracing skills.

\begin{figure}
    \centering
    \includegraphics[width=.9\linewidth]{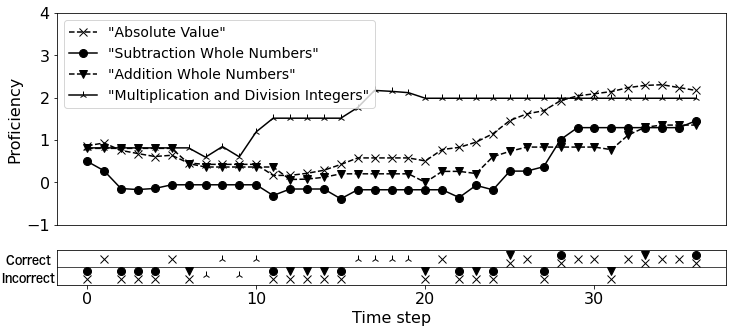}
    \caption{
    Visualization of skill tracing for a learner obtained by the
    proposed dynamical non-compensatory MIRT model on the ASSISTments
    2009-2010 dataset. Bottom panel shows responses to the questions,
    where successes and failures correspond to the upward and downward
    positions of markers, and the required skill for each question is
    denoted as the marker type.
    }
    \label{pic:skill_tracing}
\end{figure}

To enable accurate and real-time skill tracing in situations wherein
the non-compensatory assumption holds,
we propose the dynamical extension of non-compensatory MIRT models.
Our proposed model is a combination of a linear dynamical system and a
non-compensatory model.
This results in a complicated posterior of the latent skills; therefore,
we approximate it with a Gaussian distribution by minimizing
the Kullback–Leibler (KL) divergence between the approximated posterior
and the true posterior.
The estimation method for the model parameters is derived
through the Monte Carlo Expectation Maximization (EM) algorithm
(Wei and Tanner 1990).
Simulation studies verify that the proposed method is able to reproduce
latent skills accurately, whereas the dynamical compensatory model suffers
from significant underestimation errors. Furthermore, experiments on an
actual dataset demonstrate that our dynamical non-compensatory model
can infer practical skill tracing as shown in Figure \ref{pic:skill_tracing}
and clarify differences in skill tracing between non-compensatory and
compensatory models.

The remainder of this paper is organized as follows.
Section \ref{sec:model} briefly reviews the compensatory
and non-compensatory MIRT models and
the linear dynamical system, followed by our proposed dynamical non-compensatory
MIRT model. Section \ref{sec:inference} describes the inference of the
posterior of latent skills
and the model parameters. Section \ref{sec:eval_skill} evaluates
the accuracy of the skill estimation on simulation data
as well as the model parameters. The accuracy of predicting responses
on simulation data is evaluated in Section \ref{sec:prediction_simulation}.
The Gaussian approximation of the posterior is evaluated in
Section \ref{sec:eval_posterior}. Section \ref{sec:real_data} shows
experimental results on actual data.
Finally, we discuss the advantages of the proposed model
in Section \ref{sec:discussion} along with possible future extensions.

\section{Model}\label{sec:model}

Our proposed model is based on MIRT and linear dynamical systems
(LDS; Kalman 1960; Ghahramani and Hinton 1996; Bishop 2006).
We briefly describe MIRT and LDS before introducing the
dynamical non-compensatory MIRT model.

\subsection{Multidimensional Item Response Theory}\label{subsec:mirt}

MIRT is a multi-skill extension of IRT that was used to estimate the proficiency
of examinees and the difficulty of questions.
MIRT measures proficiency and difficulty for multiple skills from an examination
result; it is typically expressed as binary matrix $Y$ whose $(i,j)$-th entry 
$y_{i,j} \in \{0,1\}$ represents the correct or incorrect answer to question $i$
of examinee $j$. The questions in the examination are assumed to require
multiple skills.

MIRT involves compensatory models and non-compensatory models.
Compensatory models assume that each skill can complement other skills.
In the compensatory model, the probability that examinee $j$ can answer
question $i$ correctly is defined as
\begin{equation}\label{eqn:comp_irt}
    p(y_{i,j}=1|\bm{z_j}, \bm{a_i}, b_i) = 
    \frac{1}{1+\exp(-(\sum_{k=1}^K a_{i,k}z_{j,k} - b_i))},
\end{equation}
where $\bm{z}_j$ is a $K$-dimensional proficiency vector for skills of 
examinee $j$, $K$ is the number of skills, $k$ is the skill index,
$\bm{a}_i$ is a $K$-dimensional vector of the discrimination parameters
of question $i$, and $b_i$ represents the difficulty parameter of question $i$.
This is called the two-parameter logistic model by the discrimination and
difficulty parameters. 
The proficiency is summed up; thus, each skill could complement other skills.
Figure \ref{pic:cm_ncm_diff} (a) shows a two-dimensional example of
Eq. \eqref{eqn:comp_irt}.
In contrast, non-compensatory models assume that individual skills
cannot complement other skills. In the non-compensatory model,
the probability that examinee $j$ can answer question $i$ correctly is defined as
\begin{equation}\label{eqn:ncomp_irt}
    p(y_{i,j}=1|\bm{z_j}, \bm{a_i}, \bm{b_i}) = 
    \prod_{k=1}^{K} \frac{1}{1+\exp(-a_{i,k} (z_{j,k} - b_{i,k}))},
\end{equation}
where difficulty $\bm{b}_i$ for question $i$ is now a $K$-dimensional vector
whose element denotes the difficulty for each skill.
Here, the proficiency is multiplied; thus, individual skills cannot complement
other skills. Figure \ref{pic:cm_ncm_diff} (b) shows a two-dimensional example
of Eq. \eqref{eqn:ncomp_irt}. Our model employs the non-compensatory model
to handle situations where one skill cannot complement the other
skills.

\subsection{Linear Dynamical System}\label{subsec:lds}

\begin{figure}[tb]
  \centering
  \begin{minipage}{.45\linewidth}
    \includegraphics[width=\linewidth]{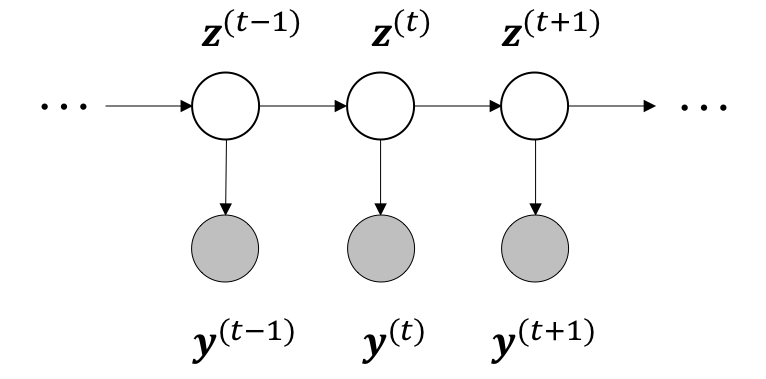}
    \subcaption{Graphical model of LDS}
  \end{minipage}
  \begin{minipage}{.5\linewidth}
    \includegraphics[width=\linewidth]{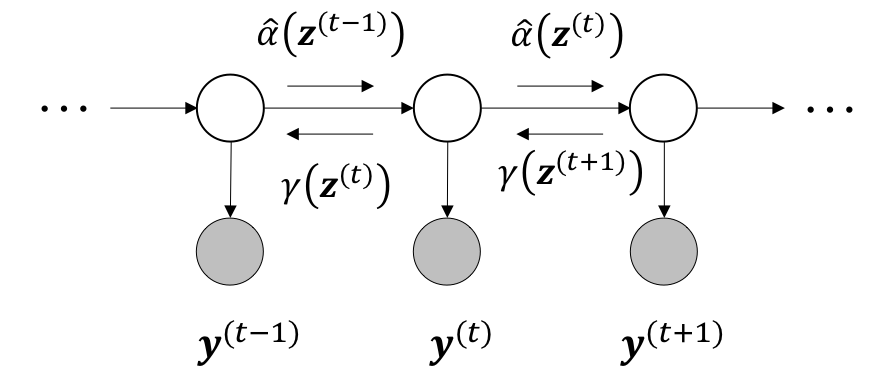}
    \subcaption{Message passing in E-step}
  \end{minipage}
\caption{
Overview of the message passing process.
(a) is a graphical representation of LDS. The transition
of the latent states only depends on the previous state and each state
emits observable data; the latent and observable variables are
represented as white and gray circles, respectively. (b) shows the
forward $\hat{\alpha}$ message passing and the backward $\gamma$
message passing to calculate the posterior of the state variable
$\bm{z}^{(t)}$.
}
\label{pic:lds}
\end{figure}

LDS has been employed to infer changes of latent
states from observations with noise.
The generative model for LDS is defined by the initial state probability,
state transition probability, and emission probability, which are given below.
\begin{empheq}[left={\empheqlbrace\,}]{align}
  p(\bm{z}^{(1)}) &= \mathcal{N}(\bm{z}^{(1)} | \bm{\mu_0}, P_0), \\
  p(\bm{z}^{(t)}|\bm{z}^{(t-1)}) 
  &= \mathcal{N}(\bm{z}^{(t)}| D\bm{z}^{(t-1)}, \Gamma), \\
  p(\bm{y}^{(t)}|\bm{z}^{(t)})
  &= \mathcal{N}(\bm{y}^{(t)}| C\bm{z}^{(t)}, \Sigma), 
\end{empheq}
where $\bm{z}^{(t)}$ and $\bm{y}^{(t)}$ denote the latent state and observed data
at time step $t$, respectively, and 
$\bm{\theta} = \{D, \Gamma, C, \Sigma, \bm{\mu}_0, P_0\}$ is the set of model
parameters.
The graphical model of LDS is shown in Figure \ref{pic:lds} (a).
The initial state $\bm{z}^{(1)}$ is generated by a Gaussian distribution 
$\mathcal{N}(\bm{z}^{(1)}| \bm{\mu_0}, P_0)$ where $\bm{\mu_0}$ denotes mean and 
$P_0$ denotes covariance.
The state transition is linear transformation $D$ with Gaussian noise of zero mean
and covariance $\Gamma$. Sample $\bm{y}^{(t)}$ is observed after linear
transformation $C$ has been applied to the current state, $\bm{z}^{(t)}$,
with Gaussian noise of zero mean and covariance $\Sigma$.
The posterior of the latent states and model parameters can be estimated
by a Kalman smoother and an EM algorithm, respectively.

\subsection{Dynamical Non-compensatory MIRT}\label{subsec:proposed}

Our proposed model is a dynamical extension of non-compensatory MIRT.
It can be considered as a combination of an MIRT model and LDS.
We describe the target data and the generative model below.

Our target dataset
$\{ (i_j^{(t)}, y_j^{(t)}) \}_{j=1, ...,N,\ t=1, ... ,T_j}$ is
a sequence of the log of the answers to the questions by different
learners over time. $N$ and $T_j$ denote the number of learners and 
the number of logs for learner $j$, respectively.
$i_j^{(t)} \in \{1,...,M\}$ denotes the index of the question answered
by learner $j \in \{1, ..., N\}$ at time step $t \in \{1,...,T_j\}$, 
also denoted by $i(j,t)$; we abbreviate it to $i$ when it is clear
from the context. $M$ denotes the number of questions.
$y_j^{(t)} \!\in \{0,1\}$ represents
whether learner $j$ answered question $i(j,t)$ correctly or not at
time step $t$.
The difference between our proposed model and the ordinary LDS is that
observation $y_j^{(t)}$ is binary associated with the index
of the question answered,
and each learner has a different sequence of questions.

The generative model of the dynamical non-compensatory MIRT is defined 
in a manner similar to that of LDS, except that the emission probability
here is a non-compensatory item response model.
\begin{empheq}[left={\empheqlbrace\,}]{align}
  p(\bm{z}_j^{(1)}) &= \mathcal{N}(\bm{z}_j^{(1)} | \bm{\mu}_0, P_0),
    \label{eqn:init} \\
  p(\bm{z}_j^{(t)}|\bm{z}_j^{(t-1)}) &=
    \mathcal{N}\left(\bm{z}_j^{(t)} \middle| D_{i(j,t-1)} \bm{z}_j^{(t-1)} 
    + \begin{bmatrix}
          \vdots \\
          \bm{\beta}_k^\top \bm{x}_{j,k}^{(t)} \\
          \vdots
        \end{bmatrix}, \Gamma_j^{(t)} \right), \label{eqn:transition}\\
  p(y_j^{(t)}|\bm{z}_j^{(t)}) &=
  \mbox{Bernoulli}\left(y_j^{(t)} \middle|
  \prod_{k: Q_{i,k}=1} \sigma\left(
    a_{i,k}(z_{j,k}^{(t)} - b_{i,k})
    \right)
  \right). \label{eqn:emit}
\end{empheq}
$\bm{z}_j^{(t)}$ denotes the latent skill state of learner $j$ at time step $t$.
$\bm{z}_j^{(t)}$ is a $K$ dimensional vector and $z_{j,k}^{(t)}$ denotes
the proficiency of the $k$-th skill.
Initially, the state of the learner is drawn from the Gaussian
with mean $\bm{\mu}_0$ and covariance $P_0$ in Eq. \eqref{eqn:init}.
It then transits by a linear transformation
by $D_i$ and $\bm{\beta}_k^\top \bm{x}_{j,k}^{(t)}$ with Gaussian noise of
zero mean and covariance $\Gamma_j^{(t)}$ in Eq. \eqref{eqn:transition};
$D_i$ is a diagonal matrix and $k$-th diagonal element $d_{i,k}$ is fixed to one 
when skill $k$ is not required for question $i$.
$\bm{x}_{j,k}^{(t)} \in \mathbb{R}^{F_k}$ is an $F_k$ dimensional covariate variable.
This is assumed to be given as extra data.
Covariance $\Gamma_j^{(t)}$ is also diagonal, and $(\Gamma_j^{(t)})_{k,k} = \gamma_k$
if question $i(j,t-1)$ or $i(j,t)$ requires skill $k$; otherwise,
$(\Gamma_j^{(t)})_{k,k} \!= 0$.
The response of a learner to question $i$ is drawn from the Bernoulli distribution
in Eq.~\eqref{eqn:emit}; $a_{i,k}$ and $b_{i,k}$ denote
item discrimination and item difficulty, respectively, and
$\mbox{Bernoulli}(y|p)$
denotes the Bernoulli distribution of random variable $y$ with head probability $p$.
$\sigma(x) = 1 / (1 + \exp(-x))$ is a sigmoid function.
$Q_{i,k} \in \{0,1\}$ denotes the question-skill mapping
whether question
$i$ requires skill $k$ and it is called the Q-matrix (Tatsuoka~1983).
We assume the Q-matrix is given in advance. In our model, the parameters of
the initial state probability, the state transition
probability, and the emission probability are listed as follows.
\begin{empheq}[left={\empheqlbrace\,}]{align}
  \bm{\theta}^{init} &= \left\{\,\bm{\mu}_0, P_0 \,\right\},
  \label{eqn:init_params} \\
  \bm{\theta}^{trans}
  &= \left\{\,d_{i,k} \middle| i \in \{1,...,M\}, k \in \{1,...,K\},
     Q_{i,k} =1 \,\right\}
     \cup \left\{\,\gamma_k, \bm{\beta}_k \middle| k \in \{1,...,K\} \,\right\},
  \label{eqn:trans_params} \\
  \bm{\theta}^{emit}
  &= \left\{\,a_{i,k}, b_{i,k} \middle| i \in \{1,...,M\}, k \in \{1,...,K\},
       Q_{i,k} =1 \,\right\}.
  \label{eqn:emit_params}
\end{empheq}

The covariate variable, $\bm{x}_{j,k}^{(t)}$, can be designed to fit state
transitions.
The bias term of the linear transformation for each question is an example of this.
Let $F_k$ be the number of questions requiring skill $k$, and let
$\bm{x}_{j,k}^{(t)}$ be a binary vector, where each element of the vector
corresponds to each question requiring skill $k$.
When learner $j$ answers question $i$ requiring skill $k$ at time step
$t-1$, let a corresponding element of $\bm{x}_{j,k}^{(t)}$ for question $i$ be one
and let the other elements be zero.
The linear transformation in Eq. \eqref{eqn:transition} can be
simplified to $D_{i(j,t-1)} \bm{z}_j^{(t-1)} + \bm{\beta}'_{i(j,t-1)}$,
where $\bm{\beta}'_i$ is a $K$ dimensional vector and
the elements corresponding to the non-required skills to question $i$ are zero.
Another example involves including a forgetting factor
in the covariates, as given by $2^{-\frac{\Delta}{h}}$, where $\Delta$ is the
elapsed time since the last practice of the skill and $h$ is half the time of
memory. Various covariate variables can be considered
(e.g., time passed since a learner studied a skill and the number of
practicing questions requiring a skill).

When we fix the discrimination parameters $a_{i,k}$ to one, our model is
considered as the combination of the LDS and the multicomponent latent trait
model (Whitely 1980). On the other hand, our model with discrimination
parameters is considered as the combination of the LDS and 2-parameter
logistic model (Sympson 1978).
Section \ref{sec:eval_skill} includes the simulation studies in both settings,
and the former combination (i.e., fixing $a_{i,k}$ to one) turns out to be
sufficient to infer the skills in most cases.

\section{Inference}\label{sec:inference}

We derive methods for estimating the posterior of the skill state and
the model parameters based on
the LDS methods: forward-backward algorithm and EM algorithm, respectively
(Kalman 1960; Dempster 1977; Ghahramani and Hinton 1996; Bishop 2006).
In the forward-backward algorithm, an $\hat{\alpha}$ message is passed forward
and a $\gamma$ message is passed backward as shown in Figure \ref{pic:lds} (b).
The $\hat{\alpha}$ message is the posterior of the latent state, which can be
exactly calculated as a Gaussian from a Gaussian likelihood and a Gaussian prior
in LDS. However, the $\hat{\alpha}$ message in our model cannot be exactly
calculated due to the non-conjugacy between the likelihood of non-compensatory
item response function in Eq. \eqref{eqn:emit} and a Gaussian prior.
To address this problem, we propose to approximate the $\hat{\alpha}$ message as
a Gaussian distribution. Once it is approximated, the $\gamma$ message
can be calculated using the same method as LDS. The estimation method for the
model parameters can be derived through the EM algorithm.
In the following sections, we introduce the Gaussian approximation of the
$\hat{\alpha}$ message, followed by the EM algorithm to estimate 
the posterior and the model parameters using the Gaussian approximation.

\subsection{Gaussian Approximation of $\hat{\alpha}$ Message}\label{subsec:gauss_apx}

We derive a Gaussian approximation of the $\hat{\alpha}$ message, which is employed
in the next section for estimating the posterior of the skill state. In our model, the 
$\hat{\alpha}$ message is proportional to the product of the likelihood of a non-compensatory item response function and a Gaussian prior.
Given likelihood $p(y|\bm{z})$ and prior $p(\bm{z})$ defined as
\begin{empheq}[left={\empheqlbrace\,}]{align}
  p(y|\bm{z}) &= \mbox{Bernoulli}\left(y \middle| \prod_{k} \sigma\left(a_k(z_k - b_k)\right)\right),
  \label{eqn:likelihood_of_gauss_apx} \\
  p(\bm{z}) &= \mathcal{N}(\bm{z}| \bm{m},G),
  \label{eqn:prior_of_gauss_apx}
\end{empheq}
here we consider the problem of finding the approximated posterior
$q(\bm{z})=\mathcal{N}(\bm{z}| \bm{\mu}, V)$ minimizing the KL-divergence between
$q(\bm{z})$ and $p(\bm{z}|y)$.

The KL-divergence is calculated as 
\begin{equation}
  \mbox{KL}\left(q(\bm{z})||p(\bm{z}|y)\right)
  = \mbox{KL}\left(q(\bm{z})||p(\bm{z})\right)
  - \int q(\bm{z}) \ln p(y|\bm{z}) d\bm{z} + \mbox{const}, \label{eqn:kl-div}
\end{equation}
where the constant term is with respect to $q(\bm{z})$.
The first term on the right-hand side of Eq. \eqref{eqn:kl-div} is the
KL-divergence between the approximated posterior and the prior, which can be calculated 
analytically as follows.
\begin{equation}
  \mbox{KL}\left(q(\bm{z})||p(\bm{z})\right)
  = \frac{1}{2} \left\{ \ln \frac{|G|}{|V|} + tr(G^{-1}V) 
    + (\bm{m}-\bm{\mu})^\top G^{-1} (\bm{m}-\bm{\mu}) \right\}.
\end{equation}
The second term is the expectation of the log likelihood with respect to 
the approximated posterior, which cannot be obtained analytically. Here, we
employ a reparameterization trick (Kingma and Welling 2013) to approximate
the second term.
\begin{equation}\label{eqn:approx_expected_log_likelihood}
  \int q(\bm{z}) \ln p(y|\bm{z}) d\bm{z} 
  \approx \frac{1}{S} \sum_{s=1}^{S} \ln p(y|\bm{z}_s),
\end{equation}
where sample $\bm{z}_s$ from the approximated posterior is defined as
\[
  \bm{z}_s=\bm{\mu} + L\bm{\epsilon}_s, \quad
  \bm{\epsilon}_s \sim \mathcal{N}(\bm{0},I).
\]
Matrix $L$ is the Cholesky decomposition of covariance matrix $V$ and
a lower triangular matrix (i.e., $V=LL^\top$). The KL-divergence is calculated with
the parameters of the approximated posterior, $\bm{\mu}$ and $L$; thus,
we can derive the gradients as follows.
\begin{align}
  \frac{\partial \mbox{KL}}{\partial \bm{\mu}}
  &= G^{-1}(\bm{\mu} - \bm{m}) - \frac{1}{S} \sum_{s=1}^{S} W_{:,s}, \\
  \frac{\partial \mbox{KL}}{\partial L} 
  &= -(L^{-1})^\top + G^{-1}L - \frac{1}{S} W E,
\end{align}
where $W$ and $E$ are the matrices whose elements are defined as
\begin{align*}
    W_{k,s}
    &= (2y - 1) a_k (1 - \sigma(a_k(z_{s,k} - b_k)))
     \frac{\prod_{k} \sigma\left(a_k(z_{s,k} - b_k)\right)}{p(y|\bm{z}_s)}, \\
    E_{s,k} &= \epsilon_{s,k},
\end{align*}
respectively, and $W_{:,s}$ represents the $s$-th column vector of $W$.
Now that we have obtained the gradients of the KL-divergence, we can apply
gradient-based optimization to find the optimal $\bm{\mu}$ and $L$.
(We employed gradient descent with armijo line search in our implementation.)

\subsection{E-step}

Given the parameters of the model and observation $\bm{y}_j^{(1:T_j)}$,
which denotes $\{y_j^{(1)}, ..., y_j^{(T_j)}\}$,
the E-step infers the posterior
of the skill state $p(\bm{z}_j^{(t)}|\bm{y}_j^{(1:T_j)})$.
We derive the E-step based on the forward-backward algorithm
employed in the LDS. By calculating forward messages and backward
messages for all the skill states of the time series, we obtain
the posterior of the skill state.

\subsubsection{Forward Message Passing}

In this section, we derive the forward $\hat{\alpha}$ message, which is
the posterior $p(\bm{z}_j^{(t)}| \bm{y}_j^{(1:t)})$, based on
the LDS method and the Gaussian approximation derived
in Section \ref{subsec:gauss_apx}.

The $\hat{\alpha}$ message is defined by the following recursive formula.
\begin{equation}
 \hat{\alpha}(\bm{z}_j^{(t)}) \propto
 p(y_j^{(t)}|\bm{z}_j^{(t)}) \int p(\bm{z}_j^{(t)}|\bm{z}_j^{(t-1)})
 \hat{\alpha}(\bm{z}_j^{(t-1)}) d\bm{z}_j^{(t-1)}. \label{eqn:forward}
\end{equation}
We can obtain the $\hat{\alpha}$ messages for all the state variables by
calculating this recursive definition from time step $1$ to time step $T_j$,
which is called forward message passing as shown in Figure~\ref{pic:lds}~(b).
Eq. \eqref{eqn:forward} can be re-written as
\begin{equation}
 \hat{\alpha}(\bm{z}_j^{(t)}) \propto
 p(y_j^{(t)}|\bm{z}_j^{(t)}) p(\bm{z}_j^{(t)}|y_j^{(1)}, \dots, y_j^{(t-1)})
 \label{eqn:forward_bayes_update}
\end{equation}
by calculating the integral. We approximate the $\hat{\alpha}$ message
as a Gaussian; thus let $\hat{\alpha}(\bm{z}_j^{(t)})$ be 
$\mathcal{N}(\bm{z}_j^{(t)}| \bm{\mu}_j^{(t)}, V_j^{(t)})$.
The inside of integral in Eq. \eqref{eqn:forward}, which is the product of
two Gaussians, is calculated as the following Gaussian
\begin{equation}
 p(\bm{z}_j^{(t)}|y_j^{(1)}, \dots, y_j^{(t-1)}) =
 \mathcal{N}(\bm{z}_j^{(t)}| \bm{m}_j^{(t)}, G_j^{(t)}),
\end{equation}
where $\bm{m}_j^{(t)}$ and $G_j^{(t)}$ is defined as
\begin{equation}
  \bm{m}_j^{(t)} = D_{i(j,t-1)} \bm{\mu}_j^{(t-1)}
  + \begin{bmatrix}
      \vdots \\
      \bm{\beta}_k^T \bm{x}_{j,k}^{(t)} \\
      \vdots
    \end{bmatrix}, \quad
  G_j^{(t)} = D_{i(j,t-1)} V_j^{(t-1)} D_{i(j,t-1)}^\top + \Gamma_j^{(t)}. \nonumber
\end{equation}
We separate state variable $\bm{z}_j^{(t)}$ into two state variables: $\bm{z}_a$ and
$\bm{z}_b$, where $\bm{z}_a$ corresponds to the skills required by question 
$i(j,t)$ and $\bm{z}_b$ to the other skills. We omit index $j,t$ to avoid
complexity here. Furthermore, the mean and covariance are separated accordingly as
\begin{eqnarray}
\bm{z}_j^{(t)} = \begin{bmatrix}
                   \bm{z}_a \\
                   \bm{z}_b
                 \end{bmatrix},
\bm{m}_j^{(t)} = \begin{bmatrix}
                   \bm{m}_a \\
                   \bm{m}_b
                 \end{bmatrix},
G_j^{(t)} = \begin{bmatrix}
              G_{a,a} & G_{a,b} \\
              G_{b,a} & G_{b,b}
            \end{bmatrix}. \nonumber
\end{eqnarray}
Eq. \eqref{eqn:forward_bayes_update} can be re-written as
\begin{equation}\label{eqn:separated_forward_bayes_update}
    \hat{\alpha}(\bm{z}_j^{(t)}) \propto p(y_j^{(t)}|\bm{z}_a) p(\bm{z}_a) p(\bm{z}_b|\bm{z}_a),
\end{equation}
where $p(\bm{z}_a)$ and $p(\bm{z}_b|\bm{z}_a)$ are defined as
\begin{align}
    p(\bm{z}_a) & = \mathcal{N}(\bm{z}_a | \bm{m}_a, G_{a,a}), \\
    p(\bm{z}_b|\bm{z}_a)
    & = \mathcal{N}(\bm{z}_b|
          \bm{m}_b + G_{b,a}G_{a,a}^{-1}(\bm{z}_a-\bm{m}_a),
          G_{b,b} - G_{b,a}G_{a,a}^{-1}G_{a,b}
        ).
\end{align}
We apply our Gaussian approximation in Section \ref{subsec:gauss_apx} to
$p(y_j^{(t)}|\bm{z}_a) p(\bm{z}_a)$
and obtain the approximation of $p(\bm{z}_a| y_j^{(t)})$ as
$\mathcal{N}(\bm{z}_a| \bm{m}'_a, G_{aa}')$. Now that
$p(\bm{z}_a| y_j^{(t)}) p(\bm{z}_b|\bm{z}_a)$ is the product of
two Gaussians, we finally obtain
$\hat{\alpha}(\bm{z}_j^{(t)})=\mathcal{N}(\bm{z}_j^{(t)}| \bm{\mu}_j^{(t)}, V_j^{(t)})$,
where $\bm{\mu}_j^{(t)}$ and $V_j^{(t)}$ are defined as
\begin{eqnarray}
  \bm{\mu}_j^{(t)} & = & 
  \begin{bmatrix}
    \bm{m}'_a \\
    \bm{m}_b + G_{b,a}G_{a,a}^{-1}(\bm{m}'_a - \bm{m}_a)
  \end{bmatrix}, \\
  V_j^{(t)} & = &
  \begin{bmatrix}
    G'_{a,a} & G'_{a,a}G_{a,a}^{-1}G_{a,b} \\
    G_{b,a}G_{a,a}^{-1}G'_{a,a} & G_{b,b} +
    G_{b,a}(G_{a,a}^{-1}G'_{a,a}G_{a,a}^{-1} - G_{a,a}^{-1})G_{a,b}
  \end{bmatrix}.
\end{eqnarray}

\subsubsection{Backward Message Passing}

The backward $\gamma$ message is the posterior
$p(\bm{z}_j^{(t)}|\bm{y}_j^{(1:T_j)})$ and its expectation is
the skill estimation in the condition that the sequence of item
responses for learner $j$ is given.
Since the $\hat{\alpha}$ message is obtained as the Gaussian distribution,
the $\gamma$ message is exactly the same as the result of the LDS.
We omit the derivation (Bishop 2006) and describe the obtained
$\gamma$ message as $\gamma(\bm{z}_j^{(t)}) = \mathcal{N}(\bm{z}_j^{(t)}| \hat{\bm{\mu}}_j^{(t)}, \hat{V}_j^{(t)})$,
where $\hat{\bm{\mu}}_j^{(t)}$ and $\hat{V}_j^{(t)}$ are defined as
\begin{eqnarray}
  \hat{\bm{\mu}}_j^{(t)} & = &
  \bm{\mu}_j^{(t)} + J_j^{(t)} \left(
    \hat{\bm{\mu}}_j^{(t+1)} -
    D_{i(j,t)} \bm{\mu}_j^{(t)} -    
    \begin{bmatrix}
      \vdots \\
      \bm{\beta}_k^\top \bm{x}_{j,k}^{(t+1)} \\
      \vdots
    \end{bmatrix}
  \right), \\
  \hat{V}_j^{(t)} & = &
  V_j^{(t)} + J_j^{(t)}(\hat{V}_j^{(t+1)} - P_j^{(t)}) {J_j^{(t)}}^\top.
\end{eqnarray}
$P_j^{(t)}$ and $J_j^{(t)}$ are also defined as
\begin{align}
  P_j^{(t)} & = D_{i(j,t)} V_j^{(t)} D_{i(j,t)}^\top + \Gamma_j^{(t+1)},\\
  J_j^{(t)} & = V_j^{(t)} D_{i(j,t)}^\top {P_j^{(t)}}^{-1}.
\end{align}
The $\gamma$ message is passed backward as shown in Figure \ref{pic:lds} (b)
to obtain the posteriors for all the time steps.

\subsection{M-step}

For the M-step, the expected complete data log-likelihood is
maximized with
respect to the model parameters. To avoid overfitting, we introduce 
the priors to each model parameter.
The maximization can be divided into three parts:
the parameters of the initial state, state transition, and emission.
For simplicity, $\mathbb{E}[\cdot]$ in this section denotes the
expectation with respect to 
$\bm{z}_j^{(1:T_j)}$ given $\bm{y}_j^{(1:T_j)}$.

\subsubsection{Initial State}
The parameters of the initial state probability are $\bm{\mu}_0$
and$P_0$.
We assume a Gaussian-Wishart distribution as a prior on these parameters.
The objective function with respect to $\bm{\mu}_0$ and $P_0$ is
\begin{align}
  Q_{\mbox{init}}(\bm{\mu_0}, P_0)
  & = \sum_{j=1}^{N} \ \mathbb{E} \left[ \ln p(\bm{z}_j^{(1)}| \bm{\mu}_0, P_0) \right]
    + \ln \mathcal{N}(\bm{\mu}_0|\bm{0}, \tau P_0)
    + \ln \mathcal{W}(P_0^{-1}|W, \nu),
\end{align}
where $\mathcal{W}(P_0^{-1}|W, \nu)$ denotes the Wishart distribution.
Setting its derivative to zero gives us the following update rules:
\begin{align}
  \bm{\mu}_0 & = \frac{1}{(N+\frac{1}{\tau})} \sum_{j=1}^{N} \hat{\bm{\mu}}_j^{(1)}, \\
  P_0   & = \frac{1}{N + \nu - K} \biggl\{
          \sum_{j=1}^{N} \left\{\hat{\bm{\mu}_j}^{(1)} \hat{\bm{\mu}_j}^{(1)\top} + \hat{V}_j^{(1)} \right\}
          + W^{-1} - (N + \frac{1}{\tau}) \bm{\mu}_0 \bm{\mu}_0^\top \biggr\}.
\end{align}

\subsubsection{State Transition}\label{sec:m_step_state_trans}

The parameters with respect to the state transition are listed in
$\bm{\theta}^{trans}$ (Eq. \eqref{eqn:trans_params}).
We assumed Gaussian-gamma distributions on these parameters.
The objective function with respect to these parameters is
\begin{align}
  Q_{\mbox{trans}}(\bm{\theta}^{trans})
  & = \sum_{j=1}^{N} \sum_{t=2}^{T_j} \mathbb{E} \biggl[ 
    \ln p(\bm{z}_j^{(t)}|\bm{z}_j^{(t-1)}, \bm{\theta}^{trans})
    \biggl] 
    + \sum_{k=1}^K \sum_{i:Q_{i,k}=1} \ln \mathcal{N}(d_{i,k}|\mu_d, \sigma_d^2\gamma_k)
    \nonumber \\
  & \quad + \sum_{k=1}^K \ln \mathcal{N}(\bm{\beta_k}| \bm{0},
                                   \sigma_{\beta}^2 \gamma_k \mbox{I})
    + \sum_{k=1}^K \ln \mbox{Gam}(\gamma_k^{-1}|\phi, \psi),
  \label{eqn:objective_trans}
\end{align}
where $\mbox{Gam}(\gamma_k^{-1}|\phi, \psi)$ denotes the gamma distribution.
The probability of state transition is a Gaussian with diagonal covariance;
therefore, the term inside the expectation above can be re-written as
\begin{align}
  \ln p(\bm{z}_j^{(t)}|\bm{z}_j^{(t-1)}, \bm{\theta}^{trans})
  & = \sum_{k:\ qq_{j,t,k} = 1*} \left\{
      -\frac{1}{2} \ln \gamma_k
      -\frac{1}{2\gamma_k} \left(
        z_{j,k}^{(t)} - d_{i(j,t-1),k} z_{j,k}^{(t-1)}
        - \bm{\beta}_k^\top \bm{x}_{j,k}^{(t)}
      \right)^2
    \right\} \nonumber \\
  & \quad + \sum_{k:\ qq_{j,t,k}= 01} \left\{
      -\frac{1}{2} \ln \gamma_k
      -\frac{1}{2\gamma_k} \left(
        z_{j,k}^{(t)} - z_{j,k}^{(t-1)} - \bm{\beta}_k^\top \bm{x}_{j,k}^{(t)}
      \right)^2
    \right\} + \mbox{const},
\end{align}
where $qq_{j,t,k}$ denotes the pair of values $Q_{i(j,t-1),k}$ and $Q_{i(j,t),k}$
(i.e., $00$, $01$, $10$, and $11$). $qq_{j,t,k}=1*$ represents $10$ and $11$.
The maximization of Eq. \eqref{eqn:objective_trans} can be seen as an 
independent optimization
for each $\bm{\theta}_k^{trans}$, where $\bm{\theta}_k^{trans}$ is defined as
\[
  \bm{\theta}_k^{trans} = \{d_{i,k}| i \in \{1,...,M\}, Q_{i,k}=1\} \cup \{\gamma_k, \bm{\beta}_k\}.
\]
We focus on the optimization with respect to $\bm{\theta}_k^{trans}$:
\begin{align}
  Q_{\mbox{trans},k}(\bm{\theta}_k^{trans})
  & = -\frac{\ln \gamma_k}{2} \sum_{j,t:\ qq_{j,t,k} \ne 00} 1 \nonumber \\
  & \quad -\frac{1}{2\gamma_k} \sum_{j,t:\ qq_{j,t,k}=1*}
    \mathbb{E} \left[ 
      \left( z_{j,k}^{(t)} - d_{i(j,t-1),k} z_{j,k}^{(t-1)} - \bm{\beta}_k^\top \bm{x}_{j,k}^{(t)} \right)^2
    \right] \nonumber \\
  & \quad -\frac{1}{2\gamma_k} \sum_{j,t:\ qq_{j,t,k}=01}
    \mathbb{E} \left[ 
      \left( z_{j,k}^{(t)} - z_{j,k}^{(t-1)} - \bm{\beta}_k^\top \bm{x}_{j,k}^{(t)} \right)^2
    \right] \nonumber \\
  & \quad 
    - \frac{1}{2 \gamma_k \sigma_d^2} \sum_{i:\ Q_{i,k}=1}||d_{i,k}-\mu_d||^2
    - \frac{1}{2 \gamma_k \sigma_{\beta}^2}||\bm{\beta}_k||^2
    - (\phi - 1) \ln \gamma_k - \frac{\psi}{\gamma_k}.
  \label{eqn:objective_trans_k}
\end{align}
We first maximize $Q_{\mbox{trans},k}(\bm{\theta}_k^{trans})$ with respect to 
the parameters, except for $\gamma_k$. Considering the parameters except $\gamma_k$
yields the following objective function to be minimized.
\begin{align}
  & \sum_{j,t:\ qq_{j,t,k}=1*}
  \mathbb{E} \left[ 
    \left(z_{j,k}^{(t)} - d_{i(j,t-1),k} z_{j,k}^{(t-1)}
          - \bm{\beta}_k^\top \bm{x}_{j,k}^{(t)} \right)^2
  \right] \nonumber \\
  & \quad + \sum_{j,t:\ qq_{j,t,k}=01}
  \mathbb{E} \left[ 
    \left(z_{j,k}^{(t)} - z_{j,k}^{(t-1)} - \bm{\beta}_k^\top \bm{x}_{j,k}^{(t)} \right)^2
  \right]
  + \frac{1}{\sigma_d^2} \sum_{i: Q_{i,k}=1} ||d_{i,k}-\mu_d||^2
  + \frac{1}{\sigma_{\beta}^2}||\bm{\beta}_k||^2.
  \label{eqn:objective_trans_k_d_beta}
\end{align}
Eq. \eqref{eqn:objective_trans_k_d_beta} is of the quadratic form; thus, we can
re-write it as
\begin{equation}
  \eqref{eqn:objective_trans_k_d_beta}  
  = \bm{v}^\top
    \left(
      L_{d,\beta}^k + \sum_{j,t:\ qq_{j,t,k}=1*} E_{j,t}^{k,1*}
      + \sum_{j,t:\ qq_{j,t,k}=01} E_{j,t}^{k,01}
    \right)
    \bm{v} 
  \quad (\coloneqq \bm{v}^\top R^k \bm{v})
\end{equation}
where $\bm{v}$, $E_{j,t}^{k,1*}$, $E_{j,t}^{k,01}$, and $L_{d,\beta}^k$ are defined as
\begin{eqnarray}
  \bm{v} & = & 
  \begin{bmatrix}
    1 & d_{i_k^{-1}(1),k} & ... & d_{i_k^{-1}(M_k),k} & \bm{\beta}_k^\top
  \end{bmatrix}^\top, \\
  E_{j,t}^{k,1*} & = &
  \begin{bmatrix}
    \mathbb{E}[(z_{j,k}^{(t)})^2] & ... & -\mathbb{E}[z_{j,k}^{(t)} z_{j,k}^{(t-1)}] &
      ... & -\mathbb{E}[z_{j,k}^{(t)}] \bm{x}_{j,k}^{(t)\top} \\
    \vdots & & \vdots & & \vdots \\
    -\mathbb{E}[z_{j,k}^{(t)}z_{j,k}^{(t-1)}] & ... & \mathbb{E}[(z_{j,k}^{(t-1)})^2] &
      ... & \mathbb{E}[z_{j,k}^{(t-1)}] \bm{x}_{j,k}^{(t)\top} \\
    \vdots & & \vdots & & \vdots \\
    -\mathbb{E}[z_{j,k}^{(t)}]\bm{x}_{j,k}^{(t)} & ... & \mathbb{E}[z_{j,k}^{(t-1)}]\bm{x}_{j,k}^{(t)} &
      ... & \bm{x}_{j,k}^{(t)} \bm{x}_{j,k}^{(t)\top}
  \end{bmatrix}, \\
  E_{j,t}^{k,01} & = &
  \begin{bmatrix}
    \mathbb{E}[(z_{j,k}^{(t)}-z_{j,k}^{(t-1)})^2] & ... & 
      -\mathbb{E}[z_{j,k}^{(t)} - z_{j,k}^{(t-1)}] \bm{x}_{j,k}^{(t)\top} \\
    \vdots & & \vdots \\
    -\mathbb{E}[z_{j,k}^{(t)} - z_{j,k}^{(t-1)}] \bm{x}_{j,k}^{(t)} & ... & 
      \bm{x}_{j,k}^{(t)} \bm{x}_{j,k}^{(t)\top}
  \end{bmatrix}, \\
  L_{d,\beta}^k & = &
  \begin{bmatrix}
    \frac{M_k \mu_d^2}{\sigma_d^2} & -\frac{\mu_d}{\sigma_d^2} \bm{1}^{\top} &  \\
    -\frac{\mu_d}{\sigma_d^2} \bm{1} & \frac{1}{\sigma_d^2} \mbox{I} &  \\
      & &  \frac{1}{\sigma_{\beta}^2} \mbox{I}
  \end{bmatrix}. \label{eqn:trans_quad_prior}
\end{eqnarray}
Here, we let $i_k(\cdot)$ be a function that takes the index of each
question and returns the index among questions requiring skill $k$;
for example, $i_k(10) = \sum_{i=1}^{10} Q_{i,k}$ if the tenth 
question requires skill $k$.
Furthermore, $i_k^{-1}(\cdot)$ denotes its inverse function. 
Element $\mathbb{E}[(z_{j,k}^{(t-1)})^2]$ in $E_{j,t}^{k,1*}$ is located in the
$1+i_k(i(j,t-1))$-th row and column. The unspecified elements of $E_{j,t}^{k,1*}$
and $E_{j,t}^{k,01}$ are zero, and these expectations can be calculated using
the following equations.
\begin{equation}
  \mathbb{E} \left[ z_{j,k}^{(t)} z_{j,k}^{(t-1)} \right] =
  (J_j^{(t-1)}\hat{V}_j^{(t)})_{k,k} + \hat{\mu}_{j,k}^{(t-1)}\hat{\mu}_{j,k}^{(t)}, \quad 
  \mathbb{E} \left[ (z_{j,k}^{(t)})^2 \right] = 
  (\hat{V}_j)_{k,k} + (\hat{\mu}_{j,k}^{(t)})^2.
  \nonumber
\end{equation}
$\frac{1}{\sigma_d^2} \mbox{I}$ in $L_{d,\beta}^k$ has $M_k$ rows
and columns, where $M_k (= \sum_i Q_{i,k})$ denotes the number of
questions requiring skill $k$, and $\frac{1}{\sigma_{\beta}^2} \mbox{I}$
has $F_k$ rows and columns.
The sizes of $E_{j,t}^{k,1*}$ and $E_{j,t}^{k,01}$ are the same as that 
of $L_{d,\beta}^k$. 

Note that matrix $R^k$ is positive definite. We write $R^k$ as
\[
R^k = \begin{bmatrix}
 R_{1,1} & R_{1,2} \\
 R_{2,1} & R_{2,2} \\
\end{bmatrix},
\]
where $R_{1,1}$ is a one-by-one matrix. Completing the square with respect to
$d_{i,k}$ and $ \bm{\beta}_k$ gives us the optimized parameters
\begin{eqnarray}
  \begin{bmatrix}
   d_{i_k^{-1}(1),k} &  ... & d_{i_k^{-1}(M_k),k} & \bm{\beta}_k^\top
  \end{bmatrix}^\top = -(R_{2,2})^{-1}R_{2,1}.
\end{eqnarray}
Also, its minimized value is $R_{1,1} - R_{1,2} R_{2,2}^{-1} R_{2,1}$.

Next, we maximize Eq. \eqref{eqn:objective_trans_k} with respect to $\gamma_k$.
Substituting the minimized value into Eq.~\eqref{eqn:objective_trans_k} yields
the following objective to be maximized.
\begin{equation}
  Q_{\mbox{trans},k}(\gamma_k) =
  -\frac{\ln \gamma_k}{2} T_{k,\lnot 00}
  -\frac{1}{2\gamma_k} (R_{1,1} - R_{1,2} R_{2,2}^{-1} R_{2,1}) 
  - (\phi - 1) \ln \gamma_k - \frac{\psi}{\gamma_k}. 
\end{equation}
Here, we defined $T_{k, \lnot 00} \coloneqq \sum_{j,t:\ qq_{j,t,k} \neq 00} 1$.
Taking the derivative with respect to $\gamma_k$ and setting it to zero
optimizes $\gamma_k$, as shown below.
\begin{equation}
  \gamma_k =
  \frac{(R_{1,1} - R_{1,2} R_{2,2}^{-1} R_{2,1}) + 2\psi}
  {T_{k,\lnot 00} + 2\phi - 2}.
\end{equation}

\subsubsection{State Transition with a Joint Prior on Slope and Bias}
\label{sec:mstep_trans_joint_prior}

The previous section introduced the optimization method for state
transition parameters and assumed the independent prior distributions
on the slope parameter $d_{i,k}$ and the bias parameter $\bm{\beta}_k$
of the linear transformation.
This section introduces an extension for a joint prior distribution
on these parameters in the case where linear transformation for each
skill has a question-specific bias term. In actual data, the method
in the previous section may estimate the parameters which lead to an
unexpectedly high or low skill. Setting an appropriate joint prior
distribution avoids such unexpected cases and controls skills within
a reasonable range. This section focuses on the optimization with the
joint prior distribution. Details of how the joint prior is determined
are discussed in Section \ref{sec:real_data_joint_prior}.

The linear transformation with a question specific bias term is
obtained as discussed in Section \ref{subsec:proposed}.
Let us define the covariate variable
$\bm{x}_{j,k}^{(t)}$ as a binary vector with $M_k$ elements,
i.e., number of questions requiring skill $k$.
The $i_k(i(j,t))$-th element is set to one and the other elements
are set to zero. The linear transformation is now represented as
\begin{equation}
    z_{j,k}^{(t+1)} = d_{i(j,t),k} z_{j,k}^{(t)} + \beta_{k,i_k(i(j,t))}.    
\end{equation}
This equation implies that the skill of a learner asymptotically
approaches $\beta_{k,i_k(i(j,t))} / (1 - d_{i(j,t),k})$ by
answering this question many times (applying the linear transformation
many times) when $0 < d_{i(j,t),k} < 1$. The asymptotic
value of skill should be within a reasonable range. By introducing 
a joint prior on $d_{i,k}$ and $\beta_{k, i_k(i)}$, we are able to apply
a prior knowledge of the asymptotic value to the skill inference.
Here, we consider a multivariate Gaussian prior
$\mathcal{N}([\mu_d, \mu_{\beta}]^{\top}, \gamma_{k} \Lambda_{d,\beta}^{-1})$
on $d_{i,k}$ and $\beta_{k,i_k(i)}$ where the precision matrix
$\Lambda_{d,\beta}$ is given by
\begin{eqnarray}
    \Lambda_{d,\beta} & = &
    \begin{bmatrix}
        \lambda_d & \lambda_{d,\beta} \\
        \lambda_{d,\beta} & \lambda_{\beta}
    \end{bmatrix}.
\end{eqnarray}
The prior terms of $d_{i,k}$ and $\bm{\beta}_k$ in the 
objective function in Eq. \eqref{eqn:objective_trans} is
rewritten as
\[
    \sum_{k=1}^K \sum_{i: Q_i,k=1}
    \mathcal{N}\left(
    [d_{i,k}, \beta_{k, i_k(i)}]^{\top}
    \middle|
    [\mu_d, \mu_{\beta}]^{\top},
    \gamma_k \Lambda_{d,\beta}^{-1}
    \right).
\]
To optimize the objective function with the joint prior,
only Eq. \eqref{eqn:trans_quad_prior} is required to
modify as 
\begin{eqnarray}
  L_{d,\beta}^k & = &
  \left[
  \begin{array}{c:c:c:c:c:c:c}
    \lambda_{1,1}^k & \lambda_{1,d} & \cdots & \lambda_{1,d} &
    \lambda_{1,\beta} & \cdots & \lambda_{1,\beta} \\
    \hdashline
    \lambda_{1,d} & \lambda_{d} & & & \lambda_{d,\beta} & & \\
    \hdashline
    \vdots & & \ddots & & & \ddots & \\
    \hdashline
    \lambda_{1,d} & & & \lambda_{d} & & & \lambda_{d, \beta}\\
    \hdashline
    \lambda_{1,\beta} & \lambda_{d,\beta} & & & \lambda_{\beta} & & \\
    \hdashline
    \vdots & & \ddots & & & \ddots & \\
    \hdashline
    \lambda_{1,\beta} & & & \lambda_{d, \beta}& & & \lambda_{\beta} \\
  \end{array}
  \right],
\end{eqnarray}
where $\lambda_{1,1}^k$, $\lambda_{1,d}$, and $\lambda_{1,\beta}$
are defined as
\begin{empheq}[left={\empheqlbrace\,}]{align}
  \lambda_{1,1}^k &= 
  M_k \left(
    \lambda_d \mu_d^2
    + 2 \lambda_{d,\beta} \mu_d \mu_{\beta}
    + \lambda_{\beta} \mu_{\beta}^2
  \right)
  \\
  \lambda_{1,d} &=  -\lambda_d \mu_d - \lambda_{d,\beta} \mu_{\beta} \\
  \lambda_{1,\beta} &=  -\lambda_{d,\beta} \mu_d - \lambda_{\beta} \mu_{\beta}
  .
\end{empheq}
The unspecified elements in $L_{d,\beta}^k$ are zero.


%

\subsubsection{Emission}
The parameters with respect to the emission are listed in $\bm{\theta}^{emit}$ (Eq. \eqref{eqn:emit_params}).
We assume log normal and Gaussian distributions as priors on $a_{i,k}$
and $b_{i,k}$, respectively. The objective function with respect to
these parameters is
\begin{align}
  Q_{\mbox{emit}}(\bm{\theta}^{emit})
  & = \sum_{j=1}^{N} \sum_{t=1}^{T_j}
    \mathbb{E} \left[
      \ln p\left(y_j^{(t)} \middle| \bm{z}_{j}^{(t)}, \bm{\theta}^{emit}\right)
    \right] \nonumber \\
  & \quad + \sum_{i,k:\ Q_{i,k}=1}
    \left\{ \ln \mbox{logN}(a_{i,k}|\mu_a, \sigma_a^2)
    + \ln \mathcal{N}(b_{i,k}|0, \sigma_b^2) \right\}, \label{eqn:objective_emit}
\end{align}
where $\mbox{logN}(a_{i,k}|\mu_a, \sigma_a^2)$ denotes the log normal distribution.
The expectation above can be approximated by sampling as
\begin{equation}
  \mathbb{E} \left[
      \ln p\left(y_j^{(t)} \middle| \bm{z}_{j}^{(t)}, \bm{\theta}^{emit}\right)
  \right] \approx
  \frac{1}{H} \sum_{h=1}^{H}
  \ln p\left(y_j^{(t)}\middle| \bm{z}_{j,h}^{(t)}, \bm{\theta}^{emit}\right),
\end{equation}
where $\bm{z}_{j,h}^{(t)}$ denotes a sample from the $\gamma$ message
and $h \in \{1, ..., H\}$ is the index of the sample.
We derive the derivatives of Eq. \eqref{eqn:objective_emit}
with respect to $a_{i,k}$ and $b_{i,k}$ as
\begin{eqnarray}
  \frac{\partial Q_{emit}}{\partial a_{i,k}}
  & \approx & \frac{1}{H} \sum_{j,t:\ i(j,t)=i} \sum_{h=1}^{H}
    (2y_j^{(t)} - 1) 
    \frac{p(y_j^{(t)}=1 | \bm{z}_{j,h}^{(t)})}{p(y_j^{(t)} | \bm{z}_{j,h}^{(t)})}
    \notag \\
  & & \left\{1 - \sigma\left(a_{i,k} (z_{j,h,k}^{(t)} - b_{i,k})\right) \right\}
    (z_{j,h,k}^{(t)} - b_{i,k})
    - \frac{1}{a_{i,k}}\left\{1 + \frac{1}{\sigma_a^2}(\ln a_{i,k} - \mu_a) \right\}, 
      \label{eqn:opt_a_in_mstep} \\
  \frac{\partial Q_{emit}}{\partial b_{i,k}} 
  & \approx & \frac{1}{H} \sum_{j,t:\ i(j,t) = i} \sum_{h=1}^{H}
    (2y_j^{(t)} - 1) 
    \frac{p(y_j^{(t)}=1 | \bm{z}_{j,h}^{(t)})}{p(y_j^{(t)} | \bm{z}_{j,h}^{(t)})}
    \notag \\
  & & \left\{1 - \sigma\left(a_{i,k} (z_{j,h,k}^{(t)} - b_{i,k})\right) \right\}
    (- a_{i,k}) - b_{i,k}/\sigma_b^2.
    \label{eqn:opt_b_in_mstep}
\end{eqnarray}
The gradient-based optimization can be used to obtain the optimal values for
$a_{i,k}$ and $b_{i,k}$.

\subsection{Smoothing Extension in E-step}\label{subsec:estep_window}

Herein, we introduce a smoothing technique as an extension of the E-step.
Since the observation
$y_j^{(t)}$ is zero or one, and its ability to specify the skill state $\bm{z}_j^{(t)}$
is smaller than that of the LDS. We attempt to extract
information to
specify the skill states as much as possible from the observations. When we consider
a certain time step $t$, observations near the time step $t$, e.g., $y_j^{(t-1)}$
and $y_j^{(t+1)}$, can be useful for inferring the skill state $\bm{z}_j^{(t)}$.
Therefore, we extend the likelihood $p(y_j^{(t)}|\bm{z}_a)$ in Eq.
\eqref{eqn:separated_forward_bayes_update}
to the weighted products of the likelihood functions within a time window.
The example of 
the width $\pm 1$ is
\begin{equation}
    p(y_j^{(t-1)}|\bm{z}_a)^{w_{-1}} p(y_j^{(t)}|\bm{z}_a)^{w_0}
    p(y_j^{(t+1)}|\bm{z}_a)^{w_{+1}},
\end{equation}
where $w_{-1}$, $w_{0}$, and $w_{+1}$ denote the weights for each likelihood,
respectively,
and $\bm{z}_a$ is now a vector whose skills are required by the questions
$i(j,t-1)$, $i(j,t)$, and $i(j,t+1)$.
The Gaussian approximation of the $\hat{\alpha}$ message can
be derived in the same way as described in Section \ref{subsec:gauss_apx}
by replacing the likelihood
in Eq.~\eqref{eqn:approx_expected_log_likelihood} with the new one.
In the following evaluations, we experiment the effect of smoothing with
$\pm 1$ width.

\section{Evaluation of Skill Inference}\label{sec:eval_skill}

We demonstrate the accuracy of skill and parameter estimation on six
types of simulation data. The data were generated from our generative model.
We compare the accuracy of the skill estimation among three models:
our proposed dynamical non-compensatory MIRT model, a dynamical compensatory
MIRT model, and a non-compensatory MIRT model.

\subsection{Simulation Design}

We generated six types of simulation data via our generative model;
these six types include three aspects: whether the number of latent
skills is 2 or 100, whether the discrimination parameters in the generative
model are fixed to one or randomly-generated, and whether the skills in
initial state correlate or not.
Subsequently, the latent skills were estimated on
each type of data using the dynamical non-compensatory MIRT (dnMIRT)
model, a dynamical compensatory MIRT (dcMIRT) model, and a non-compensatory
MIRT (nMIRT) model. The dnMIRT model is the proposed model.
The dcMIRT model is the compensatory counterpart of
the proposed model, which is obtained by replacing the non-compensatory
model in Eq. \eqref{eqn:ncomp_irt} with the compensatory model
in Eq. \eqref{eqn:comp_irt}.
The parameter inference for dcMIRT can also be derived in the same manner.
The nMIRT model is the two-parameter product model in Eq. \eqref{eqn:ncomp_irt},
and we employed the mirt package in R (Chalmers~2012) for this evaluation.
This is not a dynamical model; thus, first, we estimated the
parameters for item discrimination and difficulty using all item responses.
Subsequently, we estimated the latent skills for each time step using the 
item responses that were within the sliding window. Window sizes of five and
ten were used in the following experiments.

\subsection{Generation of Data}\label{sec:simulation_data}

Here, we describe the generation of six types of simulation data: (A)-(F).
The six types corresponding to our model with different settings, and five
datasets for each data type were generated with different random seeds.
(A) was from a model whose latent state of skills was two-dimensional,
the item discrimination parameters were fixed to one, and the initial state
$\bm{z}_j^{(1)}$ was drawn from $\mathcal{N}(\bm{0},I)$.
(B) was from a model
whose latent state of skills was two-dimensional, the item discrimination
parameters were randomly generated from $\mbox{logN}(0, 0.25^2)$,
and the initial state
$\bm{z}_j^{(1)}$ was drawn from $\mathcal{N}(\bm{0},I)$.
The generated discrimination parameters (in a 
random seed) ranged from $0.59$ to $1.85$ and the mean was $1.05$.
(C) was from a model whose latent state of skills was hundred-dimensional,
the item discrimination parameters were fixed to one, 
and the initial state $\bm{z}_j^{(1)}$ was drawn from
$\mathcal{N}(\bm{0},I)$.
(D) is from a model whose latent state of skills was hundred-dimensional,
the item discrimination parameters were randomly generated from
$\mbox{logN}(0, 0.25^2)$,
and the initial state $\bm{z}_j^{(1)}$ was drawn from
$\mathcal{N}(\bm{0},I)$. The generated discrimination parameters (in a
random seed) ranged from $0.38$ to $2.32$ and the mean was $1.04$.
(E) is from a model whose latent state of skills was hundred-dimensional,
the item discrimination parameters were randomly generated from
$\mbox{logN}(0.4, 0.2^2)$,
and the initial state $\bm{z}_j^{(1)}$ was drawn from
$\mathcal{N}(\bm{0},I)$. The generated discrimination parameters (in a 
random seed) ranged from $0.71$ to $2.69$ and the mean was $1.53$.
(F) is from a model whose latent state of skills was
hundred-dimensional, the item discrimination parameters were randomly
generated from $\mbox{logN}(0, 0.25^2)$,
and the initial state $\bm{z}_j^{(1)}$ was drawn from
$\mathcal{N}(\bm{0},\Sigma^F)$, where the diagonal elements in $\Sigma^F$
is 1.0 and otherwise 0.2. The generated discrimination parameters (in a 
random seed) ranged from $0.46$ to $2.10$ and the mean was $1.04$.
Table \ref{tab:dataset} shows the basic information of 2-skill (A and B)
and 100-skill (C, D, E, and F) data types.

Other options for the model were common among the six data types. The
parameters $b_{i,k}$, and $d_{i,k}$ were drawn from the 
following distributions:
$b_{i,k} \sim \mathcal{N}(0,1)$ and $d_{i,k} \sim \mbox{U}(0.6,0.8)$,
where $\mbox{U}$ represents the uniform distribution.
The variance of the Gaussian noise in skill transition $\gamma_k$ was
fixed to $0.1^2$. We defined the covariate variable $\bm{x}_{j,k}^{(t)}$ as
the bias term of the linear transformation for each question introduced
in Section \ref{subsec:proposed}.
Therefore, the state transition for skill $k$ when answering question
$i$ in time step $t$ was given by a linear transformation:
$d_{i,k} z_{j,k}^{(t)} + \beta'_{i,k}$ with Gaussian noise.
The parameter $\beta'_{i,k}$ was determined deterministically from the
parameters $a_{i,k}$, $b_{i,k}$ and $d_{i,k}$.
We assumed that learners could reach a particular skill level after
mastering a question; that skill level was assumed that the learners could
correctly answer the question with a $0.9$ probability.
Since $d_{i,k}$ was positive and less than one, the level of skill $k$
converged to $\beta'_{i,k} / (1 - d_{i,k})$ after applying that linear
transformation to a skill state infinitely many times.
This value was set to the level of skill that learners could answer
question $i$ with a $0.9$ probability: i.e., $\sigma^{-1}(0.9)/a_{i,k} + b_{i,k}$.
Solving for $\beta'_{i,k}$ yields the following equation.
\begin{equation}
  \beta'_{i,k} = (1-d_{i,k}) \left\{ \frac{\sigma^{-1}(0.9)}{a_{i,k}} + b_{i,k} \right\}.
\end{equation}

The Q-matrix was created differently for the 2-skill and 100-skill datasets.
The Q-matrix of the 2-skill datasets indicated that ten questions required 
skill 1, and
another ten questions required skill 2; the rest of the 25 questions required
both skills. For the Q-matrix of the 100-skill datasets, we introduced 20
skill categories and each category included five skills (100 skills in total).
Fifty questions were created for each skill category (1,000 questions
in total). The skill requirement for a question in a skill
category was determined as follows: first, we drew the number of skills the
question required from the multinomial
distribution: $\{(1, 0.5), (2, 0.3), (3, 0.15), (4, 0.03), (5, 0.02)\}$,
where each pair denotes the number of skills and its probability.
Second, we randomly chose the selected number of skills from the skill set
in the corresponding category. The question sequence for a learner in
the 2-skill datasets was determined by randomly choosing 20 questions from
45 questions. The following rule determined the sequence of questions
for a learner in the 100-skill datasets. First, we randomly chose two skill
categories for the learner. Subsequently, 100 questions
from the two categories were randomly sampled with replacement to
reflect that a learner may study the same question multiple times.

\begin{table}[tb]
    \centering
    \caption{Basic information of the datasets.}
    \begin{tabular}{lcccc}
    \toprule
    & Learner & Question & Skill & Time step \\
    \midrule
    2-skill   & 1,000 & 45    & 2   & 20 \\
    100-skill & 3,000 & 1,000 & 100 & 100 \\
    \bottomrule
    \end{tabular}
    \label{tab:dataset}
\end{table}

\subsection{Inference Settings}\label{sec:inf_setting_simulation}

Here, we describe some of the settings used to perform inference.
We fixed the model parameters $\bm{\mu}_0 = 0$ and $P_0=I$.
The average of $b_{i,k}$ over $i$ was then set to zero by
subtracting the mean after each M-step in the dcMIRT and dnMIRT
models. These were for identifiability.
$\gamma_k$ was fixed to $0.2^2$ or estimated with hyperparameters
$(\phi, \psi) = (10, 1)$. When we inferred skills with $\gamma_k$
estimated, it tended to overfit to data. Therefore, we used the
additional validation datasets, which were generated in the same way
as the training datasets; the number of users in validation datasets
was 20\% of training datasets. We predicted the next response by
conducting sequential binary classification on the validation
datasets and stopped EM iterations when the prediction accuracy
became the best. (We used the average precision for incorrect
answers as the accuracy metric.) We did not combine the smoothing
extension and the estimation of $\gamma_k$ because preliminary
experiments did not perform well.
The hyperparameters were set as follows:
$\sigma_b = 1.0$,
$\sigma_d = 1.0$, and $\sigma_{\beta} = 1.0$.
Multiple values of the hyperparameter $\sigma_a$ were used to find
an optimal value.
The smoothing extension in the E-step required width and weight.
We used the $\pm 1$ width window with the weights
$w_{-1} = 0.25$, $w_0 = 1.0$, and $w_{+1} = 0.25$. 
The Gaussian distribution $\mathcal{N}(\bm{0}, I)$ was set to the
prior of skill states in the nMIRT model.

\subsection{Results}

We show the estimation results for the six data types.
The estimation error was evaluated with the mean average
error (MAE) and Pearson's correlation (Corr). With regards to the errors of
skills, we calculated them both at all the time steps and at the last time step.
Owning to the fact that we used
five datasets for each type, the following results are the average of those
five datasets.

Table \ref{tab:skill2_no_slope} presents the results of data set A:
2-skill, $a_{i,k}=1$, and no skill correlation.
With regards to the skills,
we can see that three settings in dnMIRT resulted in much smaller MAEs than
the other models. When we compare the results of dnMIRT with $\gamma_k$ fixed
and with it estimated, the MAE was slightly better when fixing it because
near optimal $\gamma_k$ was used.
When we compare the results of dnMIRT with and without smoothing,
dnMIRT with smoothing yielded a smaller MAE,
highlighting the effectiveness of the smoothing extension.
This trend is also observed in all types of data except E.
The MAEs of nMIRT are significantly inferior to those of dnMIRT;
this is because of the large estimation error of the difficulty parameters.
nMIRT does not model the change of latent skills; thus, the parameter
estimation using the whole time-series data introduces significant errors.
The MAE of dcMIRT is the worst among the three models even though its correlation
is high. The underestimation of skills can be observed in the scatter plot below.
With regards to the parameters,
the item difficulty $b_{i,k}$ and the bias of the linear transition of skill
$\beta'_{i,k}$ had small MAEs and high correlations. The slope of linear
transition $d_{i,k}$ had a small MAE; however, the correlation was low.
This implied that the estimated values were present in the same area as the
actual values i.e., from $0.6$ to $0.8$, but there was no linearity.
The variance of noise in skill transition $\gamma_k$ tended to be overestimated.

\begin{figure}
  \begin{minipage}{0.32\linewidth}
    \centering
    \includegraphics[width=\linewidth]{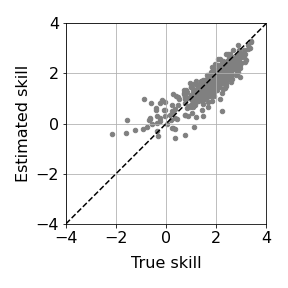}
    \subcaption{dnMIRT}
    \label{fig:scatterplot_dnc_mirt}
  \end{minipage}
    \begin{minipage}{0.32\linewidth}
    \centering
    \includegraphics[width=\linewidth]{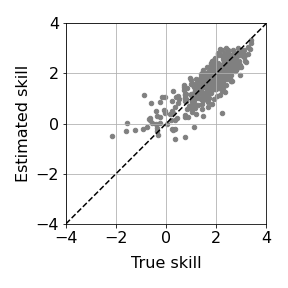}
    \subcaption{dnMIRT (estimate $\gamma_k$)}
    \label{fig:scatterplot_dn_mirt_est_g}
  \end{minipage}
  \begin{minipage}{0.32\linewidth}
    \centering
    \includegraphics[width=\linewidth]{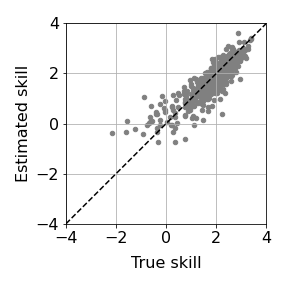}
    \subcaption{dnMIRT (smoothing)}
    \label{fig:scatterplot_dnc_mirt_win}
  \end{minipage}
  \\
  \begin{minipage}{0.32\linewidth}
    \centering
    \includegraphics[width=\linewidth]{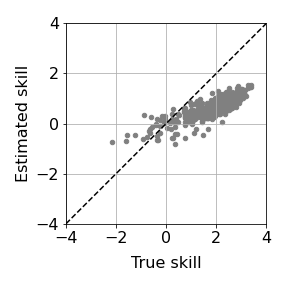}
    \subcaption{dcMIRT}
    \label{fig:scatterplot_dc_mirt}
  \end{minipage}
  \begin{minipage}{0.32\linewidth}
    \centering
    \includegraphics[width=\linewidth]{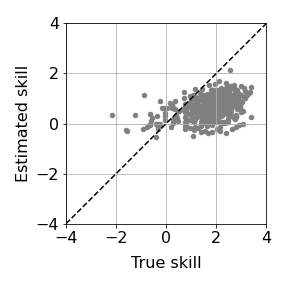}
    \subcaption{nMIRT (window=5)}
    \label{fig:scatterplot_nc_mirt_w05}
  \end{minipage}
  \begin{minipage}{0.32\linewidth}
    \centering
    \includegraphics[width=\linewidth]{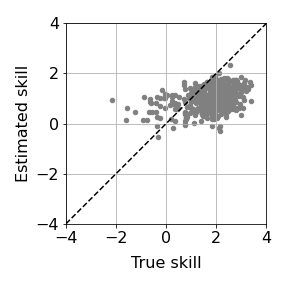}
    \subcaption{nMIRT (window=10)}
    \label{fig:scatterplot_nc_mirt_w10}
  \end{minipage}
  \caption{
  Scatter plots of the true skill and the estimated skill in the 2-skill data set A.
  (a)~-~(c) show that dnMIRT can reproduce the true skill adequately.
  (d) shows that dcMIRT tends to
  underestimate the skills, however correlation is high.
  (e) and (f) show that nMIRT suffers from high MAEs and low correlations.
  }
  \label{fig:scatterplot}
\end{figure}

\begin{figure}
 \centering
 {\tabcolsep=0pt
 \begin{tabular}{ccc}
  \begin{minipage}{0.32\linewidth}
  \includegraphics[width=\linewidth]{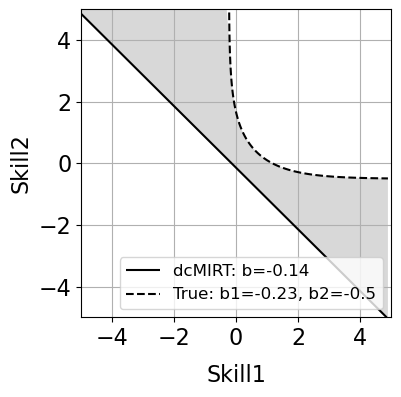}
  \end{minipage}
  &
  \begin{minipage}{0.32\linewidth}
  \includegraphics[width=\linewidth]{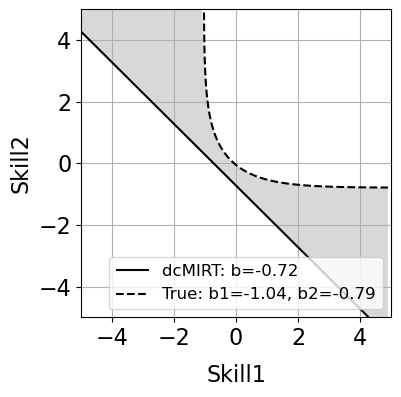}
  \end{minipage}
  &
  \begin{minipage}{0.32\linewidth}
  \centering
  \includegraphics[width=\linewidth]{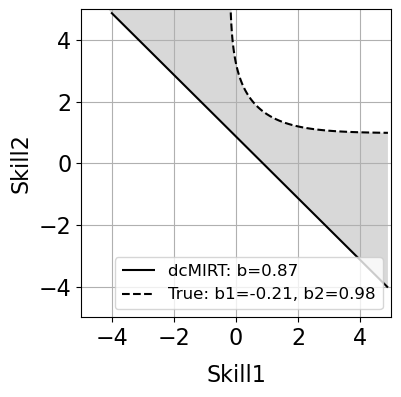}
  \end{minipage}\\
  (a) & (b) & (c) \\
  \end{tabular}
  }
\caption{
Comparisons of contours ($p=0.5$) of item response surfaces obtained
from dcMIRT and the true model on data set A: 2-skill and $a_{i,k}=1$.
The contour of dcMIRT is always below the true model.
This causes the significant underestimation of skills in dcMIRT.
}
\label{pic:irs_diff}
\end{figure}

Figure \ref{fig:scatterplot} shows the scatter plots of the true skill and
the estimated skill for each method in data set A.
We can confirm that the three results of dnMIRT have small MAEs and high
correlations; the result of dcMIRT has a large MAE and a high correlation, and the
two results of nMIRT have large MAEs and low correlations.
Figure \ref{fig:scatterplot} (d) clearly shows the underestimation of dcMIRT.
We also compare the contours ($p=0.5$) of the item response
surfaces obtained from dcMIRT and those of the true model in
Figure \ref{pic:irs_diff}.
The contours obtained from dcMIRT are always below those of the true models in the
three questions and this trend is observed for most of the questions.
We believe that this fitting trend causes the significant underestimation of
skills in dcMIRT.

Table \ref{tab:skill2_slope} presents the results of data set B: 
2-skill, $a_{i,k} \sim \mbox{logN}(0, 0.25^2)$, and no skill correlation.
The second type of data was drawn from the model with item
discrimination parameters. We estimated those with two prior settings
$\sigma_a=\{0.25, 0.1\}$, and also fixed the discrimination parameters
to one instead of estimating them.
The same trends as Table \ref{tab:skill2_no_slope} can be observed;
the best MAE was obtained from dnMIRT (dnMIRT with smoothing is
better than without smoothing); dcMIRT had large MAEs and high correlations.
Item discrimination parameters were added to be estimated, thus, the errors of
the skills and parameters were larger than those in the case without
discrimination parameters.
With regard to the results of dnMIRT in different settings, case $a_{i,k}=1$
with smoothing yielded the best MAE. dnMIRT with $\gamma_k$ estimated resulted
in better MAE than fixed $\gamma_k$.
The result of $\sigma_a=0.25$ without
smoothing, which is the same setting as the generative model, was the worst
among the dnMIRT results.
Owing to the fact that the lower variance in the prior of discrimination
parameters provided better skill estimation,
it can be deduced that the overestimation of the discrimination
parameters raised the estimation error.

Table \ref{tab:skill100_no_slope} presents the results of data set C:
100-skill, $a_{i,k} = 1$, and no skill correlation.
We could not obtain the results of nMIRT in this
setting because the implementation of nMIRT required significant memory
for the 100-skill estimation.
The same trends in the results of data set A can also be observed here.
As a new trend, the gap of correlation in skill inference between dnMIRT
and dcMIRT became large.
Because the number of skills and parameters to be estimated increased
significantly, the estimation errors increased compared to the results
of data set A.

Table \ref{tab:skill100_slope} presents the results of data set D:
100-skill, $a_{i,k} \sim \mbox{logN}(0, 0.25^2)$, and no skill correlation.
dnMIRT with $\sigma_a=0.1$ and smoothing provides the best MAE for the skill
of the last time step, whereas dnMIRT with $a_{i,k}=1$ and smoothing provides
the best MAE for the skill of the all time steps.
The gap of correlation in skill inference between dnMIRT and dcMIRT is large.
Since the discrimination parameters were added to be estimated, the estimation
errors increased compared to the results of data set C.

Table \ref{tab:skill100_slope2} presents the results of data set E:
100-skill, $a_{i,k} \sim \mbox{logN}(0.4, 0.2^2)$, and no skill correlation.
dnMIRT with $\mu_a=0, \sigma_a=0.25$ provides the best MAE of skill inference.
In this data set,
smoothing did not contribute to reducing the MAE of skill inference. 
dnMIRT with $\gamma_k$ estimated had worse MAE in skill inference; however,
the correlation was high.
The gap of correlation in skill inference between dnMIRT and dcMIRT is large.
Data set E has a larger scale of discrimination parameters than data set D.
In comparison with the results of data set D, the MAE of skill inference
became better, whereas the correlation became worse. The MAE of discrimination
parameters became worse and the correlation became worse.

Table \ref{tab:skill100_slope_corr} presents the results of data set F:
100-skill, $a_{i,k} \sim \mbox{logN}(0, 0.25^2)$, and skill correlation is 0.2.
dnMIRT with $a_{i,k}=1$ and $\gamma_k$ estimated provides the best MAE of skill
inference. 
The gap of correlation in skill inference between dnMIRT and dcMIRT is large.
It became even larger than the gap in the results of data set D.

The standard deviations for Tables \ref{tab:skill2_no_slope} to
\ref{tab:skill100_slope_corr} are shown in Tables
\ref{tab:skill2_no_slope_std} to \ref{tab:skill100_slope_corr_std} in Appendix.
Almost all the standard deviations are small, showing the results are stable.
Only MAEs of skill inference in data set B (2-skill) have relatively
large standard deviations; however, correlations of skill inference have
small standard deviations.

\begin{figure}
  \centering
  {\tabcolsep=0pt
  \begin{tabular}{cc}
    \begin{minipage}{0.49\linewidth}
    \includegraphics[width=\linewidth]{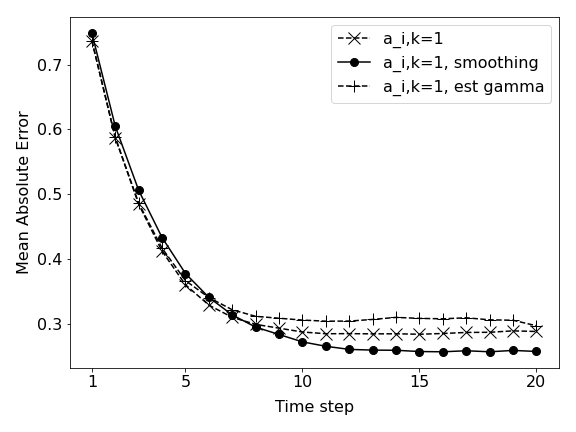}
    \subcaption{Data set A: 2-skill, $a_{i,k}=1$}
    \label{pic:mae_time_skill2_no_slope}
    \end{minipage}
    &
    \begin{minipage}{0.49\linewidth}
    \includegraphics[width=\linewidth]{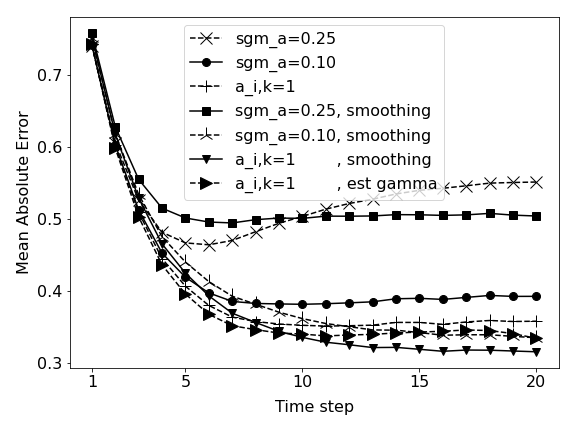}
    \subcaption{Data set B: 2-skill and $a_{i,k} \sim \mbox{logN}(0, 0.25^2)$}
    \label{pic:mae_time_skill2_slope}
    \end{minipage}\\
    \begin{minipage}{0.49\linewidth}
    \includegraphics[width=\linewidth]{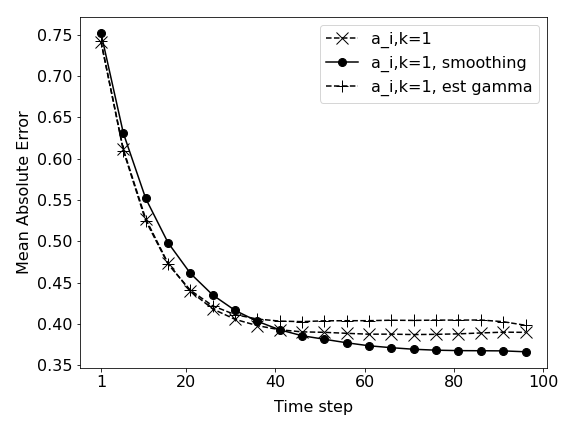}
    \subcaption{Data set C: 100-skill, $a_{i,k}=1$}
    \label{pic:mae_time_skill100_no_slope}
    \end{minipage}
    &
    \begin{minipage}{0.49\linewidth}
    \includegraphics[width=\linewidth]{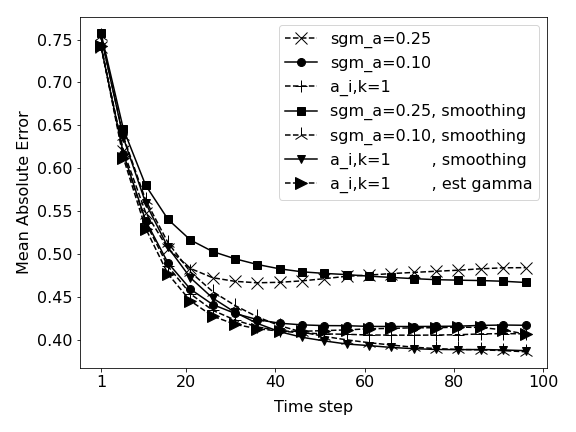}
    \subcaption{Data set D: 2-skill, $a_{i,k} \sim \mbox{logN}(0, 0.25^2)$}
    \label{pic:mae_time_skill100_slope}
    \end{minipage}\\
  \end{tabular}
  }
  \caption{
  Changes in the skill estimation errors over time. The errors decrease over time and
  saturate at approximately ten-time steps in the 2-skill data sets A and B, whereas
  fifty-time steps in the 100-skill data sets C and D.
  }
  \label{pic:mae_time}
\end{figure}

Figures \ref{pic:mae_time} (a) and (b) show the changes in MAEs over time for
the results of the data set A and B, respectively.
We only show the results of dnMIRT. It can be
seen that the MAEs decrease over time and saturate at approximately
ten-time steps. When we compare the results with smoothing and without
smoothing, it can be seen that dnMIRT without smoothing has a lower error
at the beginning of the time steps while dnMIRT with smoothing has a
lower error at the end. This trend can also be seen in the 100-skill
data sets C and D.

Figures \ref{pic:mae_time} (c) and (d) show the changes in MAEs over time for
the results of data sets C and D, respectively. The MAEs decrease over time,
similar to the 2-skill datasets, and saturate at approximately fifty-time steps.
A learner studies
ten different skills for the 100-skill datasets; therefore, approximately
five logs of
answering questions per skill are required to obtain stable skill
estimates. This is the same as the 2-skill data sets A and B.

\begin{sidewaystable}
    \begin{center}
    \small
    \caption{
    Accuracy of skill inference for data set A:
    2-skill, $a_{i,k}=1$, and no skill correlation.
    }
    \label{tab:skill2_no_slope}
    \begin{tabular}{llrrrrrrrrrrr}
    \toprule
           &        & \multicolumn{2}{c}{Skills (MAE)} & \multicolumn{4}{c}{Parameters (MAE)} & \multicolumn{2}{c}{Skills (Corr)} & \multicolumn{3}{c}{Parameters (Corr)} \\
    \cmidrule(r){3-4} \cmidrule(r){5-8} \cmidrule(r){9-10} \cmidrule{11-13}
    Model  & Settings &          All &   Last &      $b_{i,k}$ &      $d_{i,k}$ &   $\beta'_{i,k}$ &     $\gamma_k$ &      All &   Last &      $b_{i,k}$ &      $d_{i,k}$ &   $\beta'_{i,k}$ \\
    \midrule
    dnMIRT & & 0.348 & 0.289 & 0.312 & 0.080 & 0.176 & 0.030 & 0.864 & 0.797 & 0.922 & 0.264 & 0.832 \\
           & est. $\gamma_k$ & 0.363 & 0.298 & 0.325 & 0.084 & 0.164 & 0.211 & 0.842 & 0.727 & 0.913 & 0.304 & 0.837 \\
           & sm. & 0.339 & 0.258 & 0.319 & 0.095 & 0.201 & 0.030 & 0.854 & 0.788 & 0.916 & 0.268 & 0.791 \\
    dcMIRT & & 1.187 & 1.429 & - & - & - & - & 0.811 & 0.765 & - & - & - \\
    nMIRT  & window=5 & 1.148 & 1.364 & 0.964 & - & - & - & 0.382 & 0.147 & 0.513 & - & - \\
           & window=10 & 0.892 & 0.998 & 0.964 & - & - & - & 0.334 & 0.118 & 0.513 & - & - \\
    \bottomrule
    \multicolumn{12}{l}{
    Smoothing is abbreviated as ``sm.''. Estimate $\gamma_k$ is abbreviated as
    ``est. $\gamma_k$''.
    }
    \end{tabular}
    \end{center}
\end{sidewaystable}

\begin{sidewaystable}
    \centering
    \small
    \caption{
    Accuracy of skill inference for data set B:
    2-skill, $a_{i,k} \sim \mbox{logN}(0, 0.25^2)$, and no skill correlation.
    }
    \label{tab:skill2_slope}
    \begin{tabular}{llrrrrrrrrrrrrr}
    \toprule
           &        & \multicolumn{2}{c}{Skills (MAE)} & \multicolumn{5}{c}{Parameters (MAE)} & \multicolumn{2}{c}{Skills (Corr)} & \multicolumn{4}{c}{Parameters (Corr)} \\
    \cmidrule(r){3-4} \cmidrule(r){5-9} \cmidrule(r){10-11} \cmidrule{12-15}
    Model  & Settings &          All &   Last &                $a_{i,k}$ &      $b_{i,k}$ &      $d_{i,k}$ &   $\beta'_{i,k}$ & $\gamma_k$ &         All &   Last &                 $a_{i,k}$ &      $b_{i,k}$ &      $d_{i,k}$ &   $\beta'_{i,k}$ \\
    \midrule
    dnMIRT & $\sigma_a=0.25$ &        0.531 &  0.551 &            0.225 &  0.371 &  0.074 &  0.221 & 0.030 &     0.878 &  0.835 &             0.572 &  0.884 &  0.301 &  0.855 \\
           & $\sigma_a=0.1$ &        0.428 &  0.393 &            0.190 &  0.395 &  0.075 &  0.211 & 0.030 &        0.876 &  0.819 &             0.589 &  0.861 &  0.288 &  0.846 \\
           & $a_{i,k}=1$ &        0.404 &  0.358 &            0.209 &  0.429 &  0.076 &  0.209 & 0.030 &     0.872 &  0.803 &               - &  0.841 &  0.284 &  0.840 \\
           & $a_{i,k}=1$, est. $\gamma_k$ & 0.392 &     0.335 &  0.209 &  0.441 &  0.083 &     0.185 &      0.230 &    0.850 &     0.735 &    - &  0.838 &  0.319 &     0.845 \\
           & $\sigma_a=0.25$, sm. &        0.525 &  0.504 &            0.279 &  0.385 &  0.084 &  0.259 & 0.030 & 0.850 &  0.769 &             0.506 &  0.867 &  0.293 &  0.817 \\
           & $\sigma_a=0.1$, sm. &        0.409 &  0.336 &            0.191 &  0.409 &  0.091 &  0.239 & 0.030 & 0.858 &  0.781 &             0.546 &  0.854 &  0.288 &  0.812 \\
           & $a_{i,k}=1$, sm. &        0.390 &  0.316 &            0.209 &  0.450 &  0.093 &  0.234 & 0.030 & 0.855 &  0.765 &               - &  0.830 &  0.285 &  0.803 \\
    dcMIRT & $\sigma_a=0.25$ &        1.364 &  1.644 &              - &    - &    - &    - & - &     0.817 &  0.813 &               - &    - &    - &    - \\
           & $\sigma_a=0.1$ &        1.315 &  1.580 &              - &    - &    - &    - & - &        0.825 &  0.805 &               - &    - &    - &    - \\
           & $a_{i,k}=1$ &        1.307 &  1.569 &              - &    - &    - &    - & - & 0.824 &  0.797 &               - &    - &    - &    - \\
    nMIRT & window=5 &        1.331 &  1.598 &            0.296 &  1.112 &    - &    - & - & 0.387 &  0.134 &             0.015 &  0.503 &    - &    - \\
          & window=10 &        1.041 &  1.189 &            0.296 &  1.112 &    - &    - & - &        0.341 &  0.110 &             0.015 &  0.503 &    - &    - \\
    \bottomrule
    \multicolumn{12}{l}{
    Smoothing is abbreviated as ``sm.''. Estimate $\gamma_k$ is abbreviated as
    ``est. $\gamma_k$''.
    }
    \end{tabular}
\end{sidewaystable}

\begin{sidewaystable}
    \centering
    \small
    \caption{
    Accuracy of skill inference for data set C:
    100-skill, $a_{i,k} = 1$, and no skill correlation.
    }
    \label{tab:skill100_no_slope}
    \begin{tabular}{llrrrrrrrrrrr}
    \toprule
           &        & \multicolumn{2}{c}{Skills (MAE)} & \multicolumn{4}{c}{Parameters (MAE)} & \multicolumn{2}{c}{Skills (Corr)} & \multicolumn{3}{c}{Parameters (Corr)} \\
    \cmidrule(r){3-4} \cmidrule(r){5-8} \cmidrule(r){9-10} \cmidrule{11-13}
    Model  & Settings &          All &   Last &      $b_{i,k}$ &      $d_{i,k}$ &   $\beta'_{i,k}$ & $\gamma_k$ &    All &   Last &      $b_{i,k}$ &      $d_{i,k}$ &   $\beta'_{i,k}$ \\
    \midrule
    dnMIRT &        &        0.426 &  0.390 &            0.444 &  0.079 &  0.209 & 0.030 &        0.801 &  0.563 &             0.809 &  0.172 &  0.697 \\
           & est. $\gamma_k$ & 0.436 & 0.392 &  0.452 &  0.087 &     0.189 &      0.212 &    0.778 &     0.543 &    0.805 &  0.211 &     0.714 \\
           & sm. &        0.424 &  0.365 &            0.459 &  0.101 &  0.232 & 0.030 &        0.787 &  0.567 &             0.796 &  0.197 &  0.666 \\
    dcMIRT &        &        1.413 &  1.682 &              - &    - &    - & - &        0.587 &  0.439 &               - &    - &    - \\
    \bottomrule
    \multicolumn{12}{l}{
    Smoothing is abbreviated as ``sm.''. Estimate $\gamma_k$ is abbreviated as
    ``est. $\gamma_k$''.
    }
    \end{tabular}
\end{sidewaystable}

\begin{sidewaystable}
    \centering
    \small
    \caption{
    Accuracy of skill inference for data set D:
    100-skill, $a_{i,k} \sim \mbox{logN}(0, 0.25^2)$, and no skill correlation.
    }
    \label{tab:skill100_slope}
    \begin{tabular}{llrrrrrrrrrrrrr}
    \toprule
           &              & \multicolumn{2}{c}{Skills (MAE)} & \multicolumn{5}{c}{Parameters (MAE)} & \multicolumn{2}{c}{Skills (Corr)} & \multicolumn{4}{c}{Parameters (Corr)} \\
    \cmidrule(r){3-4} \cmidrule(r){5-9} \cmidrule(r){10-11} \cmidrule{12-15}
    Model  & Settings & All & Last &                $a_{i,k}$ &      $b_{i,k}$ &      $d_{i,k}$ &   $\beta'_{i,k}$ & $\gamma_k$ &        All &   Last &                 $a_{i,k}$ &      $b_{i,k}$ &      $d_{i,k}$ &   $\beta'_{i,k}$ \\
    \midrule
    dnMIRT & $\sigma_a=0.25$ &        0.496 &  0.486 &            0.194 &  0.459 &  0.080 &  0.233 & 0.030 &        0.797 &  0.559 &             0.461 &  0.791 &  0.178 &  0.734 \\
           & $\sigma_a=0.1$ &        0.450 &  0.418 &            0.192 &  0.473 &  0.080 &  0.229 & 0.030 &       0.814 &  0.622 &             0.473 &  0.779 &  0.170 &  0.734 \\
           & $a_{i,k}=1$ &        0.442 &  0.408 &            0.204 &  0.488 &  0.081 &  0.229 & 0.030 &        0.810 &  0.609 &               - &  0.767 &  0.167 &  0.728 \\
           & $a_{i,k}=1$, est. $\gamma_k$ & 0.443 &     0.400 &  0.204 &  0.495 &  0.090 &     0.198 &      0.233 &    0.794 &     0.614 &    - &  0.771 &  0.212 &     0.745 \\
           & $\sigma_a=0.25$, sm. &        0.504 &  0.466 &            0.229 &  0.475 &  0.095 &  0.260 & 0.030 &        0.775 &  0.544 &             0.401 &  0.773 &  0.206 &  0.703 \\
           & $\sigma_a=0.1$, sm. &        0.443 &  0.385 &            0.193 &  0.483 &  0.100 &  0.250 & 0.030 &        0.799 &  0.613 &             0.433 &  0.770 &  0.196 &  0.711 \\
           & $a_{i,k}=1$, sm. &        0.440 &  0.387 &            0.204 &  0.505 &  0.101 &  0.247 & 0.030 &        0.794 &  0.598 &               - &  0.756 &  0.190 &  0.703 \\
    dcMIRT & $\sigma_a=0.25$ &        1.432 &  1.704 &              - &    - &    - &    - & - &         0.606 &  0.451 &               - &    - &    - &    - \\
           & $\sigma_a=0.1$ &        1.435 &  1.708 &              - &    - &    - &    - & - &        0.603 &  0.447 &               - &    - &    - &    - \\
           & $a_{i,k}=1$ &        1.439 &  1.714 &              - &    - &    - &    - & - &         0.604 &  0.448 &               - &    - &    - &    - \\
    \bottomrule
    \multicolumn{12}{l}{
    Smoothing is abbreviated as ``sm.''. Estimate $\gamma_k$ is abbreviated as
    ``est. $\gamma_k$''.
    }
    \end{tabular}
\end{sidewaystable}

\begin{sidewaystable}
    \centering
    \small
    \caption{
    Accuracy of skill inference for data set E:
    100-skill, $a_{i,k} \sim \mbox{logN}(0.4, 0.2)$, and no skill correlation.
    }
    \label{tab:skill100_slope2}
\footnotesize
\begin{tabular}{llrrrrrrrrrrrrr}
\toprule
           &              & \multicolumn{2}{c}{Skills (MAE)} & \multicolumn{5}{c}{Parameters (MAE)} & \multicolumn{2}{c}{Skills (Corr)} & \multicolumn{4}{c}{Parameters (Corr)} \\
    \cmidrule(r){3-4} \cmidrule(r){5-9} \cmidrule(r){10-11} \cmidrule{12-15}
    Model  & Settings & All & Last &                $a_{i,k}$ &      $b_{i,k}$ &      $d_{i,k}$ &   $\beta'_{i,k}$ & $\gamma_k$ &    All &   Last &                 $a_{i,k}$ &      $b_{i,k}$ &      $d_{i,k}$ &   $\beta'_{i,k}$ \\
\midrule
dnMIRT & $\mu_a=0.4, \sigma_a=0.2$  &        0.438 &  0.449 &            0.253 &  0.394 &  0.121 &  0.165 & 0.030 &        0.749 &  0.635 &             0.342 &  0.855 &  0.154 &  0.831 \\
       & $\mu_a=0, \sigma_a=0.1$ &        0.385 &  0.359 &            0.493 &  0.441 &  0.089 &  0.177 & 0.030 &         0.756 &  0.635 &             0.323 &  0.835 &  0.104 &  0.823 \\
       & $\mu_a=0, \sigma_a=0.25$ &        0.379 &  0.353 &            0.358 &  0.401 &  0.090 &  0.172 & 0.030 &       0.742 &  0.596 &             0.323 &  0.848 &  0.124 &  0.825 \\
       & $a_{i,k}=1$ & 0.402 &     0.386 &  0.526 &  0.464 &  0.090 &     0.180 & 0.030 &   0.749 &     0.620 &    - &  0.829 &  0.099 &     0.819 \\
       & $a_{i,k}=1$, est. $\gamma_k$ & 0.642 &     0.685 &  0.526 &  0.514 &  0.097 &     0.221 &      0.242 &    0.750 &     0.670 &    - &  0.837 &  0.237 &     0.818 \\
       & $\mu_a=0.4, \sigma_a=0.2$, sm. &        0.417 &  0.391 &            0.312 &  0.404 &  0.082 &  0.175 & 0.030 &        0.736 &  0.623 &             0.273 &  0.842 &  0.218 &  0.807 \\
       & $\mu_a=0, \sigma_a=0.1$, sm. &        0.427 &  0.434 &            0.485 &  0.448 &  0.101 &  0.166 & 0.030 &        0.757 &  0.656 &             0.300 &  0.843 &  0.153 &  0.809 \\
       & $\mu_a=0, \sigma_a=0.25$, sm. &        0.397 &  0.371 &            0.338 &  0.411 &  0.089 &  0.169 & 0.030 &        0.734 &  0.596 &             0.287 &  0.847 &  0.186 &  0.800 \\
       & $a_{i,k}=1$, sm. & 0.454 &     0.482 &  0.526 &  0.475 &  0.105 &     0.168 & 0.030 &    0.753 &     0.648 &    - &  0.837 &  0.145 &     0.804 \\
dcMIRT & $\mu_a=0.4, \sigma_a=0.2$ &        1.093 &  1.279 &              - &    - &    - &    - & - &        0.555 &  0.497 &               - &    - &    - &    - \\
       & $\mu_a=0, \sigma_a=0.1$ &        1.053 &  1.232 &              - &    - &    - &    - & - &         0.525 &  0.471 &               - &    - &    - &    - \\
       & $\mu_a=0, \sigma_a=0.25$ &        1.047 &  1.224 &              - &    - &    - &    - & - &        0.536 &  0.482 &               - &    - &    - &    - \\
    \bottomrule
    \multicolumn{12}{l}{
    Smoothing is abbreviated as ``sm.''. Estimate $\gamma_k$ is abbreviated as
    ``est. $\gamma_k$''.
    }
\end{tabular}
\normalsize
\end{sidewaystable}

\begin{sidewaystable}
    \centering
    \small
    \caption{
    Accuracy of skill inference for data set F: 
    100-skill, $a_{i,k} \sim \mbox{logN}(0, 0.25)$, and skill correlation is 0.2.
    }
    \label{tab:skill100_slope_corr}
\begin{tabular}{llrrrrrrrrrrrrr}
\toprule
           &              & \multicolumn{2}{c}{Skills (MAE)} & \multicolumn{5}{c}{Parameters (MAE)} & \multicolumn{2}{c}{Skills (Corr)} & \multicolumn{4}{c}{Parameters (Corr)} \\
    \cmidrule(r){3-4} \cmidrule(r){5-9} \cmidrule(r){10-11} \cmidrule{12-15}
    Model  & Settings & All & Last &                $a_{i,k}$ &      $b_{i,k}$ &      $d_{i,k}$ &   $\beta'_{i,k}$ & $\gamma_k$ &       All &   Last &                 $a_{i,k}$ &      $b_{i,k}$ &      $d_{i,k}$ &   $\beta'_{i,k}$ \\
\midrule
dnMIRT & $\sigma_a=0.25$ &        0.559 &  0.583 &            0.210 &  0.462 &  0.076 &  0.251 & 0.030 &         0.806 &  0.582 &             0.434 &  0.799 &  0.105 &  0.765 \\
       & $\sigma_a=0.1$ &        0.494 &  0.482 &            0.189 &  0.473 &  0.077 &  0.251 & 0.030 &         0.819 &  0.627 &             0.444 &  0.785 &  0.096 &  0.761 \\
       & $a_{i,k}=1$ &        0.478 &  0.458 &            0.200 &  0.485 &  0.078 &  0.251 & 0.030 &         0.814 &  0.609 &               - &  0.775 &  0.093 &  0.755 \\
       & $a_{i,k}=1$, est. $\gamma_k$ & 0.440 &     0.401 &  0.200 &  0.493 &  0.085 &     0.197 &      0.216 &    0.795 &     0.597 &    - &  0.775 &  0.217 &     0.753 \\
       & $\sigma_a=0.25$, sm. &        0.531 &  0.511 &            0.251 &  0.474 &  0.085 &  0.268 & 0.030 &        0.788 &  0.575 &             0.371 &  0.782 &  0.150 &  0.737 \\
       & $\sigma_a=0.1$, sm. &        0.455 &  0.400 &            0.190 &  0.476 &  0.093 &  0.261 & 0.030 &         0.808 &  0.633 &             0.405 &  0.781 &  0.134 &  0.741 \\
       & $a_{i,k}=1$, sm. &        0.444 &  0.389 &            0.200 &  0.494 &  0.096 &  0.259 & 0.030 &         0.803 &  0.617 &               - &  0.769 &  0.129 &  0.734 \\
dcMIRT & $\sigma_a=0.25$ &        1.540 &  1.861 &              - &    - &    - &    - & - &         0.558 &  0.418 &               - &    - &    - &    - \\
       & $\sigma_a=0.1$ &        1.539 &  1.862 &              - &    - &    - &    - & - &         0.549 &  0.406 &               - &    - &    - &    - \\
       & $a_{i,k}=1$ &        1.540 &  1.864 &              - &    - &    - &    - & - &         0.550 &  0.407 &               - &    - &    - &    - \\
    \bottomrule
    \multicolumn{12}{l}{
    Smoothing is abbreviated as ``sm.''. Estimate $\gamma_k$ is abbreviated as
    ``est. $\gamma_k$''.
    }
    \end{tabular}
\end{sidewaystable}

\section{Evaluation of Prediction}\label{sec:prediction_simulation}
 
We evaluate the prediction performance of the dynamical non-compensatory
MIRT (dnMIRT) model and the dynamical compensatory MIRT (dcMIRT) model
on the simulation data described in Section \ref{sec:simulation_data}.
Because the data set consists of sequences of correct (y=1) or incorrect
(y=0) answers for learners, this is a sequential binary classification
problem.

We used data sets D and F and employed a 5-fold cross-validation
to measure prediction performance; therefore, 80\% of learners'
logs were used for training and 20\% were used for testing.
In the 20\% test data, the forward message was used to predict
whether a learner can correctly answer a question or not, one by one.
We excluded the first log for each learner from the evaluation 
in the same way as Piech et al. (2015).
The evaluation metrics we used are AUC, average precision for 
correct answers, and average precision for incorrect answers.
The hyperparameter settings were the same as Section
\ref{sec:inf_setting_simulation}. We tried three cases of
hyperparameters of the discrimination parameters: 
$\sigma_a = \{0.1, 0.25\}$ and $a_{i,k}=1$ (fixed to one).
$\gamma_k$ was only fixed to $0.2^2$.

Table \ref{tab:prediction_simulation} shows the prediction accuracy
of dnMIRT and dcMIRT for data sets D and F. In both data sets, we can see
that dnMIRT is better than dcMIRT in three metrics. The gap in average
precision for incorrect answers between dnMIRT and dcMIRT is larger than
the one for correct answers, meaning that dnMIRT is much better at
predicting incorrect answers. Informing users questions that the
users are estimated to answer the questions incorrectly is more common;
therefore, prediction models with high average precision for incorrect
answers are useful. When we look at the effect of the hyperparameters, 
we can see that the prediction performance is not sensitive to the
settings of the discrimination parameters.
The difference between data sets D and F is whether skills in initial
states correlate or not. The performance gap between dnMIRT and
dcMIRT in data set F is a little smaller than the gap in data set D.

\begin{table}[p]
    \centering
    \caption{
    Accuracy of prediction on the data set D (100-skill,
    $a_{i,k} \sim \mbox{logN}(0, 0.25)$, and no skill correlation) and
    F (100-skill, $a_{i,k} \sim \mbox{logN}(0, 0.25)$, and skill correlation is 0.2).
    dnMIRT is better than dcMIRT in both data sets.
    }
    \label{tab:prediction_simulation}
\begin{tabular}{lllrrr}
\toprule
Data & Method & Settings & AUC & Average Precision (y=1) & Average Precision (y=0)\\
\midrule
D & dnMIRT & $a=1$ &  0.791 &  0.892 &  0.631 \\
  &    & $\sigma_a=0.1$ &  0.791 &  0.892 &  0.631 \\
  &    & $\sigma_a=0.25$ &  \textbf{0.792} &  \textbf{0.893} &  \textbf{0.632} \\
  & dcMIRT & $a=1$ &  0.765 &  0.878 &  0.586 \\
  &    & $\sigma_a=0.1$ &  0.766 &  0.878 &  0.587 \\
  &    & $\sigma_a=0.25$ &  0.769 &  0.880 &  0.592 \\
\midrule
F & dnMIRT & $a=1$ &  0.789 &  0.895 &  0.616 \\
  &     & $\sigma_a=0.1$ &  0.789 &  0.895 &  \textbf{0.617} \\
  &     & $\sigma_a=0.25$ &  \textbf{0.790} &  \textbf{0.895} &  \textbf{0.617} \\
  & dcMIRT & $a=1$ &  0.764 &  0.881 &  0.573 \\
  &     & $\sigma_a=0.1$ &  0.765 &  0.882 &  0.574 \\
  &     & $\sigma_a=0.25$ &  0.768 &  0.884 &  0.579 \\
\bottomrule
\end{tabular}
\end{table}

\section{Evaluation of $\hat{\alpha}$ Message Approximation}\label{sec:eval_posterior}

In this section, we evaluate the Gaussian approximation of the $\hat{\alpha}$ message
introduced in Section \ref{subsec:gauss_apx}. First, we visually confirm that our
approximated posterior adequately approximates the true posterior by showcasing
two-dimensional simulation results.
Second, the approximation errors are quantitatively
compared between our method and the Laplace approximation varying the number of
dimensions in the latent state. Third, the relationship between
the approximation error and the number of samples used in the approximation
is investigated to clarify the sufficient sample size.

\subsection{Visual Evaluation of Approximated Posterior}

To confirm that the proposed Gaussian approximation
adequately approximates the true posterior ($\hat{\alpha}$-message),
we visualize some simulation results
in two dimensions and compare the proposed method with the Laplace approximation
(MacKay 2003; Bishop 2006).

\begin{figure}
 \centering
 {\tabcolsep=0pt
 \begin{tabular}{cccc}
  & Likelihood and Prior & Posterior (Proposed) & Posterior (Laplace) \\
  (a)
  &
  \begin{minipage}{0.3\linewidth}
  \includegraphics[width=\linewidth]{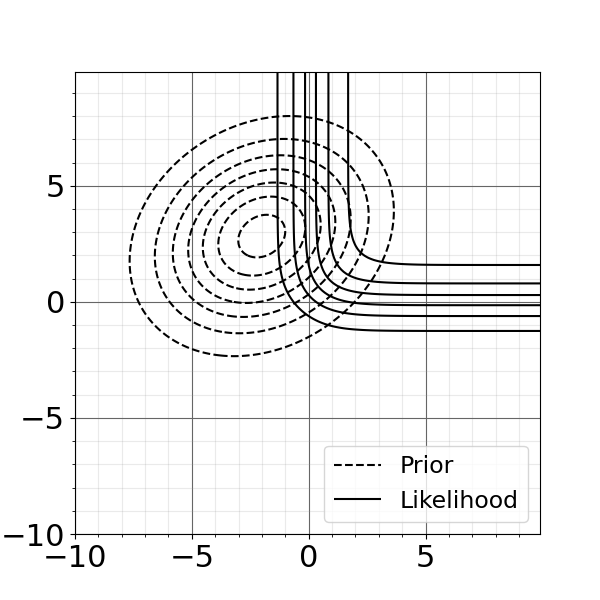}
  \end{minipage}
  &
  \begin{minipage}{0.3\linewidth}
  \includegraphics[width=\linewidth]{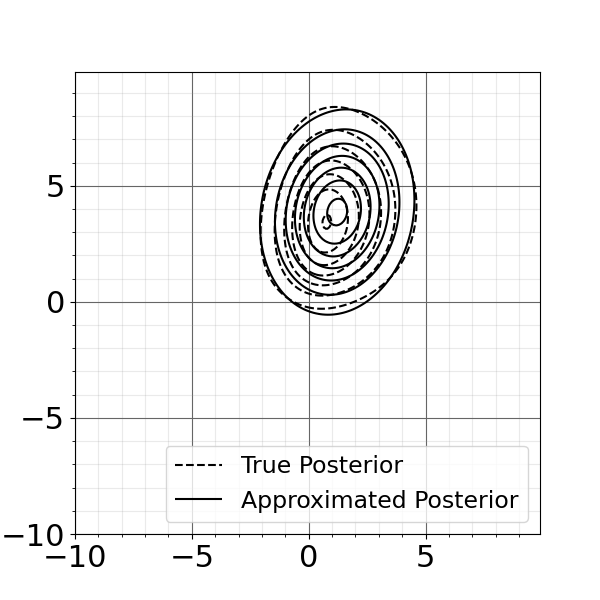}
  \end{minipage}
  &
  \begin{minipage}{0.3\linewidth}
  \includegraphics[width=\linewidth]{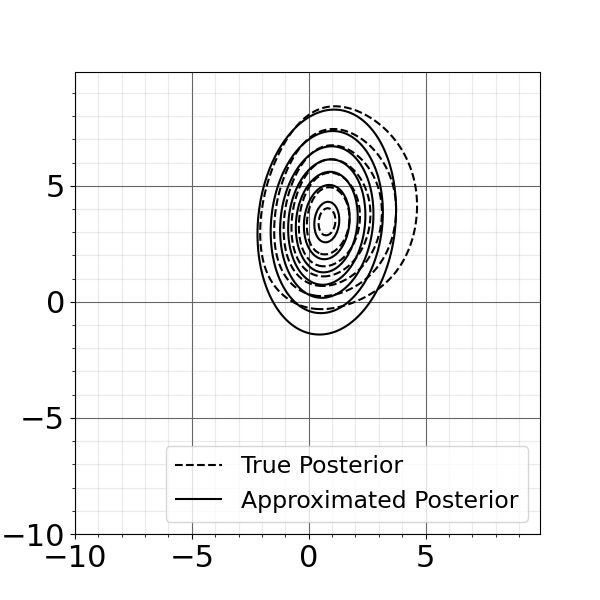}
  \end{minipage}\\
  (b)
  &
  \begin{minipage}{0.3\linewidth}
  \includegraphics[width=\linewidth]{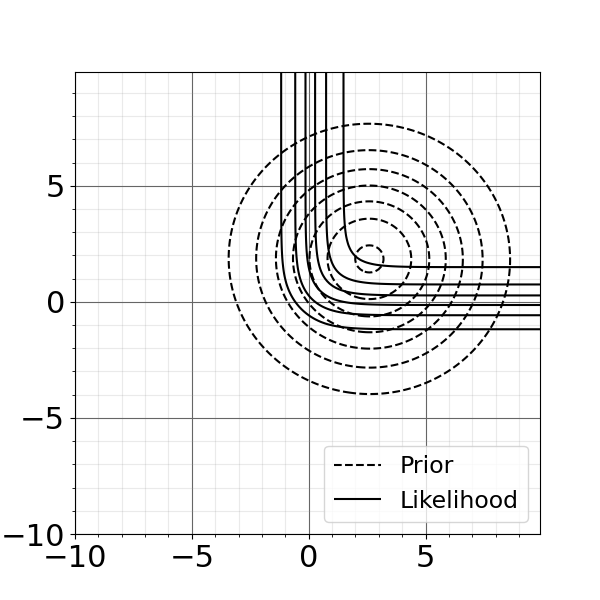}
  \end{minipage}
  &
  \begin{minipage}{0.3\linewidth}
  \includegraphics[width=\linewidth]{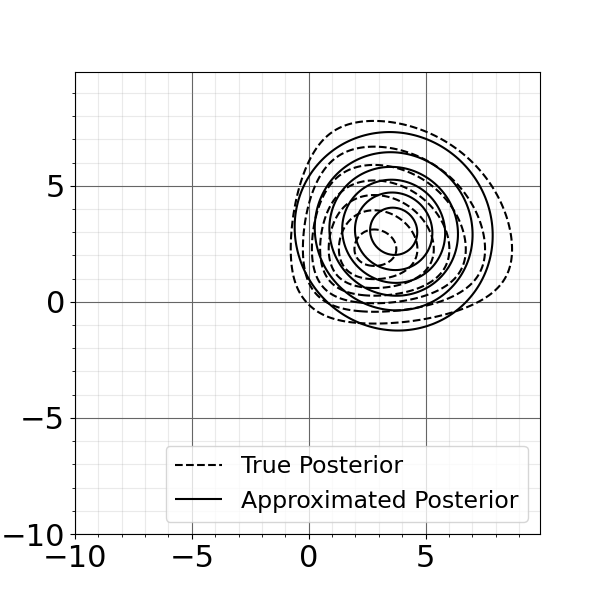}
  \end{minipage}
  &
  \begin{minipage}{0.3\linewidth}
  \includegraphics[width=\linewidth]{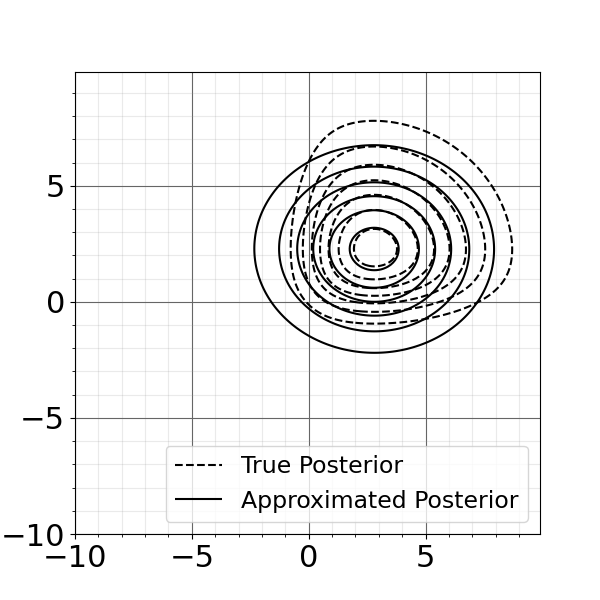}
  \end{minipage}\\
  (c)
  &
  \begin{minipage}{0.3\linewidth}
  \includegraphics[width=\linewidth]{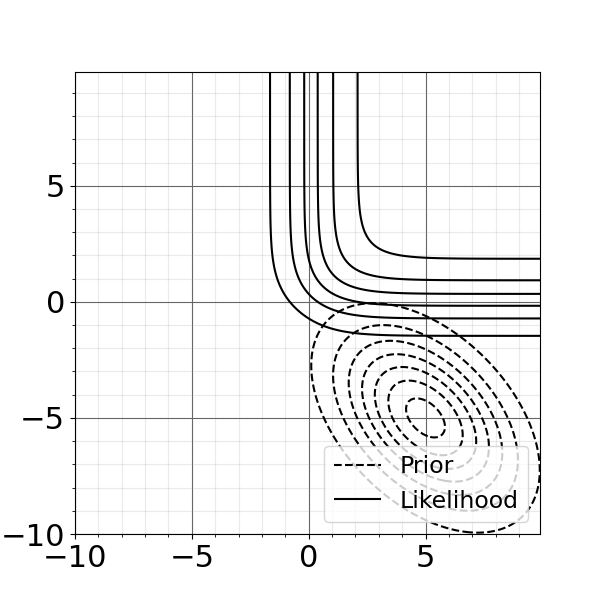}
  \end{minipage}
  &
  \begin{minipage}{0.3\linewidth}
  \includegraphics[width=\linewidth]{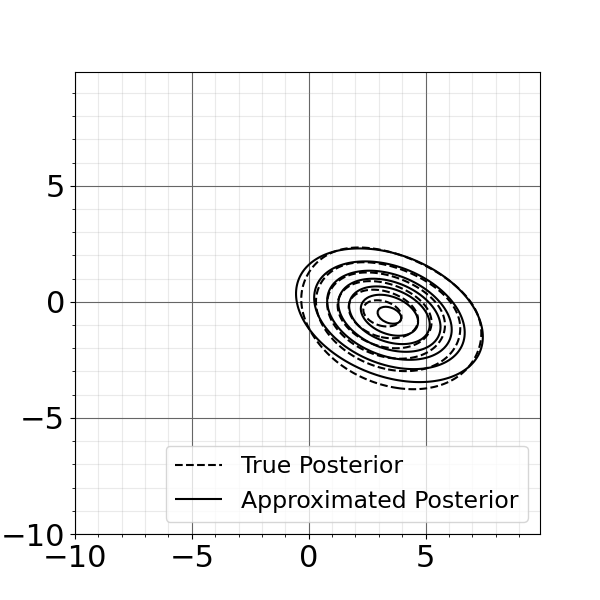}
  \end{minipage}
  &
  \begin{minipage}{0.3\linewidth}
  \includegraphics[width=\linewidth]{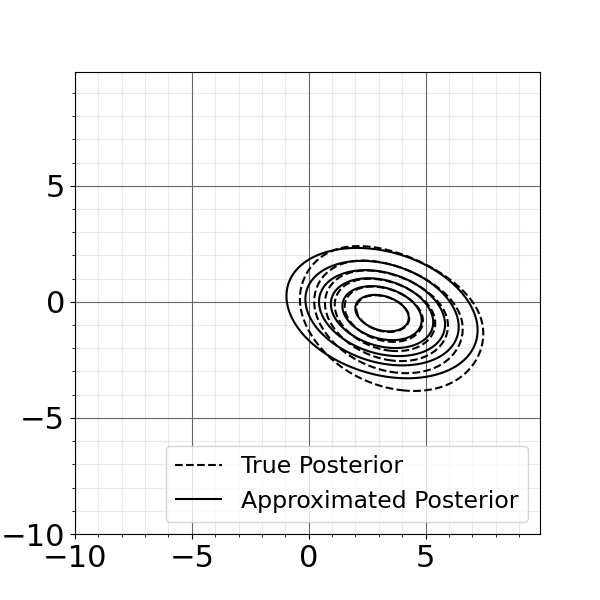}
  \end{minipage}
  \end{tabular}
  }
\caption{Comparison between the true posterior and approximated posterior
for correct answers $y=1$. The posterior of the proposed
method approximates the true posterior more adequately than the Laplace
approximation.
}\label{pic:alpha_message_apx_y1}
\end{figure}

Figure \ref{pic:alpha_message_apx_y1} shows three cases (a)-(c) for
correct answers: i.e., $y=1$ in Eq. \eqref{eqn:likelihood_of_gauss_apx}. 
Given the Gaussian priors $p(\bm{z})$ (dashed line) and the non-compensatory
likelihood functions $p(y=1|\bm{z})$ (solid line) in the left column,
we display the approximated
posteriors of the proposed method (solid line) and the true posteriors (dashed line)
in the middle column.
For comparison, we also display the posteriors obtained from the Laplace
approximation (solid line), and true posteriors (dashed line) in the right column.
We can see that the proposed method adequately approximates the true posteriors
in case (a)-(c) from the middle column. In contrast, the Laplace approximation
underestimates the variance in x-axis and overestimates the variance in y-axis
in case (a), and extends to the lower left in case (b).
Owning to the fact that the Laplace approximation only considers the
maximum point of the true posterior and the Hessian matrix, the error tends
to be large as it extends far from the maximum point.

\begin{figure}
 \centering
 {\tabcolsep=0pt
 \begin{tabular}{cccc}
   & Likelihood and Prior & Posterior (Proposed) & Posterior (Laplace) \\
  (d)
  &
  \begin{minipage}{0.32\linewidth}
  \includegraphics[width=\linewidth]{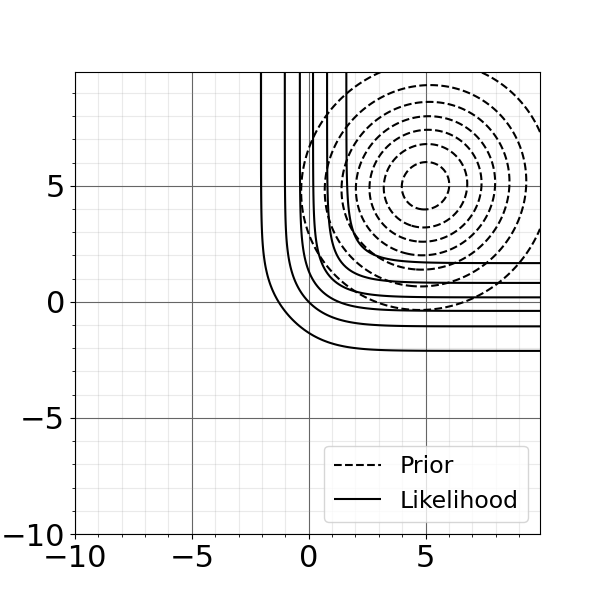}
  \end{minipage}
  &
  \begin{minipage}{0.32\linewidth}
  \includegraphics[width=\linewidth]{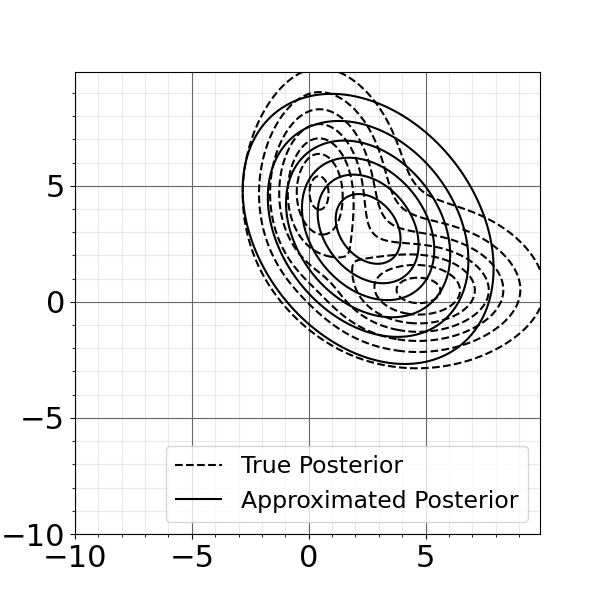}
  \end{minipage}
  &
  \begin{minipage}{0.32\linewidth}
  \includegraphics[width=\linewidth]{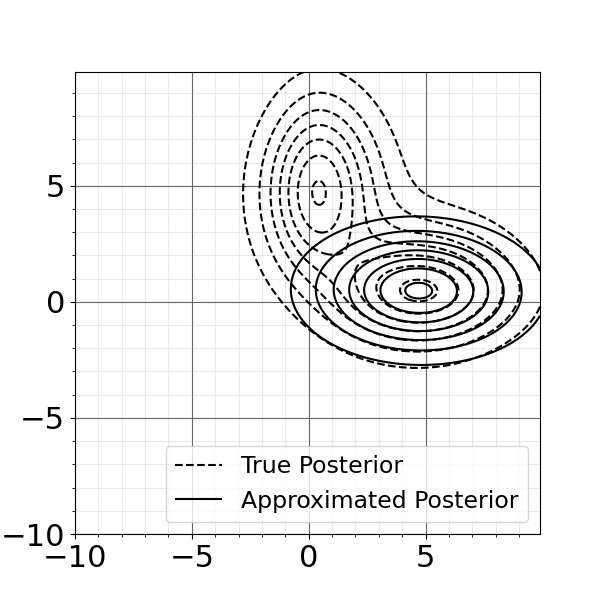}
  \end{minipage}\\
  (e)
  &
  \begin{minipage}{0.32\linewidth}
  \includegraphics[width=\linewidth]{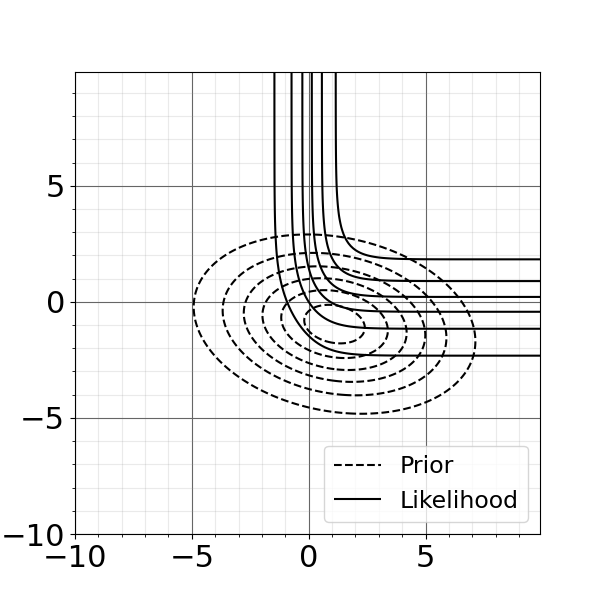}
  \end{minipage}
  &
  \begin{minipage}{0.32\linewidth}
  \includegraphics[width=\linewidth]{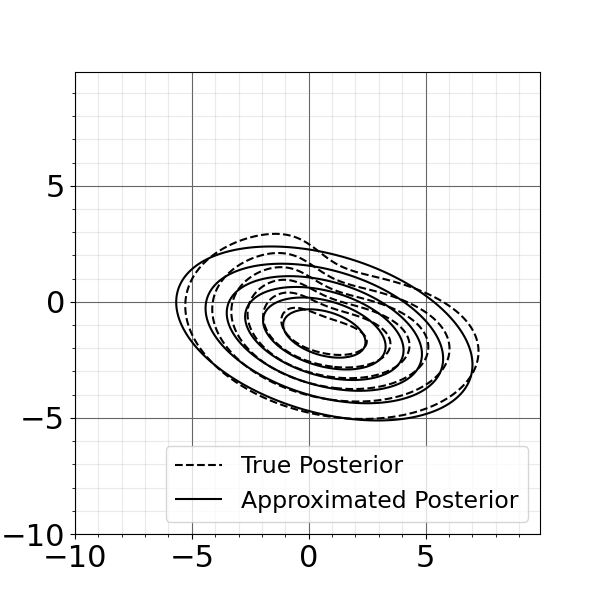}
  \end{minipage}
  &
  \begin{minipage}{0.32\linewidth}
  \includegraphics[width=\linewidth]{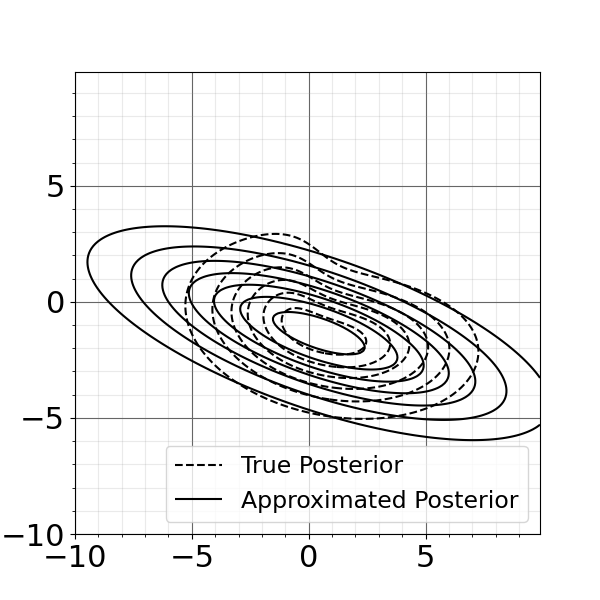}
  \end{minipage}\\
  (f)
  &
  \begin{minipage}{0.32\linewidth}
  \includegraphics[width=\linewidth]{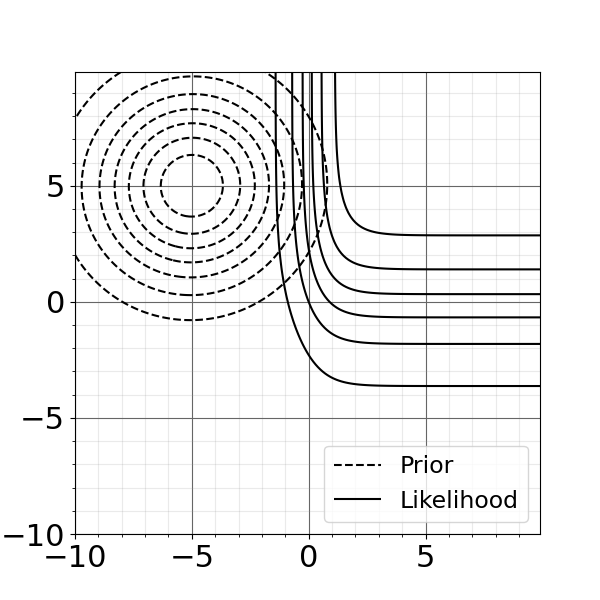}
  \end{minipage}
  &
  \begin{minipage}{0.32\linewidth}
  \includegraphics[width=\linewidth]{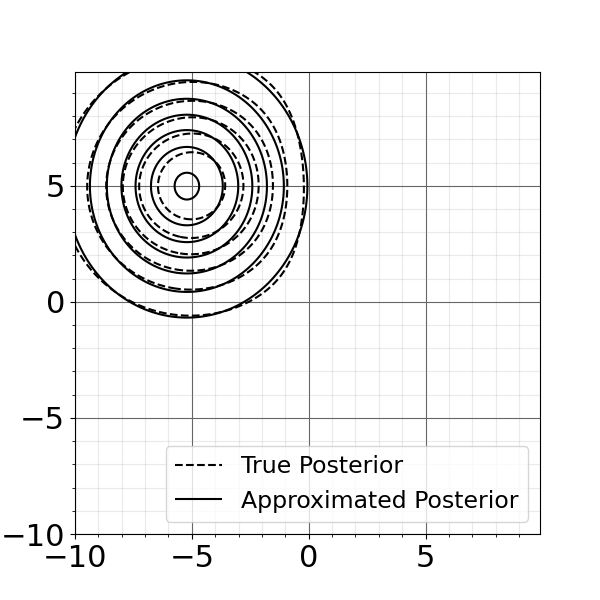}
  \end{minipage}
  &
  \begin{minipage}{0.32\linewidth}
  \includegraphics[width=\linewidth]{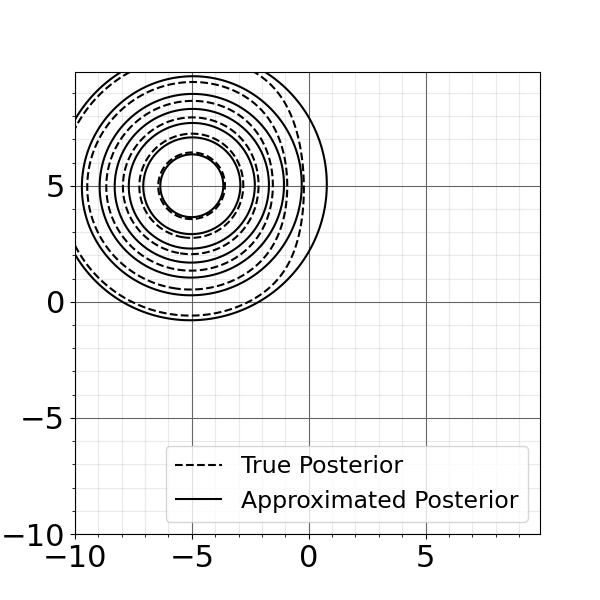}
  \end{minipage}
  \end{tabular}
  }
\caption{Comparison between the true posterior and approximated posterior
for incorrect answers $y=0$. True posteriors are more complicated
than the cases of correct answers. The posterior of the proposed
method approximates the true posterior adequately for the three cases.}
\label{pic:alpha_message_apx_y0}
\end{figure}

Figure \ref{pic:alpha_message_apx_y0} shows three cases (d)-(f) for
incorrect answers, i.e., $y=0$.
The true posteriors for the cases of incorrect answers tend to be more complex
than in the cases of correct answers. In case (d), the proposed method
covers the two peaks of the true posterior, whereas the Laplace approximation
only covers one of the peaks. In cases (e) and (f), the proposed method
covers the true posterior correctly; however, the Laplace approximation
overestimates the variance.

\subsection{Quantitative Evaluation of Approximated Posterior}
\label{subsec:dimension_to_kl}

Here, we quantitatively compare the approximation errors of
our method and the Laplace approximation varying the number of
dimensions in the latent state.
We randomly generated
1,000 likelihood functions $p(y|\bm{z})$ and the priors $p(\bm{z})$ for 
each number of dimensions (one to five). The discrimination parameters of 
the likelihood functions were drawn from the uniform distribution
$\mbox{U}(0.5, 2.0)$, and the difficulty parameters
were fixed to zero. The mean parameters of the priors were drawn
from the uniform distribution $\mbox{U}(-1, 1)$,
and the covariance parameters were generated by the formula $PDP^T$,
where $P$ is the randomly generated orthonormal matrix and $D$ is the
diagonal matrix whose diagonal elements are drawn from the uniform
distribution $\mbox{U}(0, 10)$. The proposed approximation
method and the Laplace approximation were applied to the
abovementioned likelihood
functions and priors, and the approximation error was subsequently
computed by
the KL-divergence between the approximated posterior and the true posterior.
Then, the average of the KL-divergence for each number of dimensions
was computed.

\begin{figure}[tb]
  \centering
  \begin{minipage}{.48\linewidth}
    \includegraphics[width=.9\linewidth]{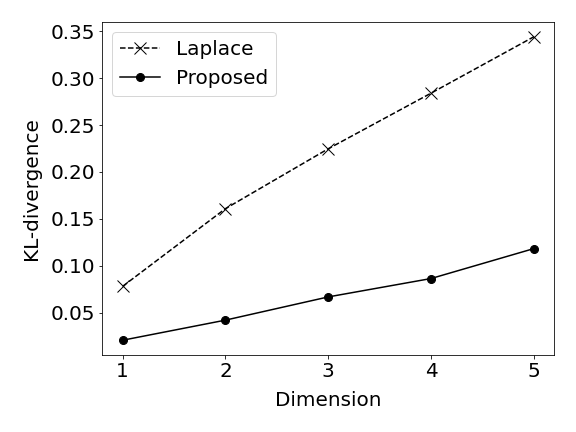}
    \subcaption{Correct answer case: $y=1$}
  \end{minipage}
  \begin{minipage}{.48\linewidth}
    \includegraphics[width=.9\linewidth]{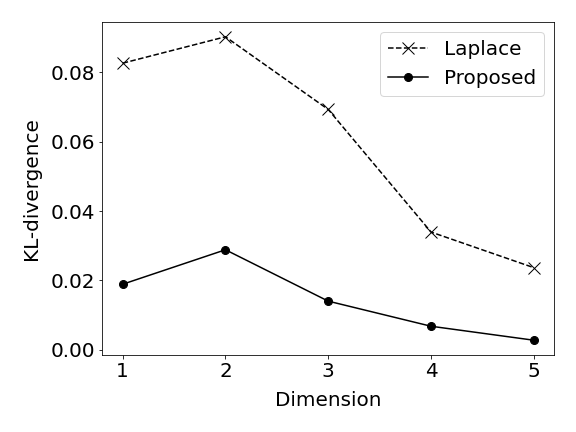}
    \subcaption{Incorrect answer case: $y=0$}
  \end{minipage}
\caption{Comparison of the approximation errors between the proposed method and
the Laplace approximation.
The proposed method shows smaller errors than the Laplace method for both cases.
The error increases when the number of dimensions increases for the case of correct
answers, whereas it decreases for the case of incorrect answers.
}
\label{pic:dimension_to_kl}
\end{figure}

Figure \ref{pic:dimension_to_kl} shows the results for a correct
answer $y=1$ and an incorrect answer $y=0$. We observe that the
KL-divergence of the proposed method is always smaller than that
of the Laplace approximation for both cases. Furthermore, it can be
observed that the
KL-divergence increases as the number of dimensions increases for
the former case, whereas it decreases for the latter case.
The shape of
the posterior for an incorrect answer can be considered to be 
a hyper-sphere (Gaussian prior) from which a rounded-shaped
hyper-cube (non-compensatory likelihood) is subtracted.
By observing the true posterior in Figure \ref{pic:alpha_message_apx_y0},
we can see that the low probability
area of the likelihood is subtracted from the Gaussian prior.
As the number of dimensions increases, the part where the hyper-cube
intersects the hyper-sphere decreases; therefore, the
approximation error decreases.

\subsection{Relationship between Approximation Error and Sample Size}

Here, we investigate the relationship between the approximation error
and the number of samples used in the reparameterization trick, namely
$S$ in Eq. \eqref{eqn:approx_expected_log_likelihood}.
We randomly generated 1,000 likelihood functions $p(y|\bm{z})$ and
the priors $p(\bm{z})$ for each dimension (one to five) with the same
procedure used in Section \ref{subsec:dimension_to_kl}. The proposed
method was applied to these likelihood functions and priors by
changing the number of samples, and the approximation error
was subsequently computed by employing the KL-divergence between
the approximated posterior
and the true posterior. Then, the average of the KL-divergence for
each number of samples and dimensions was computed.

\begin{figure}[tb]
  \centering
  \begin{minipage}{.48\linewidth}
    \includegraphics[width=.9\linewidth]{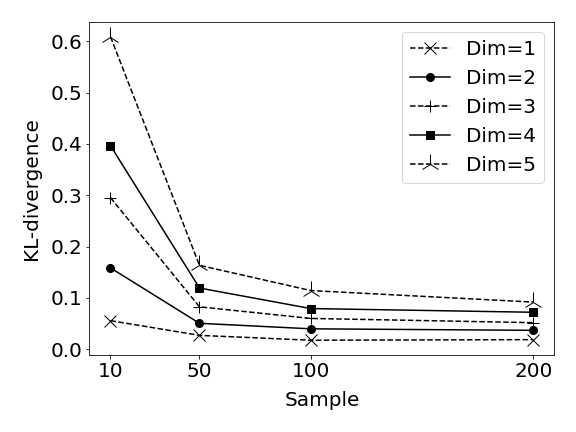}
    \subcaption{Correct answer case: $y=1$}
  \end{minipage}
  \begin{minipage}{.48\linewidth}
    \includegraphics[width=.9\linewidth]{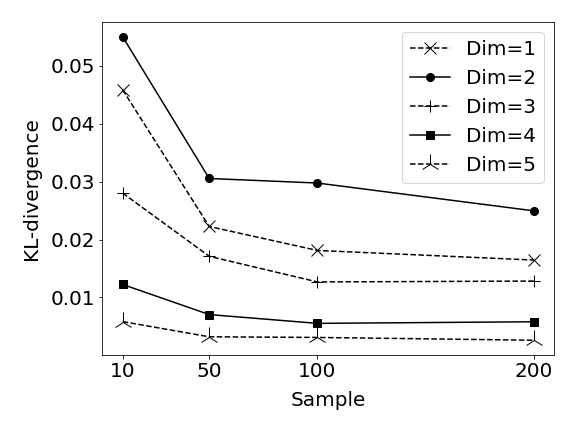}
    \subcaption{Incorrect answer case: $y=0$}
  \end{minipage}
\caption{Relationship between the number of samples and
the approximation error. 
The error decreases as the number of samples increases, and it
saturates at approximately 100 samples in both cases.
}
\label{pic:sample_to_kl}
\end{figure}

Figure \ref{pic:sample_to_kl} shows the results for the cases of 
a correct answer and an incorrect answer. We observe that the
KL-divergence
decreases as the number of samples increases for each case.
The decrease
of the KL-divergence saturates at approximately 100 samples.
Based on these
results, we used 50 samples in the simulations of skill inference in
Section \ref{sec:eval_skill} by considering the trade-off
between the approximation error and the computation cost.

\section{Application to Real Data}\label{sec:real_data}

In this section, we demonstrate how our model is used in actual
data by using the ASSISTments 2009-2010 data (Feng et al. 2009).
First, we show the prediction accuracy of whether a user is
able to solve the next problem or not, which can be used in 
personalized recommendations of the next problem a user should
solve. Second, we visualize skill tracing, i.e., how user's
skills change over time, which benefits users to understand
their skill growth.

\subsection{Data}\label{sec:assistment2009}

The ASSISTments 2009-2010 skill builder data are the logs where
learners solved math problems on the web-based learning system.
The logs consist of sequences of correct or incorrect answers with
order id, user id, and question id.
In the system, learners must answer three questions correctly in a row
to complete the assignment. If a learner uses the tutoring
(``Hint'' or ``Break this Problem Into Steps''), the question will be
marked incorrect. Learners will know immediately if they answered
the question correctly. Therefore, they are the records of the learning
process which are essentially different from logs of tests.
One or multiple skill tags are attached to a question manually.
The data
consist of 4,217 learners, 26,688 questions, and 123 skills.
Table \ref{tab:skills_in_assistment2009} in Appendix shows the first
30 skill tags. Because a question is able to have scaffolding questions,
we filtered out the logs for scaffolding questions leaving the logs for
main questions. We also filtered out the logs for questions answered
by less than ten learners. The processed data comprised 224,905 records,
4,106 learners, 8,112 questions, and 106 skills. The maximum number of 
learning records for a learner was 786.

\subsection{Joint Prior on Slope and Bias}\label{sec:real_data_joint_prior}

We set a joint prior on the slope and bias parameters to infer skills within
a reasonable range, which was introduced in Section
\ref{sec:mstep_trans_joint_prior}. Without this joint prior,
both non-compensatory and compensatory models over-fitted to the real data,
and unexpectedly large or small values of skill, which are impossible to
measure by item difficulty parameters distributed by $\mathcal{N}(0,1)$,
were inferred. Here, we introduce the heuristics we used in real data analysis
to determine the joint distribution.

\begin{figure}[tb]
  \centering
  \begin{minipage}{.32\linewidth}
    \includegraphics[width=.9\linewidth]{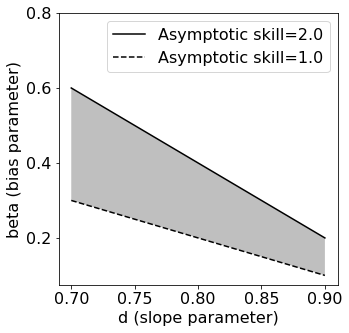}
    \subcaption{}
  \end{minipage}
  \begin{minipage}{.32\linewidth}
    \includegraphics[width=.9\linewidth]{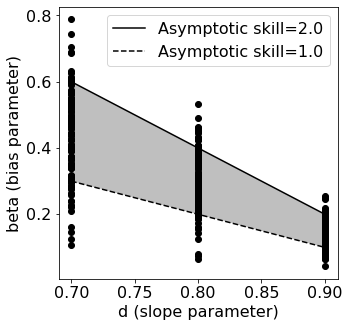}
    \subcaption{}
  \end{minipage}
  \begin{minipage}{.32\linewidth}
    \includegraphics[width=.9\linewidth]{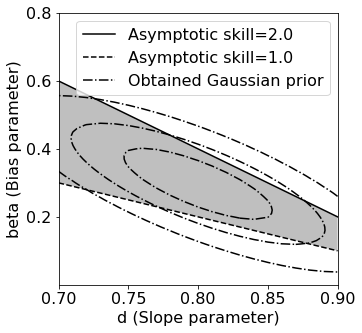}
    \subcaption{}
  \end{minipage}
  \caption{
  (a)-(c) show the heuristics that obtain the Gaussian prior on the slope and bias
  parameters. The gray area in (a) represents values of slope and bias whose
  asymptotic skill is within the range of one and two. (b) shows the samples
  used to obtain the Gaussian distribution which covers the gray area.
  (c) represents the obtained Gaussian prior distribution on the slope and
  bias parameters.
  }
\label{pic:obtain_prior_d_beta}
\end{figure}

First, we considered the reasonable range of the asymptotic value of skill.
Let us assume that the prior distribution of the difficulty parameter is
$\mathcal{N}(0, 1)$.
In this case, questions whose difficulty is larger than three are rare
and skill values larger than three are not almost measurable. 
We considered the reasonable range of the asymptotic value to be from one
to two approximately. Figure~\ref{pic:obtain_prior_d_beta} (a) shows two
lines whose asymptotic skills are two and one, respectively. When 
the slope parameter and the bias parameter are in the gray area, the 
asymptotic skill takes a value ranging from one to two.
Second, we designed a Gaussian distribution which approximately covers
the gray area. We drew samples from 
$\mathcal{N}((\beta_{\mbox{max}} + \beta_{\mbox{min}}) / 2, ((\beta_{\mbox{max}} - \beta_{\mbox{min}}) / 2)^2)$
for the slope parameters $(0.7, 0.8, 0.9)$, 
where $\beta_{\mbox{max}}$ and $\beta_{\mbox{min}}$ denote the value of bias parameter
whose asymptotic skills are two and one, respectively.
(Figure \ref{pic:obtain_prior_d_beta} (b))
Then, we obtained the Gaussian distribution as shown in
Figure \ref{pic:obtain_prior_d_beta} (c)
by calculating the mean and the covariance matrix from those samples.
The obtained mean and covariance matrix were
\begin{eqnarray}
    [\mu_d, \mu_{\beta}]^{\top} & = & [0.8, 0.297]^{\top} \label{eqn:mu_d_mu_beta} \\
    \Lambda_{d,\beta}^{-1} & = &
    \begin{bmatrix}
        6.68 \times 10^{-3}& -9.91 \times 10^{-3} \\
        -9.91 \times 10^{-3} & 2.56 \times 10^{-2}
    \end{bmatrix}. \label{eqn:Lam_d_beta}
\end{eqnarray}
We set this Gaussian distribution to the joint prior of the slope
and bias parameters in the following experiments.

\subsection{Performance of Prediction}\label{sec:prediction_real_data}

We conducted experiments for predicting whether a learner can
answer the next question correctly or not, comparing
the dynamical compensatory MIRT (dcMIRT) model, the dynamical
non-compensatory MIRT (dnMIRT) model, and dnMIRT with the smoothing
extension. Because the target variable can either be correct or
incorrect, this is considered as a
sequential binary classification task.
An accurate predictive model can be used in personalized
recommendation of the next question that should be attempted by
the learner.

We employed five-fold cross-validation and calculated AUC
(Area Under the Curve), average precision for predicting
correct answers ($y=1$), and average precision for predicting
incorrect answers ($y=0$) as performance metrics. A record in
the first-time step for each learner was excluded from the
calculation of the metrics.
The hyperparameters $\mu_a$ and $\sigma_b$ were set as
$\mu_a=0$ and $\sigma_b=1.0$. 
The hyperparameters $[\mu_d, \mu_{\beta}]^{\top}$ and
$\Lambda_{d,\beta}$ were set as Eq. \eqref{eqn:mu_d_mu_beta}
and Eq. \eqref{eqn:Lam_d_beta}, respectively.
The rest of hyperparameters were selected by choosing the
best AUC model in the following grid: 
$\sigma_a = \{0.1, 1.0\}$, 
$(\phi, \psi) = \{(100, 100), (1000, 1000)\}$, and
$(\nu - K, \tau) = \{(100, 0.01), (1000, 0.001)\}$.
All the three models, dcMIRT, dnMIRT, and dnMIRT
with smoothing, employ the forward message to predict whether
the next response is correct or not one by one. However, 
the smoothing extension leads to the target leakage.
Therefore, we applied the smoothing extension in the training
phase and did not apply smoothing in the prediction phase for
dnMIRT with smoothing.

Table \ref{tab:prediction_accuracy_real_data} shows the prediction
accuracy for each model.
We can see that dnMIRT is slightly better than dcMIRT in the three
metrics.
The Wilcoxon signed-rank test verified that there is a difference
in three metrics between dnMIRT and dcMIRT at a confidence
level of 5\%. (This means that there is a
difference; however, the gap is small.)
In the comparison between dnMIRT and dnMIRT with smoothing,
we could confirm differences in AUC and average precision for
incorrect answers by the Wilcoxon signed-rank test.
The best result for each model was obtained in the same
hyperparameter settings:
$\sigma_a=0.1$, $(\phi, \psi) = (100, 100)$, 
and $(\nu - K, \tau) = (100, 0.01)$.

\begin{table}[tb]
    \centering
    \caption{
    Comparison of prediction accuracy among dcMIRT, dnMIRT, and
    dnMIRT with smoothing. The asterisk denotes significance by
    the Wilcoxon signed-rank test at a confidence level of 5\%. 
    dnMIRT is shown to be significantly better than dcMIRT.
    (Smoothing is abbreviated as ``sm.''.)
    }
    \label{tab:prediction_accuracy_real_data}
    \begin{tabular}{lclclcl}
        \toprule
        Method &  AUC & & Average precision (y=1) & & Average Precision (y=0) &\\
        \midrule
        dcMIRT & 0.760 & \multirow{2}{*}{$\Bigr]*$} & 0.838 & \multirow{2}{*}{$\Bigr]*$} & 0.635 & \multirow{2}{*}{$\Bigr]*$} \\
        dnMIRT & 0.762 & \multirow{2}{*}{$\Bigr]*$} & 0.842 & \multirow{2}{*}{$\Bigr]$} & 0.637 & \multirow{2}{*}{$\Bigr]*$} \\
        dnMIRT sm. & 0.763 & & 0.843 & & 0.640 & \\
        \bottomrule
    \end{tabular}
\end{table}

\subsection{Skill Tracing}

This section showcases the visualization of skill tracing 
from the two models of dnMIRT and dcMIRT obtained in the experiments
of Section \ref{sec:prediction_real_data}.
By comparing the results of dnMIRT and
dcMIRT, we clarify the difference in skill tracing between dnMIRT
and dcMIRT. The visualization of skill tracing benefits learners
because they can see their current skill immediately after they
solve a problem. That quick feedback enables learners to arrange
their study plans effectively.

We show the skill tracing of four learners (referred to as A, B, C,
and D below) in ASSISTments 2009-2010. For learners A-C, we show
short periods of skill tracing and compare dnMIRT and dcMIRT. For
learner D, a long period of skill tracing is shown.
In the following,
it is shown that the direction that a skill state changes depends on the
current state in dnMIRT, however, it does not in dcMIRT.

\begin{figure}
  \centering
  {\tabcolsep=0pt
  \begin{tabular}{c}
    \begin{minipage}{0.65\linewidth}
    \includegraphics[width=\linewidth]{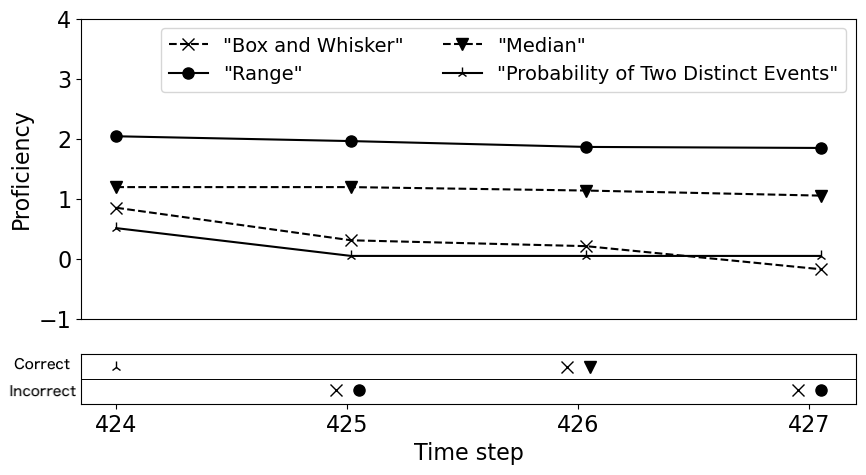}
    \end{minipage}
    \begin{minipage}{0.30\linewidth}
    \includegraphics[width=\linewidth]{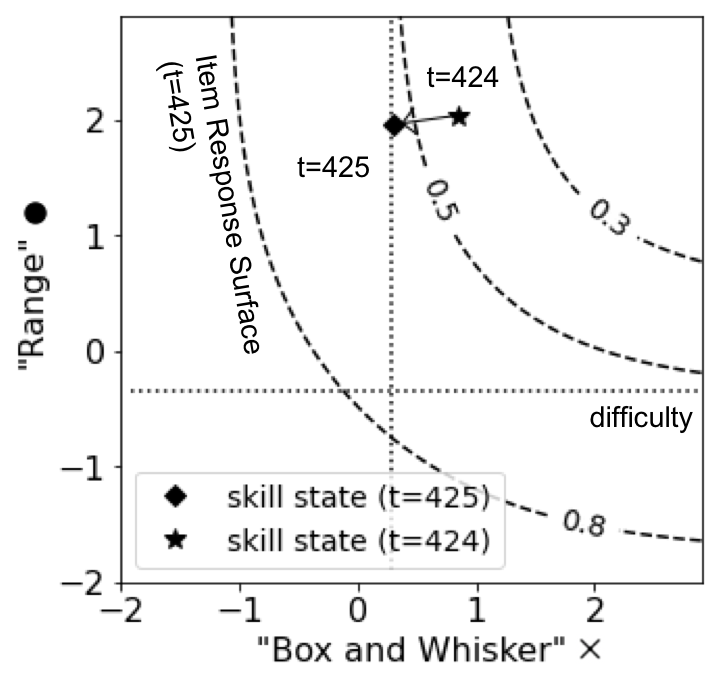}
    \end{minipage}
    \\
    (a) Skill tracing in dnMIRT\\
    \\
    \begin{minipage}{0.65\linewidth}
    \includegraphics[width=\linewidth]{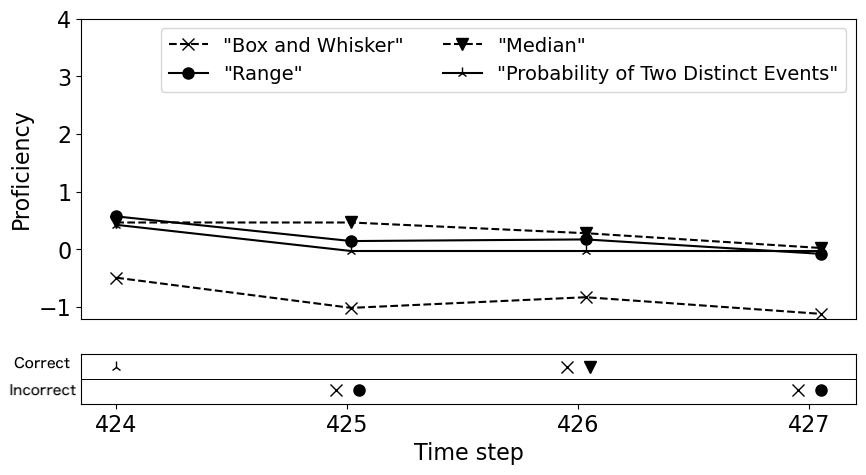}
    \end{minipage}
    \begin{minipage}{0.30\linewidth}
    \includegraphics[width=\linewidth]{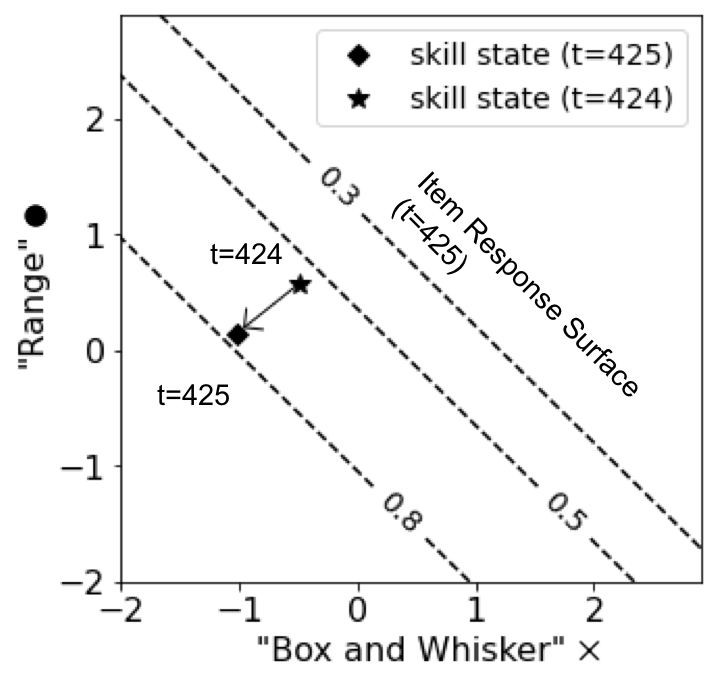}
    \end{minipage}
    \\
    (b) Skill tracing in dcMIRT
  \end{tabular}
  }
  \caption{
  The visualization of skill tracing in dnMIRT and dcMIRT for learner A.
  From 424-time steps to 425-time steps,
  the ``Range'' skill does not change so much,
  whereas the ``Box and Whisker''
  skill decreased in dnMIRT. In contrast, both skills decreased in dcMIRT.
  }
  \label{pic:trace_01}
\end{figure}

Figures \ref{pic:trace_01} (a) left and (b) left show the visualization
of skill tracing
for learner A from 424-time steps to 427-time steps which was obtained
by dnMIRT and dcMIRT, respectively. The bottom panels in (a) left and (b)
left show responses to the questions, where successes and failures correspond
to the upward and downward positions of markers, and the required skill
for each question is denoted as the marker type.
Learner~A correctly answered
a question requiring ``Probability of Two Distinct Events'' skill at
424-time steps and then incorrectly answered a question requiring both
``Box and Whisker'' and ``Range'' skills at 425-time steps. 
In dnMIRT, it can be seen that the ``Range'' skill does not change
so much from 424-time steps to 425-time steps, whereas the 
"Box and Whisker" skill decreased.
In contrast, both skills decreased in dcMIRT.
Figures \ref{pic:trace_01} (a) right and (b) right show the skill states
of learner
A at 424 and 425-time steps with the item response surface of the
question at 425-time steps (dashed line) obtained by dnMIRT and dcMIRT,
respectively. Figure \ref{pic:trace_01} (a) right additionally shows the
difficulty of the question at 425-time steps with the dotted
line. dnMIRT estimated that the ``Box and Whisker'' skill of learner A is much
weaker than the ``Range'' skill according to the relationship
between the skill state and difficulty of the question
in Figure \ref{pic:trace_01} (a) right. dnMIRT explained the incorrect answer
at 425-time steps by decreasing the ``Box and Whisker'' skill much more than 
the ``Range'' skill. It implies that the ``Box and Whisker'' skill was 
inferred as the cause of the incorrect answer by dnMIRT. 
Meanwhile, dcMIRT explained the incorrect answer at 425-time
steps by decreasing both skills. Approximately, the skills are
inferred along the way of the normal vector of the contour in both dnMIRT
and dcMIRT. Since the contour is not linear in dnMIRT, the direction
that a skill state changes depends on the current skill state.

\begin{figure}
  \centering
  {\tabcolsep=0pt
  \begin{tabular}{c}
    \begin{minipage}{0.65\linewidth}
    \includegraphics[width=\linewidth]{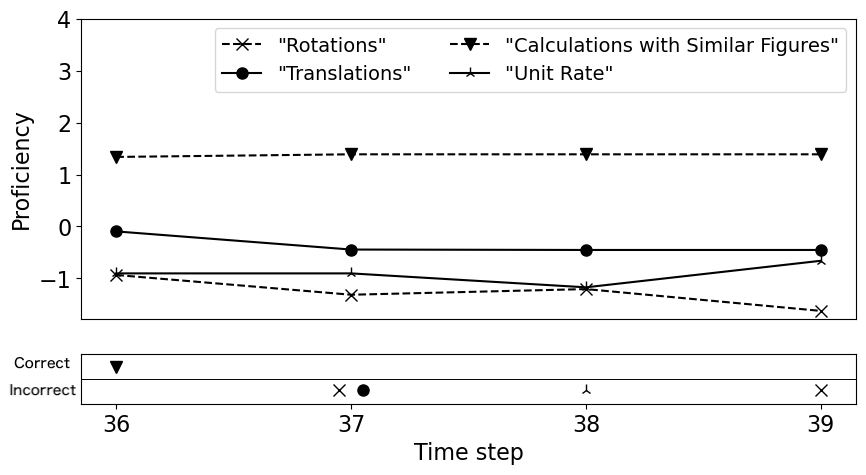}
    \end{minipage}
    \begin{minipage}{0.30\linewidth}
    \includegraphics[width=\linewidth]{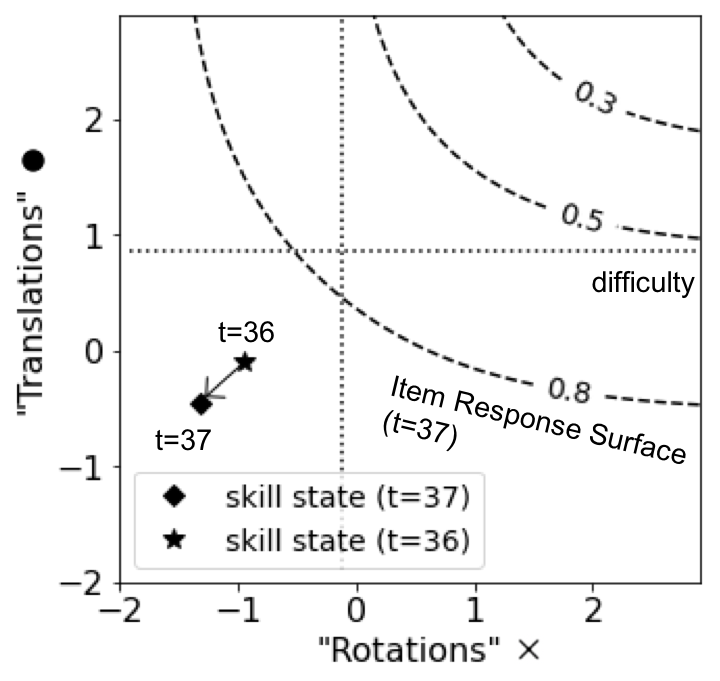}
    \end{minipage}
    \\
    (a) Skill tracing in dnMIRT
    \\
    \\
    \begin{minipage}{0.65\linewidth}
    \includegraphics[width=\linewidth]{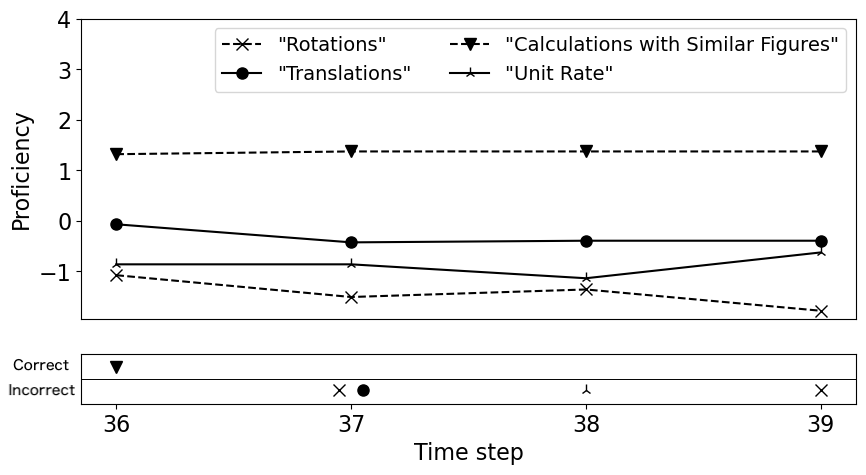}
    \end{minipage}
    \begin{minipage}{0.30\linewidth}
    \includegraphics[width=\linewidth]{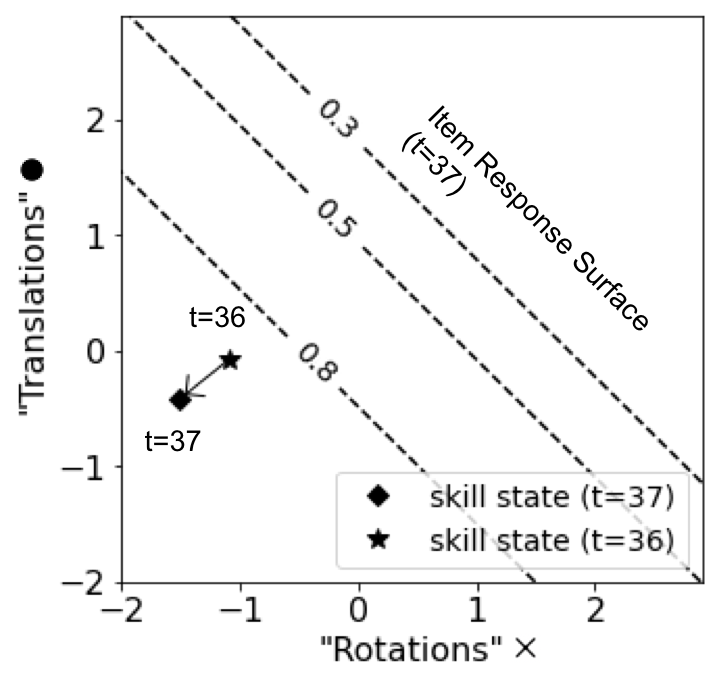}
    \end{minipage}\\
    (b) Skill tracing in dcMIRT
  \end{tabular}
  }
  \caption{
  The visualization of skill tracing in dnMIRT and dcMIRT for learner B.
  From 36-time steps to 37-time steps,
  both ``Rotations'' and ``Translations'' skills decrease in dnMIRT and dcMIRT.
  }
  \label{pic:trace_02}
\end{figure}

Figures \ref{pic:trace_02} (a) left and (b) left show the visualization
of skill tracing
for learner B from 36-time steps to 39-time steps which was obtained
by dnMIRT and dcMIRT, respectively. 
Learner B correctly answered
a question requiring ``Calculations with Similar Figures'' skill at
36-time steps and then incorrectly answered a question requiring both
``Rotations'' and ``Translations'' skills at 37-time steps.
In dnMIRT and dcMIRT, it can be seen that both ``Rotations'' and
``Translations'' skills decrease from 36-time steps to 37-time steps.
Figures \ref{pic:trace_02} (a) right and (b) right show the skill
states of learner B at 36 and 37-time steps with the same procedure
as the Figures
\ref{pic:trace_01} (a) right and (b) right.
dnMIRT estimated that the both ``Rotations'' and ``Translations''
skills of learner B were weak according to the relationship
between the skill state and difficulty of the question
in Figure \ref{pic:trace_02} (a) right.
dnMIRT explained the incorrect answer at 37-time steps by decreasing
both skills. It implies that both
skills were inferred as the cause of the incorrect answer by dnMIRT.
dcMIRT also explained the incorrect answer at 37-time steps by
decreasing both skills.

\begin{figure}
  \centering
  {\tabcolsep=0pt
  \begin{tabular}{c}
    \begin{minipage}{0.65\linewidth}
    \includegraphics[width=\linewidth]{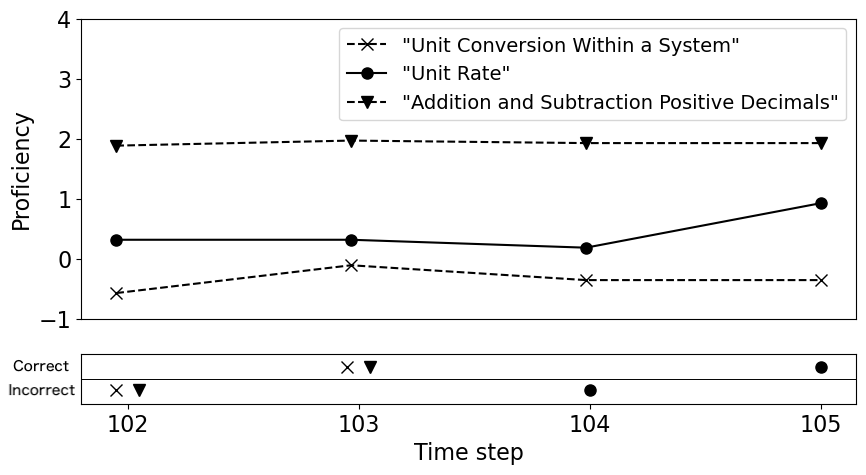}
    \end{minipage}
    \begin{minipage}{0.30\linewidth}
    \includegraphics[width=\linewidth]{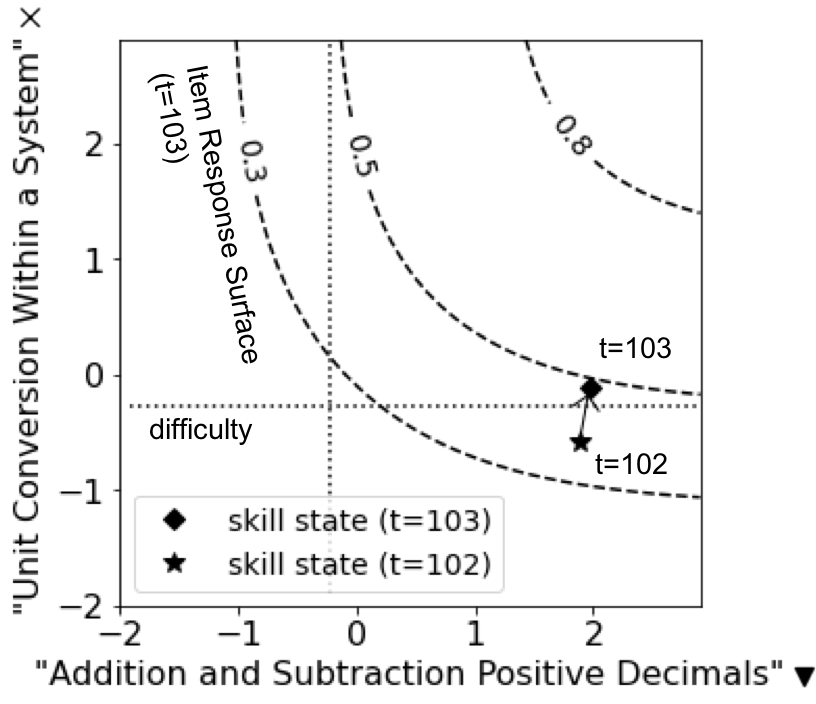}
    \end{minipage}
    \\
    (a) Skill tracing in dnMIRT
    \\
    \\
    \begin{minipage}{0.65\linewidth}
    \includegraphics[width=\linewidth]{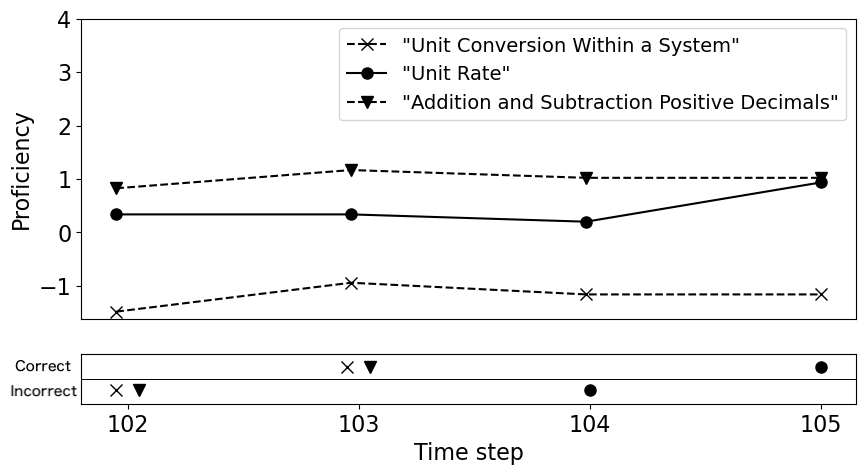}
    \end{minipage}
    \begin{minipage}{0.30\linewidth}
    \includegraphics[width=\linewidth]{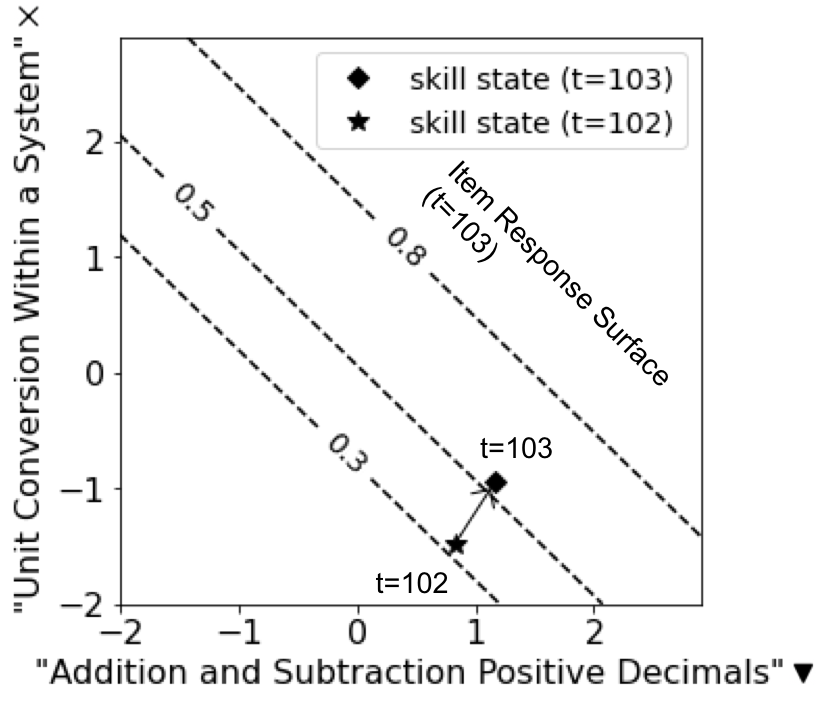}
    \end{minipage}
    \\
    (b) Skill tracing in dcMIRT
  \end{tabular}
  }
  \caption{
  The visualization of skill tracing in dnMIRT and dcMIRT for learner C.
  From 102-time steps to 103-time steps,
  the ``Addition and Subtraction Positive Decimals'' skill does not change
  so much,
  whereas the ``Unit Conversion Within a System'' skill increased in dnMIRT.
  In contrast, both skills increased in dcMIRT.
  }
  \label{pic:trace_03}
\end{figure}

Figures \ref{pic:trace_03} (a) left and (b) left show the visualization
of skill tracing
for learner~C from 102-time steps to 105-time steps which was obtained
by dnMIRT and dcMIRT, respectively. 
Learner~C incorrectly answered
a question requiring both ``Unit Conversion Within a System'' and 
``Addition and Subtraction Positive Decimals'' skills at 
102-time steps and then correctly answered a question requiring both
``Unit Conversion Within a System'' and
``Addition and Subtraction Positive Decimals'' skills at 37-time steps.
In dnMIRT, it can be seen that the
``Addition and Subtraction Positive Decimals'' skill
does not change so much from 102-time steps to 103-time steps,
whereas the ``Unit Conversion Within a System'' skill increased.
In contrast, both skills increased in dcMIRT.
Figures \ref{pic:trace_03} (a) right and (b) right show the skill
states of learner~C
at 102 and 103-time steps with the same procedure as the above.
dnMIRT estimated that the ``Unit Conversion Within a System'' skill
of learner~C was much weaker than the ``Addition and Subtraction'' skill
according to the relationship between the skill state and difficulty
of the question
in Figure \ref{pic:trace_03} (a) right. dnMIRT explained the correct
answer
at 103-time steps by increasing the ``Unit Conversion Within a System''
skill much more than the ``Addition and Subtraction Positive Decimals''
skill. It implies that the skill increase of
``Unit Conversion Within a System'' was inferred as the cause of the
correct answer by dnMIRT. 
Meanwhile, dcMIRT explained the correct answer at 103-time
steps by increasing both skills.

Figure \ref{pic:skill_tracing} presents a longer period of skill tracing
than the previous three learners. Learner~D studied problems of the 
``Absolute Value'', ``Subtraction Whole Numbers'', ``Addition Whole Numbers'',
and ``Multiplication and Division Integers'' skills. We can see that
the ``Multiplication and Division Integers'' skill grew first and 
the other three skills grew later. The 
``Absolute Value'', ``Subtraction Whole Numbers'', and 
``Addition Whole Numbers'' skills were more difficult for learner D
to acquire than the ``Multiplication and Division Integers'' skill.

\section{Discussion}\label{sec:discussion}

We proposed a dynamical non-compensatory MIRT model to provide accurate and real-time skill tracing. The proposed model combined an LDS and a non-compensatory MIRT model, resulting in a complicated posterior. We approximated the posterior as a Gaussian distribution by minimizing the KL-divergence between the approximated posterior and the true posterior. The estimation method for the model parameters was derived through Monte Carlo EM.

Our simulation studies evaluated (a) inferring latent skills, (b) predicting the next response, and (c) approximating the $\hat{\alpha}$ message. In (a), we verified that the proposed method reproduced the true latent skills for the 2-skill and 100-skill simulation data, and the dynamical compensatory model suffered from significant underestimation errors. Also, it turned out that the proposed model with fixed discrimination parameters (i.e., $a_{i,k}=1$) was sufficient to infer the skills in most cases. In (b), it was shown that the proposed model performs better than its compensatory counterpart in predicting the next response. Finally, in (c), we demonstrated that the proposed Gaussian approximation was better than the Laplace approximation at approximating the true posterior through visualizations and quantitative approximation error analyses.

Experiments on actual data presented the supposed use case of the proposed model. The results demonstrated that the dynamical non-compensatory MIRT model could infer practical skill tracing, as shown in Figure \ref{pic:skill_tracing}. We also clarified differences in skill tracing between non-compensatory and compensatory models, where the non-compensatory model considered which skill is the cause of success or failure. In predicting the responses, we showed that the proposed model was slightly better than its compensatory version.

To understand our model better, it is important to discuss how it
differs from the closest model, SPARFA-Trace (Lan et al. 2014).
SPAFA-Trace is a combination of an LDS and a compensatory MIRT
model; therefore, the main difference is the type of MIRT model,
meaning that the target data are different.
The authors approximated the posterior in the Gaussian distribution
through the expectation propagation method (EP; Minka 2001)
because their compensatory model employed the probit function and
yielded analytical solutions using the EP method.
This is the case only for their compensatory model;
non-compensatory models do not provide analytical solutions by
employing the EP method.
In contrast, we employed the objective function which is the same as
variational Bayes (VB; Bishop 2006).
Our approximation method is more general and can be applied to both
compensatory and non-compensatory models.
SPARFA-Trace is the method used for exploratory analysis; it can infer
Q-matrix, whereas our model is used for confirmatory analysis that
assumes that Q-matrix is given. Its extension to
exploratory analysis remains as future work.

As shown in the results of the comparison between compensatory and
non-compensatory models in non-dynamical settings
(Buchholz and Hartig 2018),
the dynamical compensatory MIRT model caused significant underestimation
errors. Figure \ref{fig:scatterplot} (d) clearly shows that the
underestimation of the skills and the fitting trend, shown in
Figure \ref{pic:irs_diff}, causes the underestimation.
When the dimensions of latent skills increased to 100 skills,
the correlation for the dynamical compensatory model dropped
significantly, as compared to the dynamical non-compensatory model.
The reason is that higher dimensions increase the gap between the
two models. 
This result stresses the importance of applying the dynamical
non-compensatory MIRT model to the data of time-series and
non-compensatory type.

One of our purposes was real-time skill tracing. 
We mainly implemented the inference algorithm in Python, which runs in parallel
to fully utilize many cores. Only the Gaussian approximation of $\hat{\alpha}$
message was implemented in C++.
Training our model with one hyperparameter setting on 100-skill data took
about one to two days by using a machine with 36 cores.
However, inferring skills based on the trained model is fast.
Before tracing skills on the fly, we need to train the model on available data to
obtain the parameters of the model.
Once we train the model, we are ready to trace the skills of students. Every
time we get one response data from a student, we execute forward message passing
for one time step to infer the current skills of that student. It took 0.0052 sec
for 100-skill data. Therefore, we think that real-time skill tracing is actually
possible.

Although we approximated the posterior as a Gaussian distribution,
another approximation can be employed using sampling,
such as the sequential Monte Carlo (Kitagawa 1993).
The advantage of our approach is its scalability of the dimensions in
the latent skill state. Our method
worked for higher dimensions, even when the number of dimensions for
the latent state was 100.
Despite the accuracy advantage of sampling approaches
(they can express any shape of the posterior), they can only function
in low-dimensional cases. Further comparisons with sampling
approaches will be beneficial to identify the border number of
dimensions; which approaches should be employed by considering
accuracy, speed, and cost.

Another dynamical extension of MIRT is longitudinal IRT
(Andrade and Tavares 2005; Bollen and Curran 2006; Duncan et al. 2006;
Paek et al. 2016; Wang and Nydic 2020).
This has been employed to analyze multiple test results and infer the
changes in latent skills among those tests. Longitudinal IRT typically handles
a short series of the latent skill states, and multiple item responses are
available for a particular skill state. In contrast, knowledge tracing typically
handles long-series, and a single item response is available for a
particular skill state. We formulated the dynamical extension of
the non-compensatory model in the knowledge tracing setting. However, our
proposed model can be easily re-formulated into the longitudinal IRT
setting by accepting multiple item responses from each time step in
the graphical model (Figure \ref{pic:lds} (a)).
A few modifications to the inference
algorithm are needed; Eq. \eqref{eqn:approx_expected_log_likelihood},
Eq. \eqref{eqn:opt_a_in_mstep}, and Eq. \eqref{eqn:opt_b_in_mstep} should
be modified to accept multiple likelihood functions.

When users apply our proposed model to their data, some
options exist for
hyperparameters and whether to use the smoothing extension.
Here, we summarize a recommended setting for the first trial.
First, the discrimination parameters should be fixed to one.
According to the simulation studies, fixing them to one gave
good results in most cases. Another benefit is that users do
not need to tune the prior on the discrimination parameters.
Second, the joint prior on slope and bias should be used.
Without the joint prior, the model may over-fit data and infer
unexpectedly high or low skill values. Users can use the same
values as Eq. \eqref{eqn:mu_d_mu_beta} and
Eq. \eqref{eqn:Lam_d_beta}.
Last, the smoothing extension should not be used. It will need
additional efforts to tune the weight parameters. We recommend
that users try the simplest setting first and move to the
detailed settings if it is needed.

We now summarize other limitations and possible avenues for future research.
First, the skill transitions were modeled by the linear transformations,
and many extensions can be considered in the transition model. When we
consider the linear transformation $z^{(t+1)} = d\cdot z^{(t)} + \beta$
with $0<d<1$, the skill asymptotically approaches $\beta / (1-d)$ by 
applying the linear transformation infinitely many times.
This means that the skill of
a learner, with respect to a question, approaches a particular goal value
as they practice the question.
This may seem natural; however, the linear transformation
behaves unnaturally when a learner with skill higher than
$\beta / (1-d)$ correctly answers the question.
The linear transformation unexpectedly
decreases the skill of the learner such that the skill is one step
closer to the goal value. This means that the skill decreases when a 
high-skill learner solves an easy problem.
One apparent solution to this is the piece-wise linear model,
i.e., adding a rule $z^{(t+1)} = z^{(t)}$ when $z^{(t)} > \beta / (1-d)$.
Furthermore, more flexible models such as the Gaussian process can be used
to express skill transitions (Wang et al. 2005; Cully and Demiris 2019).

Second, we showed that the prediction accuracy of the dynamical
non-compensatory MIRT model is slightly better than its compensatory
version by one actual data. Comparisons on various actual data sets
are required to see whether the dynamical non-compensatory
MIRT model generally performs better than its compensatory version.
We also employed heuristics to set a joint prior distribution on the slope
and bias parameters of skill transition. Without the joint prior,
both non-compensatory and compensatory models over-fitted to the data
and the test accuracy kept decreasing after some iterations.
Also, unexpectedly
a large or small value of skill, which is impossible to be measure by item
difficulty parameters distributed by $\mathcal{N}(0,1)$, was observed in
both models. We manually tried some heuristics and chose the best performing
one, however, we expect further studies will improve the skill transition
model which can be used without setting the joint prior.

Third, we did not combine the smoothing extension and the estimation of
$\gamma_k$ because preliminary experiments did not perform well. This
is a current limitation; however, we consider that further
exploration of weight parameters is needed to confirm if this combination
does not truly work or not.

\vspace{\fill}\pagebreak

\appendix
\renewcommand{\theequation}{A\arabic{equation}}
\setcounter{equation}{0}
\renewcommand{\thesection}{\Alph{subsection}}
\setcounter{section}{0}
\section*{Appendix}

\begin{table}[H]
    \centering
    \caption{
    The first 30 skill tags in ASSISTments2009-2010 skill builder.
    }
    \label{tab:skills_in_assistment2009}
    \begin{tabular}{ll}
    \toprule
    Skill: 1-15 & Skill: 16-30 \\
    \midrule
    Box and Whisker & Interior Angles Figures with More than 3 Sides \\
    Circle Graph & Interior Angles Triangle \\
    Histogram as Table or Graph &  Congruence \\
    Number Line & Complementary and Supplementary Angles \\
    Scatter Plot & Angles on Parallel Lines Cut by a Transversal \\
    Stem and Leaf Plot & Pythagorean Theorem \\
    Table & Nets of 3D Figures \\
    Venn Diagram & Unit Conversion Within a System \\
    Mean &  Effect of Changing Dimensions of a Shape Prpor... \\
    Median & Area Circle \\
    Mode & Circumference  \\
    Range & Perimeter of a Polygon \\
    Counting Methods & Reading a Ruler or Scale \\
    Probability of Two Distinct Events & Calculations with Similar Figures \\
    Probability of a Single Event & Conversion of Fraction Decimals Percents \\
\bottomrule
\end{tabular}
\end{table}

\begin{sidewaystable}
    \centering
    \small
    \caption{
    Standard deviations in the accuracy of skill inference for data set A:
    2-skill, $a_{i,k}=1$, and no skill correlation.
    }
    \label{tab:skill2_no_slope_std}
\begin{tabular}{llrrrrrrrrrrr}
    \toprule
           &        & \multicolumn{2}{c}{Skills (MAE)} & \multicolumn{4}{c}{Parameters (MAE)} & \multicolumn{2}{c}{Skills (Corr)} & \multicolumn{3}{c}{Parameters (Corr)} \\
    \cmidrule(r){3-4} \cmidrule(r){5-8} \cmidrule(r){9-10} \cmidrule{11-13}
    Model  & Settings &          All &   Last &      $b_{i,k}$ &      $d_{i,k}$ &   $\beta'_{i,k}$ & $\gamma_k$ &   All &   Last &      $b_{i,k}$ &      $d_{i,k}$ &   $\beta'_{i,k}$ \\
    \midrule
dnMIRT & &        0.016 &  0.027 &            0.058 &  0.008 &  0.017 & 0.000 &        0.019 &  0.042 &             0.017 &  0.095 &  0.017 \\
       & est. $\gamma_k$ &    0.049 &     0.063 &  0.066 &  0.004 &     0.020 &      0.040 &    0.015 &     0.037 &  0.020 &  0.149 &     0.014 \\
      & sm. &        0.025 &  0.050 &            0.072 &  0.012 &  0.020 & 0.000 &        0.018 &  0.039 &             0.019 &  0.096 &  0.037 \\
dcMIRT &        &        0.115 &  0.148 &              - &    - &    - & - &         0.015 &  0.044 &               - &    - &    - \\
nMIRT & window=5 &        0.094 &  0.128 &            0.071 &    - &    - & - &         0.022 &  0.047 &             0.122 &    - &    - \\
      & window=10 &        0.083 &  0.115 &            0.071 &    - &    - & - &         0.026 &  0.044 &             0.122 &    - &    - \\
\bottomrule
\multicolumn{12}{l}{Smoothing is abbreviated as ``sm.''. Estimate $\gamma_k$ is abbreviated as ``est. $\gamma_k$''.}
\end{tabular}
\end{sidewaystable}

\begin{sidewaystable}
    \centering
    \small
    \caption{
    Standard deviations in the accuracy of skill inference for data set B:
    2-skill, $a_{i,k} \sim \mbox{logN}(0, 0.25^2)$, and no skill correlation.
    }
    \label{tab:skill2_slope_std}
\begin{tabular}{llrrrrrrrrrrrrr}
    \toprule
           &        & \multicolumn{2}{c}{Skills (MAE)} & \multicolumn{5}{c}{Parameters (MAE)} & \multicolumn{2}{c}{Skills (Corr)} & \multicolumn{4}{c}{Parameters (Corr)} \\
    \cmidrule(r){3-4} \cmidrule(r){5-9} \cmidrule(r){10-11} \cmidrule{12-15}
    Model  & Settings &          All &   Last &                $a_{i,k}$ &      $b_{i,k}$ &      $d_{i,k}$ &   $\beta'_{i,k}$ & $\gamma_k$ &         All &   Last &                 $a_{i,k}$ &      $b_{i,k}$ &      $d_{i,k}$ &   $\beta'_{i,k}$ \\
    \midrule
dnMIRT & $\sigma_a=0.25$ &        0.152 &  0.207 &            0.030 &  0.044 &  0.003 &  0.041 &         0.000 & 0.015 &  0.048 &             0.109 &  0.034 &  0.154 &  0.034 \\
      & $\sigma_a=0.1$ &        0.102 &  0.131 &            0.014 &  0.049 &  0.003 &  0.035 &      0.000 &   0.017 &  0.051 &             0.096 &  0.040 &  0.161 &  0.037 \\
      & $a_{i,k}=1$ &        0.083 &  0.103 &            0.017 &  0.055 &  0.004 &  0.032 &         0.000 & 0.018 &  0.053 &               - &  0.048 &  0.162 &  0.039 \\
      & $a_{i,k}=1$, est. $\gamma_k$ &    0.039 &     0.050 &  0.017 &  0.047 &  0.015 &     0.030 &      0.060 &    0.020 &     0.074 &    - &  0.053 &  0.139 &     0.039 \\
      & $\sigma_a=0.25$, sm. &        0.136 &  0.173 &            0.033 &  0.054 &  0.007 &  0.048 &    0.000 &     0.023 &  0.088 &             0.145 &  0.041 &  0.084 &  0.039 \\
      & $\sigma_a=0.1$, sm. &        0.079 &  0.082 &            0.013 &  0.060 &  0.007 &  0.040 & 0.000 &        0.020 &  0.069 &             0.108 &  0.052 &  0.116 &  0.046 \\
      & $a_{i,k}=1$, sm. &        0.056 &  0.052 &            0.017 &  0.073 &  0.006 &  0.037 &    0.000 &     0.021 &  0.068 &               - &  0.064 &  0.124 &  0.050 \\
dcMIRT & $\sigma_a=0.25$ &        0.129 &  0.153 &              - &    - &    - &    - & - &         0.017 &  0.044 &               - &    - &    - &    - \\
      & $\sigma_a=0.1$ &        0.124 &  0.144 &              - &    - &    - &    - & - &        0.017 &  0.035 &               - &    - &    - &    - \\
      & $a_{i,k}=1$ &        0.123 &  0.143 &              - &    - &    - &    - & - &        0.017 &  0.028 &               - &    - &    - &    - \\
nMIRT & window=5 &        0.129 &  0.158 &            0.024 &  0.039 &    - &    - & - &         0.030 &  0.056 &             0.073 &  0.104 &    - &    - \\
      & window=10 &        0.117 &  0.149 &            0.024 &  0.039 &    - &    - & - &        0.027 &  0.040 &             0.073 &  0.104 &    - &    - \\
\bottomrule
\multicolumn{12}{l}{Smoothing is abbreviated as ``sm.''. Estimate $\gamma_k$ is abbreviated as ``est. $\gamma_k$''.}
\end{tabular}
\end{sidewaystable}

\begin{sidewaystable}
    \centering
    \small
    \caption{
    Standard deviations in the accuracy of skill inference for data set C:
    100-skill, $a_{i,k} = 1$, and no skill correlation.
    }
    \label{tab:skill100_no_slope_std}
\begin{tabular}{llrrrrrrrrrrr}
    \toprule
           &        & \multicolumn{2}{c}{Skills (MAE)} & \multicolumn{4}{c}{Parameters (MAE)} & \multicolumn{2}{c}{Skills (Corr)} & \multicolumn{3}{c}{Parameters (Corr)} \\
    \cmidrule(r){3-4} \cmidrule(r){5-8} \cmidrule(r){9-10} \cmidrule{11-13}
    Model  & Settings &          All &   Last &      $b_{i,k}$ &      $d_{i,k}$ &   $\beta'_{i,k}$ &    $\gamma_k$ &       All &   Last &      $b_{i,k}$ &      $d_{i,k}$ &   $\beta'_{i,k}$ \\
    \midrule
dnMIRT & &        0.014 &  0.019 &            0.009 &  0.002 &  0.006 & 0.000 &        0.008 &  0.039 &             0.006 &  0.022 &  0.014 \\
       & est. $\gamma_k$ &    0.014 &     0.011 &  0.006 &  0.008 &     0.005 &      0.041 &    0.004 &     0.046 &  0.007 &  0.025 &     0.012 \\
       & sm. &        0.013 &  0.016 &            0.009 &  0.002 &  0.008 & 0.000 &        0.009 &  0.038 &             0.007 &  0.025 &  0.014 \\
dcMIRT &        &        0.045 &  0.047 &              - &    - &    - &    - &      0.023 &  0.049 &               - &    - &    - \\
\bottomrule
\multicolumn{12}{l}{Smoothing is abbreviated as ``sm.''. Estimate $\gamma_k$ is abbreviated as ``est. $\gamma_k$''.}
\end{tabular}
\end{sidewaystable}

\begin{sidewaystable}
    \centering
    \small
    \caption{
    Standard deviations in the accuracy of skill inference for data set D:
    100-skill, $a_{i,k} \sim \mbox{logN}(0, 0.25^2)$, and no skill correlation.
    }
    \label{tab:skill100_slope_std}
\begin{tabular}{llrrrrrrrrrrrrr}
    \toprule
           &              & \multicolumn{2}{c}{Skills (MAE)} & \multicolumn{5}{c}{Parameters (MAE)} & \multicolumn{2}{c}{Skills (Corr)} & \multicolumn{4}{c}{Parameters (Corr)} \\
    \cmidrule(r){3-4} \cmidrule(r){5-9} \cmidrule(r){10-11} \cmidrule{12-15}
    Model  & Settings & All & Last &                $a_{i,k}$ &      $b_{i,k}$ &      $d_{i,k}$ &   $\beta'_{i,k}$ & $\gamma_k$ &          All &   Last &                 $a_{i,k}$ &      $b_{i,k}$ &      $d_{i,k}$ &   $\beta'_{i,k}$ \\
    \midrule
dnMIRT & $\sigma_a=0.25$ &        0.015 &  0.018 &            0.003 &  0.008 &  0.001 &  0.006 &    0.000 &     0.010 &  0.023 &             0.019 &  0.009 &  0.022 &  0.006 \\
       & $\sigma_a=0.1$ &        0.010 &  0.011 &            0.002 &  0.011 &  0.000 &  0.006 &     0.000 &    0.006 &  0.019 &             0.018 &  0.011 &  0.019 &  0.007 \\
       & $a_{i,k}=1$ &        0.010 &  0.010 &            0.003 &  0.011 &  0.000 &  0.006 &        0.000 & 0.006 &  0.018 &               - &  0.012 &  0.020 &  0.008 \\
       & $a_{i,k}=1$, est. $\gamma_k$ &    0.016 &     0.020 &  0.003 &  0.015 &  0.006 &     0.006 &      0.021 &    0.006 &     0.009 &    - &  0.014 &  0.016 &     0.010 \\
       & $\sigma_a=0.25$, sm. &        0.018 &  0.019 &            0.002 &  0.008 &  0.001 &  0.007 & 0.000 &         0.012 &  0.029 &             0.014 &  0.011 &  0.015 &  0.010 \\
       & $\sigma_a=0.1$, sm. &        0.012 &  0.012 &            0.003 &  0.012 &  0.001 &  0.007 &    0.000 &      0.007 &  0.020 &             0.017 &  0.013 &  0.015 &  0.010 \\
       & $a_{i,k}=1$, sm. &        0.010 &  0.007 &            0.003 &  0.011 &  0.001 &  0.007 &   0.000 &      0.005 &  0.015 &               - &  0.014 &  0.012 &  0.010 \\
dcMIRT & $\sigma_a=0.25$ &        0.027 &  0.030 &              - &    - &    - &    - & - &         0.015 &  0.026 &               - &    - &    - &    - \\
       & $\sigma_a=0.1$ &        0.027 &  0.030 &              - &    - &    - &    - & - &         0.022 &  0.032 &               - &    - &    - &    - \\
       & $a_{i,k}=1$ &        0.028 &  0.031 &              - &    - &    - &    - & - &         0.021 &  0.031 &               - &    - &    - &    - \\
\bottomrule
\multicolumn{12}{l}{Smoothing is abbreviated as ``sm.''. Estimate $\gamma_k$ is abbreviated as ``est. $\gamma_k$''.}
\end{tabular}
\end{sidewaystable}

\begin{sidewaystable}
    \centering
    \small
    \caption{
    Standard deviations in the accuracy of skill inference for data set E:
    100-skill, $a_{i,k} \sim \mbox{logN}(0.4, 0.2)$, and no skill correlation.
    }
    \label{tab:skill100_slope2_std}
\footnotesize
\begin{tabular}{llrrrrrrrrrrrrr}
\toprule
           &              & \multicolumn{2}{c}{Skills (MAE)} & \multicolumn{5}{c}{Parameters (MAE)} & \multicolumn{2}{c}{Skills (Corr)} & \multicolumn{4}{c}{Parameters (Corr)} \\
    \cmidrule(r){3-4} \cmidrule(r){5-9} \cmidrule(r){10-11} \cmidrule{12-15}
    Model  & Settings & All & Last &                $a_{i,k}$ &      $b_{i,k}$ &      $d_{i,k}$ &   $\beta'_{i,k}$ & $\gamma_k$ &          All &   Last &                 $a_{i,k}$ &      $b_{i,k}$ &      $d_{i,k}$ &   $\beta'_{i,k}$ \\
\midrule
dnMIRT & $\mu_a=0.4, \sigma_a=0.2$ &        0.013 &  0.022 &            0.006 &  0.010 &  0.003 &  0.004 & 0.000 &        0.005 &  0.013 &             0.018 &  0.008 &  0.033 &  0.005 \\
       & $\mu_a=0, \sigma_a=0.1$ &        0.006 &  0.008 &            0.005 &  0.008 &  0.001 &  0.004 & 0.000 &         0.005 &  0.021 &             0.019 &  0.012 &  0.035 &  0.005 \\
       & $\mu_a=0, \sigma_a=0.25$ &        0.003 &  0.008 &            0.005 &  0.008 &  0.001 &  0.004 & 0.000&          0.005 &  0.018 &             0.020 &  0.010 &  0.037 &  0.005 \\
       & $a_{i,k}=1$ &  0.009 &     0.011 &  0.005 &  0.008 &  0.001 &     0.004 & 0.000 &   0.006 &     0.023 &    - &  0.012 &  0.035 &     0.004 \\
       & $a_{i,k}=1$, est. $\gamma_k$ &    0.010 &     0.012 &  0.005 &  0.012 &  0.005 &     0.012 &      0.015 &    0.005 &     0.012 &    - &  0.010 &  0.032 &     0.006 \\
       & $\mu_a=0.4, \sigma_a=0.2$, sm. &        0.013 &  0.021 &            0.005 &  0.007 &  0.001 &  0.003 & 0.000 &        0.008 &  0.021 &             0.021 &  0.006 &  0.020 &  0.005 \\
       & $\mu_a=0, \sigma_a=0.1$, sm. &        0.009 &  0.015 &            0.005 &  0.006 &  0.001 &  0.003 & 0.000 &         0.005 &  0.020 &             0.020 &  0.009 &  0.031 &  0.004 \\
       & $\mu_a=0, \sigma_a=0.25$, sm. &        0.006 &  0.008 &            0.007 &  0.004 &  0.002 &  0.004 & 0.000 &         0.008 &  0.024 &             0.019 &  0.008 &  0.028 &  0.004 \\
       & $a_{i,k}=1$, sm. & 0.012 &     0.019 &  0.005 &  0.009 &  0.001 &     0.004 & 0.000 &    0.006 &     0.022 &    - &  0.010 &  0.032 &     0.004 \\
dcMIRT & $\mu_a=0.4, \sigma_a=0.2$ &        0.037 &  0.040 &              - &    - &    - &    - & - &         0.015 &  0.018 &               - &    - &    - &    - \\
       & $\mu_a=0, \sigma_a=0.1$ &        0.047 &  0.053 &              - &    - &    - &    - & - &         0.015 &  0.019 &               - &    - &    - &    - \\
       & $\mu_a=0, \sigma_a=0.25$ &        0.045 &  0.050 &              - &    - &    - &    - & - &        0.016 &  0.019 &               - &    - &    - &    - \\
\bottomrule
\multicolumn{12}{l}{Smoothing is abbreviated as ``sm.''. Estimate $\gamma_k$ is abbreviated as ``est. $\gamma_k$''.}
\end{tabular}
\normalsize
\end{sidewaystable}

\begin{sidewaystable}
    \centering
    \small
    \caption{
    Standard deviations in the accuracy of skill inference for data set F: 
    100-skill, $a_{i,k} \sim \mbox{logN}(0, 0.25)$, and skill correlation is 0.2.
    }
    \label{tab:skill100_slope_corr_std}
\begin{tabular}{llrrrrrrrrrrrrr}
\toprule
           &              & \multicolumn{2}{c}{Skills (MAE)} & \multicolumn{5}{c}{Parameters (MAE)} & \multicolumn{2}{c}{Skills (Corr)} & \multicolumn{4}{c}{Parameters (Corr)} \\
    \cmidrule(r){3-4} \cmidrule(r){5-9} \cmidrule(r){10-11} \cmidrule{12-15}
    Model  & Settings & All & Last &                $a_{i,k}$ &      $b_{i,k}$ &      $d_{i,k}$ &   $\beta'_{i,k}$ & $\gamma_k$ &           All &   Last &                 $a_{i,k}$ &      $b_{i,k}$ &      $d_{i,k}$ &   $\beta'_{i,k}$ \\
\midrule
dnMIRT & $\sigma_a=0.25$ &        0.015 &  0.022 &            0.005 &  0.011 &  0.001 &  0.005 & 0.000 &         0.010 &  0.034 &             0.009 &  0.014 &  0.023 &  0.012 \\
       & $\sigma_a=0.1$ &        0.013 &  0.018 &            0.003 &  0.012 &  0.001 &  0.005 & 0.000 &         0.006 &  0.025 &             0.011 &  0.013 &  0.023 &  0.011 \\
       & $a_{i,k}=1$ &        0.013 &  0.018 &            0.002 &  0.012 &  0.001 &  0.005 & 0.000 &         0.007 &  0.027 &               - &  0.013 &  0.023 &  0.012 \\
       & $a_{i,k}=1$, est. $\gamma_k$ &    0.014 &     0.015 &  0.002 &  0.014 &  0.007 &     0.006 &      0.030 &    0.005 &     0.015 &    - &  0.013 &  0.025 &     0.015 \\
       & $\sigma_a=0.25$, sm. &        0.017 &  0.024 &            0.005 &  0.010 &  0.001 &  0.005 & 0.000 &         0.012 &  0.039 &             0.002 &  0.015 &  0.022 &  0.013 \\
       & $\sigma_a=0.1$, sm. &        0.013 &  0.017 &            0.003 &  0.010 &  0.001 &  0.005 & 0.000 &         0.007 &  0.026 &             0.007 &  0.014 &  0.023 &  0.012 \\
       & $a_{i,k}=1$, sm. &        0.012 &  0.015 &            0.002 &  0.010 &  0.001 &  0.005 & 0.000 &         0.008 &  0.029 &               - &  0.014 &  0.024 &  0.012 \\
dcMIRT & $\sigma_a=0.25$ &        0.020 &  0.023 &              - &    - &    - &    - & - &         0.012 &  0.025 &               - &    - &    - &    - \\
       & $\sigma_a=0.1$ &        0.022 &  0.026 &              - &    - &    - &    - & - &         0.011 &  0.026 &               - &    - &    - &    - \\
       & $a_{i,k}=1$ &        0.022 &  0.027 &              - &    - &    - &    - & - &         0.010 &  0.025 &               - &    - &    - &    - \\
\bottomrule
\multicolumn{12}{l}{Smoothing is abbreviated as ``sm.''. Estimate $\gamma_k$ is abbreviated as ``est. $\gamma_k$''.}
\end{tabular}
\end{sidewaystable}




%




\end{document}